\documentclass[11pt]{article}
\usepackage{jheppub}
\usepackage{amsmath,amssymb,amsfonts,graphicx,slashed,color, mathtools, amsthm}

\newcommand{\be}{\begin{equation}}
\newcommand{\ee}{\end{equation}}
\newcommand{\bea}{\begin{eqnarray}}
\newcommand{\eea}{\end{eqnarray}}
\newcommand{\beas}{\begin{eqnarray*}}
\newcommand{\eeas}{\end{eqnarray*}}
\newcommand{\ba}{\begin{array}}
\newcommand{\ea}{\end{array}}

\newcommand{\sh}{\textnormal{sh}}
\newcommand{\ch}{\textnormal{ch}}
\newcommand{\gym}{g_{\textnormal{YM}}}

\title{Holographic and Localization Calculations \\ of Boundary F for ${\cal N}=4$ SUSY Yang-Mills Theory}

\author[]{Mark Van Raamsdonk,}
\author[]{Chris Waddell}

\affiliation[]{Department of Physics and Astronomy, University of British Columbia,\\
6224 Agricultural Road, Vancouver, B.C.\ V6T 1Z1, Canada.}

\emailAdd{mav@phas.ubc.ca, cwaddell@phas.ubc.ca}

\abstract{${\cal N} = 4$ Supersymmetric Yang-Mills (SYM) Theory can be defined on a half-space with a variety of boundary conditions preserving scale invariance and half of the original supersymmetry; more general theories with the same symmetry can be obtained by coupling to a 3D SCFT at the boundary. Each of these theories is characterized by a quantity called ``boundary $F$'', conjectured to decrease under boundary renormalization group flows. In this paper, we calculate boundary $F$ for $U(N)$ ${\cal N} = 4$ SYM theory with the most general half-supersymmetric boundary conditions arising from string theory constructions with D3-branes ending on collections of D5-branes and/or NS5-branes. We first perform the calculation holographically by evaluating the entanglement entropy for a half-ball centered on the boundary using the Ryu-Takayanagi formula in the dual type IIB supergravity solutions. For boundary conditions associated with D3-branes ending on D5 branes only or NS5-branes only, we also calculate boundary $F$ exactly by evaluating the hemisphere partition function using supersymmetric localization. The leading terms at large $N$ in the supergravity and localization results agree exactly as a function of the t' Hooft coupling $\lambda$.}


\begin{document}

\maketitle

\parskip=10pt

\section{Introduction}

Conformal field theories in various dimensions may be characterized by a parameter, sometimes known as $\tilde{F}$ or  ``generalized F,'' that characterizes the number of local degrees of freedom \cite{Zamolodchikov:1986gt,Jafferis:2011zi,Casini:2012ei, Cardy:1988cwa, Komargodski:2011vj, Giombi:2014xxa}. This is equal to the central charge $c$ for two-dimensional CFTs, and the Weyl-anomaly coefficient $a$ for four-dimensional CFTs. In general, $\tilde{F}$ may be defined from a regulator-independent term in the sphere free-energy, or alternatively from a universal term in the vacuum entanglement entropy for a ball-shaped region. The $\tilde{F}$ parameter is conjectured to decrease under renormalization group (RG) flows between conformal fixed points. This has been proven in two, three, and four dimensions as the $c$-theorem \cite{Zamolodchikov:1986gt}, $F$-theorem \cite{Jafferis:2011zi, Klebanov:2011gs, Casini:2012ei}, and $a$-theorem \cite{ Cardy:1988cwa, Komargodski:2011vj}, respectively.

A similar parameter, boundary $\tilde{F}$, may be defined for boundary conformal field theories (BCFTs) \cite{Affleck:1991tk,Friedan:2003yc,Jensen:2013lxa, Estes:2014hka}.\footnote{We recall that a BCFT is a local quantum field theory defined on a manifold with boundary such that the theory on a half-space preserves the conformal invariance of a CFT in one lower dimension (see e.g. \cite{Cardy:1984bb, McAvity:1995zd, Liendo:2012hy, Mazac:2018biw}). Each BCFT is associated with some bulk CFT which governs the short-distance behavior of local bulk correlators. Some BCFTs may be naturally understood by starting with this bulk CFT and choosing some boundary conditions for the fields. More generally, we can couple in (arbitrarily numerous) additional boundary degrees of freedom.} It can be understood as a measure of the number of local degrees of freedom associated with the boundary.\footnote{This quantity can be negative; in this case, we can understand the boundary condition as removing some of the bulk degrees of freedom near the boundary.} Boundary $\tilde{F}$ may be defined from the partition function of the BCFT on a hemisphere, or from the vacuum entanglement entropy of a half-ball centered on the boundary. It is conjectured to decrease under boundary RG-flows (where a UV BCFT is perturbed by a relevant boundary operator) \cite{Nozaki:2012qd, Estes:2014hka, Gaiotto:2014gha, Kobayashi:2018lil, Giombi:2020rmc}; this has been proven as the $g$-theorem in two dimensions \cite{Affleck:1991tk, Friedan:2003yc, Casini:2016fgb} and the $b$-theorem in three dimensions \cite{Jensen:2015swa, Casini:2018nym}, but remains a conjecture (the boundary $F$ theorem) for four-dimensional BCFTs. 

It is interesting to characterize the possible BCFTs that are associated with a particular bulk CFT, and specifically to understand which values of boundary $\tilde{F}$ are possible. This is understood for minimal model CFTs in two dimensions, but relatively few results are available for more complicated CFTs or CFTs in higher dimensions. The main goal of this paper is to investigate the possible values of boundary $F$ in a very special higher-dimensional example where we take the bulk CFT to be $U(N)$ ${\cal N}=4$ Supersymmetric Yang-Mills (SYM) theory and we constrain the BCFT to preserve half of the supersymmetry. 

\begin{figure}
\centering
\includegraphics[width=100mm]{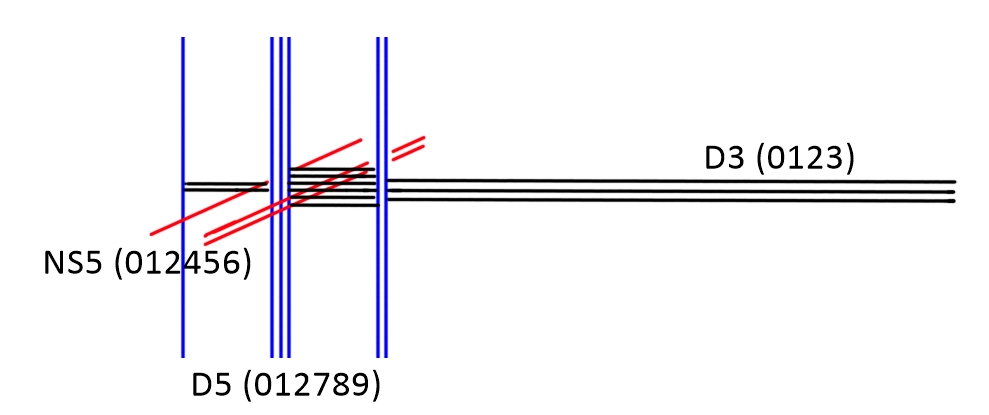}
\caption{D-brane construction of a half-supersymmetric BCFT whose bulk CFT is ${\cal N} = 4$ SYM theory.}
\label{fig:branes}
\end{figure}

This rich class of theories was classified by Gaiotto and Witten in \cite{Gaiotto:2008sa, Gaiotto:2008ak}. These theories preserve an $OSp(2,2|4)$ subgroup of the original 4D superconformal symmetry group $PSU(2,2|4)$; they are four-dimensional BCFTs with the maximum possible supersymmetry. Many of these theories describe the decoupled low-energy physics of coincident D3-branes in type IIB string theory ending in various ways on stacks of D5-branes and NS5-branes, possibly with additional D3-branes stretched between the fivebranes, as depicted in Figure \ref{fig:branes}. As for their ${\cal N}=4$ SYM parent, the associated half-supersymmetric BCFTs are holographic; their vacuum states are dual to solutions of type IIB supergravity preserving $SO(3) \times SO(3) \times SO(3,2)$ symmetry. These solutions were described in \cite{DHoker:2007zhm, DHoker:2007hhe, Aharony:2011yc,Assel:2011xz}. 

In this paper, we calculate boundary $F$ for general $OSp(2,2|4)$-symmetric BCFTs whose bulk CFT is $U(N)$ ${\cal N}=4$ SYM theory. First, we perform a holographic calculation, making use of the Ryu-Takayanagi formula to calculate the vacuum entanglement entropy for a half-ball. This was done in \cite{Estes:2014hka} for a particular type of boundary condition associated with $n k$ D3-branes ending on $k$ D5-branes;\footnote{Similar calculations were performed in \cite{Assel:2012cp} for 3D superconformal theories and in \cite{Estes:2018tnu} for 3D BCFTs.} we extend these calculations to the most general case, arising from the brane construction in Figure \ref{fig:branes} with arbitrary numbers and configurations of branes. The result is given as equation (\ref{SugraResult}) in Section 4.2. 

Next, we calculate boundary $F$ exactly by evaluating the hemisphere partition function using supersymmetric localization, for the class of boundary conditions arising from D3-branes ending on only D5-branes or only NS5-branes, in all possible ways.\footnote{Localization calculations of $F$ for related 3D SCFTs were performed in \cite{Assel:2012cp}, and
calculations of the interface entropy for supersymmetric Janus interfaces in 4D $\mathcal{N}=2$ SCFTs were performed in \cite{Goto:2020per}.} These results are given as equation (\ref{ExactLocalization}) for boundary conditions associated with NS5-branes and (\ref{ExactLocalizationD}) for boundary conditions associated with D5-branes.

We compare the localization results, which should be exact, to the supergravity calculations, which are expected to be valid at large $N$ and large 't Hooft coupling $\lambda$. The results agree precisely in a limit where a certain set of integers characterizing the theory (roughly, the number of D3-branes ending on each fivebrane in the string theory picture and the nonzero differences between these numbers) are large. Perhaps surprisingly, we find that this agreement holds exactly as a function of the 't Hooft coupling $\lambda$, suggesting a non-renormalization theorem governing the $\alpha'$ corrections in the string theory calculation.

Making use of our results, we analyze in Section 6 the distribution of possible values of boundary $F$ for various classes of boundary conditions. For the most general boundary conditions associated with D5-branes and NS5-branes, we can have arbitrarily large values of boundary $F$ for a given $N$ and $\lambda$, in accord with the fact that we can couple in an SCFT with an arbitrarily large number of degrees of freedom. For the theories associated with NS5-branes only or D5-branes only (which may be interpreted as boundary conditions for ${\cal N} = 4$ SYM theory without added degrees of freedom), we find that boundary $F$ is bounded, but can take positive or negative values. For boundary conditions associated with D5-branes only, we find that $F$ is typically negative at small 't Hooft coupling, consistent with the fact that these boundary conditions are associated with  scalar vevs that diverge near the boundary and give spatially dependent mass terms that effectively remove some of the bulk CFT degrees of freedom. For NS5-brane boundary conditions, we find that boundary $F$ is positive for small $\lambda$ but that an increasing proportion of these boundary conditions become negative as $\lambda$ grows.\footnote{For $\lambda > 4 \pi N$, we can make an S-duality transformation that maps a theory with NS5 boundary conditions to a theory with D5 boundary conditions and $\lambda < 4 \pi N$, so it is expected that the proportion of NS5 boundary conditions with negative boundary $F$ grows with $\lambda$; likewise, the proportion of D5 boundary conditions with positive boundary $F$ should grow with $\lambda$.} 

The plan for the rest of the paper is as follows. In Section 2, we review the definition of boundary $F$ and of the various BCFTs descended from ${\cal N}=4$ SYM theory that we consider. In Section 3, we review the solutions of type IIB supergravity dual to the vacuum states of these theories. In Section 4, we present our holographic calculation of boundary $F$, which we extract from the holographically-computed  vacuum entanglement entropy for a half-ball centered at the boundary. In Section 5, we present the localization calculation of boundary $F$ for the boundary conditions associated with NS5-branes and D5-branes and compare this with the supergavity results. In Section 6, we analyze the results to characterize the possible values of boundary $F$. We conclude in Section 7 with a discussion. Various technical aspects of our calculations are presented in appendices.

\section{Background}

In this section, we review some relevant background material on boundary $F$ and on half-supersymmetric BCFTs associated with the ${\cal N}=4$ SYM theory.

\subsection{Boundary Entropy and Boundary Free Energy}

In a $d$-dimensional CFT, the vacuum state entanglement entropy of a ball-shaped region of radius $R$ has the general UV divergence structure
\begin{equation}
    S [ B_{R}^{d-1} ] = a_{d-2} \left( R / \epsilon \right)^{d-2} + a_{d-4} \left( R / \epsilon \right)^{d-4} + \ldots + 
    \begin{cases}
        4 (-1)^{\frac{d-2}{2}} A \ln(R / \epsilon) & 2 \mid d \\
        (-1)^{\frac{d-1}{2}} F & 2 \nmid d
    \end{cases} \: ,
\end{equation}
where $\epsilon$ is a UV regulator. The coefficients $a_{i}$ are generally scheme-dependent, and arise from integration of local geometric quantities over the entangling surface, while the coefficients $A$ and $F$ are universal, i.e. independent of the regularization scheme. In particular, the quantity $A$ coincides with the A-type trace anomaly in even dimensions, while $F$ is the sphere free energy $F = - \ln Z[S^{d}]$; this equivalence is established by the relation
\begin{equation}
     S[B^{d-1}_{R}]_{\textnormal{univ}} = \ln Z[S^{d}_{R}]_{\textnormal{univ}}
\end{equation}
of Casini-Huerta-Myers for sphere entanglement entropy and the sphere partition function in CFT \cite{Casini:2011kv}. 
These universal terms are conjectured to be RG monotones in arbitrary dimension \cite{Myers:2010xs, Myers:2010tj, Giombi:2014xxa, Kawano:2014moa}; this has been proven in dimensions $d=2, 3$ and 4, with the results referred to as the (Zamolodchikov) $c$-theorem \cite{Zamolodchikov:1986gt}, the $F$-theorem \cite{Jafferis:2011zi, Klebanov:2011gs, Casini:2012ei}, and the $a$-theorem \cite{Cardy:1988cwa, Komargodski:2011vj} respectively. The conjectured extension to arbitrary dimension is sometimes referred to as the generalized $F$-theorem. 

\begin{figure}
\centering
\includegraphics[width=90mm]{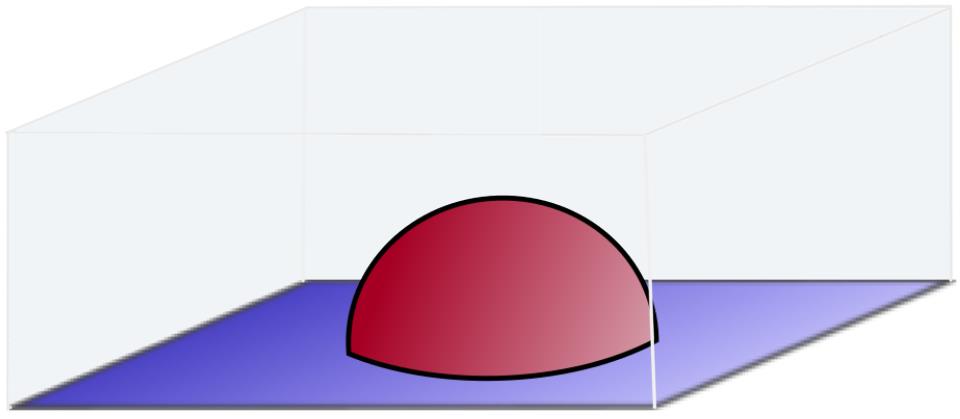}
\caption{Set-up for calculation of boundary $F$, showing the entangling surface for a half-ball region centred at the boundary for a BCFT on half of $\mathbb{R}^{3,1}$.}
\label{fig:halfball}
\end{figure}

In the BCFT case, we may instead consider the entanglement entropy of a half-ball region centred at the BCFT boundary (see Figure \ref{fig:halfball}). The entanglement entropy now has divergences of $d$-dimensional and $(d-1)$-dimensional origin, taking the form
\begin{equation}
    S[HB_{R}^{d-1}] = \tilde{a}_{d-2} \left( R / \epsilon \right)^{d-2} + \tilde{a}_{d-3} \left( R / \epsilon \right)^{d-3} + \ldots +
    \begin{cases}
        4 (-1)^{\frac{d-2}{2}} A \ln (R / \epsilon) + (-1)^{\frac{d-2}{2}} \tilde{F} & 2 \mid d \\
        4 (-1)^{\frac{d-3}{2}} \tilde{A} \ln (R / \epsilon ) + (-1)^{\frac{d-1}{2}} F & 2 \nmid d
    \end{cases} \: .
\end{equation}
The coefficient $\tilde{F}$ in this expression is not universal, insofar as the logarithmic term changes by a constant when we change regulators. However, by  analogy to the 2-dimensional case \cite{Calabrese:2009qy}, we may define the ``boundary entropy"
\begin{equation}
\label{defF1}
    S_{\partial}(R) \equiv S^{\textnormal{(BCFT)}}[HB_{R}^{d-1}] - \frac{1}{2} S^{\textnormal{(CFT)}}[B_{R}^{d-1}] \: ,
\end{equation}
where $S^{\textnormal{(CFT)}}$ denotes the entanglement entropy calculated in the ambient CFT for a region far from the boundary\footnote{In practice, $S^{\textnormal{(CFT)}}$ may be calculated in the theory without a boundary. For example, in our holographic calculation, we compute $S^{\textnormal{(CFT)}}$ using the RT formula in the AdS$_{5} \times S^{5}$ geometry. }.  
Given that the divergences with $d$-dimensional origin cancel in this subtraction, we recover boundary entropy of the form
\begin{equation}
    S_{\partial}(R) = \tilde{a}_{d-3} \left( R / \epsilon \right)^{d-3} + \tilde{a}_{d-5} \left( R / \epsilon \right)^{d-5} + \ldots + \begin{cases}
        (-1)^{\frac{d-2}{2}} \tilde{F} & 2 \mid d \\
        4 (-1)^{\frac{d-3}{2}} \tilde{A} \ln (R / \epsilon ) & 2 \nmid d
    \end{cases} \: .
\end{equation}
In particular, $\tilde{F}$ and $\tilde{A}$ are universal terms appearing in the expression for $S_{\partial}(R)$. The coefficient $\tilde{A}$ occurring for odd dimensions is related to the boundary Weyl anomaly in BCFT, using a similar argument to that of \cite{Casini:2011kv} (see also \cite{Kobayashi:2018lil, Fursaev:2016inw, Herzog:2015ioa, Casini:2018nym}). In general, as for the CFT case, the boundary entropy can be related to the logarithm of the partition function via
\begin{equation}
    S_{\partial}(R)_{\textnormal{univ}} = \left( \ln Z[HS_{R}^{d}] - \frac{1}{2} \ln Z[S_{R}^{d}] \right)_{\textnormal{univ}} \: .
\end{equation}
This quantity has also been conjectured to satisfy an RG monotonicity theorem in various dimensions \cite{Nozaki:2012qd, Estes:2014hka, Gaiotto:2014gha, Kobayashi:2018lil, Giombi:2020rmc} (see \cite{Yamaguchi:2002pa, Takayanagi:2011zk, Fujita:2011fp} for proposed holographic $g$-functions); this has been proven in dimensions $d=2$ and $d=3$, with the results referred to as the $g$-theorem \cite{Affleck:1991tk, Friedan:2003yc, Casini:2016fgb} and the $b$-theorem \cite{Jensen:2015swa, Casini:2018nym}.\footnote{In fact, the $b$-theorem establishes the monotonicity of the Weyl anomaly coefficient on a dimension-2 submanifold in arbitrary dimension.}

In this paper, we will be specifically concerned with the case $d=4$, where we have
\begin{equation}
\label{defF2}
    S_{\partial}(R) = S_{1} \frac{R}{\epsilon} + S_{\textnormal{univ}} \: . 
\end{equation}
Defining
\begin{equation}
\label{defF3}
    F_{\partial} \equiv - S_{\textnormal{univ}} \: ,
\end{equation}
the universal quantity $F_{\partial}$ appearing in the boundary entropy is referred to as ``boundary $F$" or the ``boundary free energy". The boundary free energy was conjectured to satisfy an RG monotonicity theorem in \cite{Estes:2014hka, Gaiotto:2014gha}. Note that we may extract $F_{\partial}$ from the boundary entropy by
\begin{equation}
    F_{\partial} =  \lim_{\epsilon \rightarrow 0} \left( R \frac{d}{dR} - 1 \right) S_{\partial}(R) \: .
\end{equation}
In the $d=4$ case, one finds exactly \cite{Herzog:2015ioa}
\begin{equation}
    F_{\partial} = - \lim_{\epsilon \rightarrow 0} \left( \ln Z[HS_{R}^{4}] - \frac{1}{2} \ln Z[S_{R}^{4}] \right) \: .
\end{equation}

\subsection{Half-Supersymmetric BCFTs from \texorpdfstring{${\cal N} = 4$}{} SYM} \label{sec:halfbps}

In this section, we review the boundary conformal field theories constructed from ${\cal N} = 4$ SYM that preserve half of the supersymmetry and an $OSp(2,2|4)$ subgroup of the superconformal symmetry group $PSU(2,2|4)$ of ${\cal N} = 4$ SYM. The classification of these theories is due to Gaiotto and Witten; see \cite{Gaiotto:2008sa, Gaiotto:2008ak} for details. Our conventions are similar to those of \cite{Wang:2020seq}. 

Starting with the 4-dimensional $\mathcal{N}=4$ SYM theory on $\mathbb{R}^{3,1}$, we can introduce a planar boundary at $x_{3}=0$, and consider boundary conditions preserving the subset of conformal transformations which leave this plane fixed.
Specifically, we are interested in half-BPS boundary conditions which preserve a $OSp(2, 2|4)$ superconformal subgroup of the initial superconformal group $PSU(2, 2|4)$. We will also consider the addition of extra degrees of freedom at this boundary such that the full theory preserves the same symmetry.

The bosonic sector of the residual symmetry group corresponds to 
\begin{equation}
    \mathfrak{so}(2, 3) \times \mathfrak{so}(3) \times \mathfrak{so}(3) \: .
\end{equation}
To reflect this reduction in R-symmetry, it is convenient to decompose the scalars $\Phi^{i}$ of the ${\cal N}=4$ theory as triples
\begin{equation}
    (X^{1}, X^{2}, X^{3}) \equiv (\Phi^{4}, \Phi^{5}, \Phi^{6}) \: , \qquad (Y^{1}, Y^{2}, Y^{3}) \equiv (\Phi^{7}, \Phi^{8}, \Phi^{9}) \: ,
\end{equation}
and the fermions as\footnote{Here, our notation reflects the fact that ${\cal N}=4$ SYM theory may be understood as the dimensional reduction of ten dimensional supersymmetric Yang-Mills theory. There exists a family of inequivalent $OSp(2, 2|4)$ subalgebras related by $U(1)$ outer-automorphisms of $\mathfrak{psu}(2, 2|4)$ \cite{Gaiotto:2008sa}, and we are choosing a particular one which preserves SUSY generators satisfying 
\begin{equation}
    \Gamma_{3456} \varepsilon = \varepsilon \: .
\end{equation}.}
\begin{equation}
    \Psi_{\pm} \equiv \frac{1}{2} \left( 1 \pm \Gamma_{3456} \right) \Psi \: .
\end{equation}
The 4-dimensional $\mathcal{N}=4$ vector multiplet decomposes with respect to the reduced symmetry group into two different multiplets, naturally interpreted from the perspective of the 3-dimensional $\mathcal{N}=4$ supersymmetry algebra as
\begin{equation}
    \textnormal{hyper} \: : \: \Psi_{-} \: , \: A_{3} \: , \: X^{i} \: , \qquad \textnormal{vector} \: : \: \Psi_{+} \: , \: A_{0, 1, 2} \: , \: Y^{i} \: .
\end{equation}
The various theories we consider arise from the low-energy physics of string theory configurations with D3-branes ending on and stretched between both D5-branes and NS5-branes. We consider first boundary conditions involving only D5-branes or only NS5-branes before considering the general case.

\subsubsection*{Single NS5-Brane Boundary Conditions}

For the boundary condition corresponding to D3-branes ending on a single NS5-brane in the 012789 directions, Neumann boundary conditions are imposed on the 3-dimensional vector multiplet and Dirichlet conditions on the hypermultiplet, i.e.
\begin{equation}
    \textnormal{NS5} \: : \qquad F_{3 \mu}| = X^{i}| = D_{3} Y^{i}| = 0 \: , \qquad \Psi_{-}| = 0 \: .
\end{equation}
Here, the vertical line denotes that the fields are evaluated at $x_{3} = 0$. 

\subsubsection*{D5-Brane Boundary Conditions}

For boundary conditions associated with the D3-branes ending on one or more D5-branes in the 012456 directions, we have a Dirichlet condition on the 3-dimensional vector multiplet and a (generalized) Neumann condition on the hypermultiplet,
\begin{equation}
    \textnormal{D5} \: : \qquad F_{\mu \nu}| = D_{3} X_{i}| - \frac{i}{2} \epsilon_{ijk} [X_{j}, X_{k}]| = Y_{i}| = 0 \: , \qquad \Psi_{+}| = 0 \: .
\end{equation}
This is a generalization of the Dirichlet boundary condition, sometimes referred to as a ``Nahm pole" boundary condition, since the scalar fields $X^{i}$ are seen to satisfy the Nahm equation in the vicinity of the boundary, with solution
\begin{equation}
    X^{i} = \frac{t^{i}}{x_{3}}   \: , \qquad [t^{i}, t^{j}] = i \epsilon^{ijk} t^{k} \: .
\end{equation}
Here $t^{i}$ can be $SU(2)$ generators in an arbitrary $N$-dimensional representation. Choosing the irreducible representation gives a boundary condition that corresponds to $N$ D3-branes along the (0123) directions ending on a single D5-brane. The non-commuting configuration of scalar matrices describe a non-commutative geometry corresponding to a string theory picture where the D3-branes flare out to form a ``fuzzy funnel" \cite{Constable:1999ac} as they approach the D5-brane.

Taking $t^i$ to correspond to a more general reducible representation of the $SU(2)$ with irreducible representations of size $p_i$ gives a boundary condition related to a more general brane configuration where groups of $p_i$ D3-branes each end on a single D5-brane.

\subsubsection*{General D5-NS5 Boundary Conditions}

\begin{figure}
\centering
\includegraphics[width=140mm]{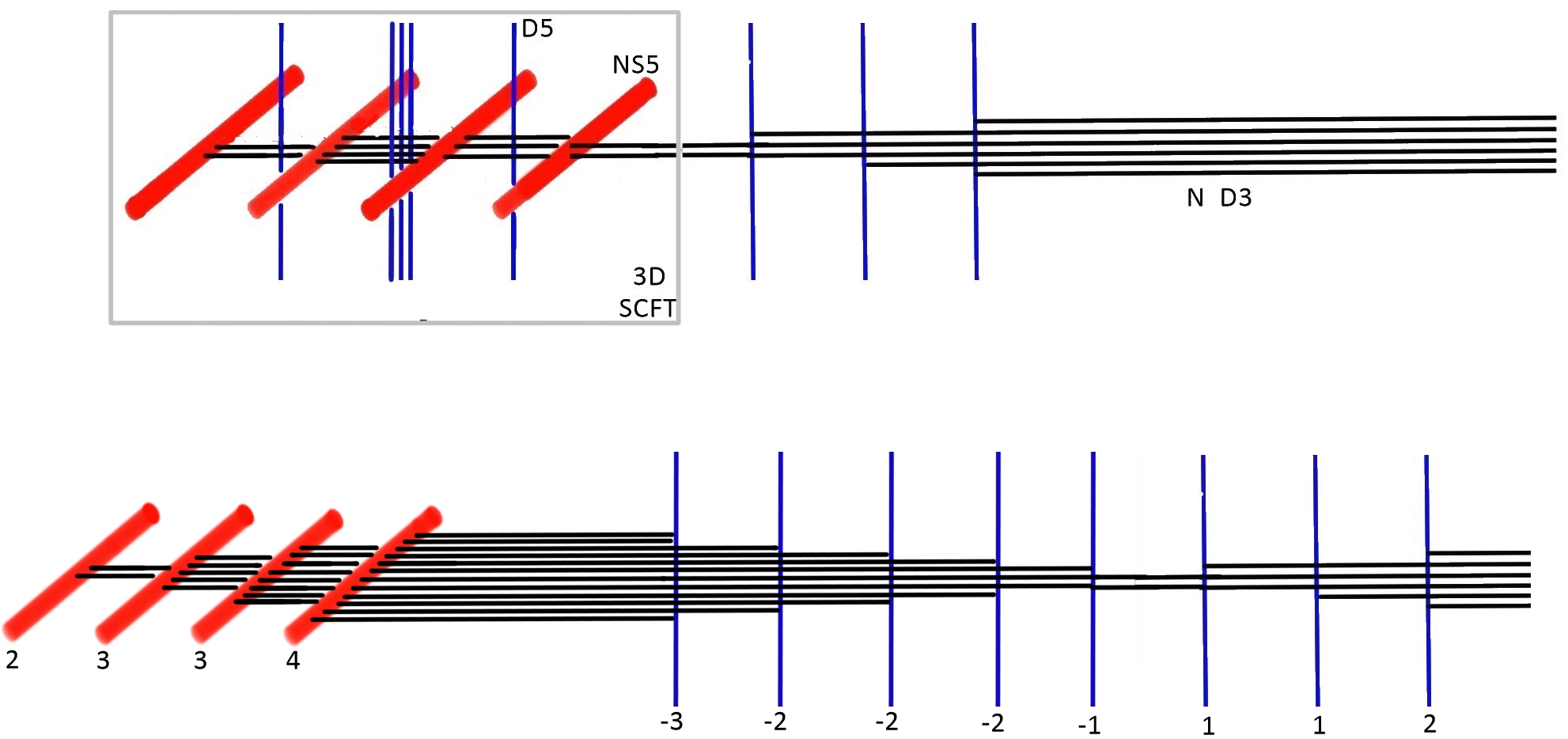}
\caption{\textbf{Top:} General brane configuration associated to a half-supersymmetric BCFT whose bulk CFT is ${\cal N} = 4$ SYM theory. For this configuration, we have $\vec{n} = (2,4,3,4)$ and $\vec{M} = (1,3,1)$. \textbf{Bottom:} the same configuration after a rearrangement of branes. Linking numbers $K_i$ and $\tilde{L}_i$ for the fivebranes are shown. (Apologies to M.C. Escher.)}
\label{fig:braneconfig}
\end{figure}

We now describe the more general theories that arise from configurations with both D5-branes and NS5-branes. It is convenient to consider first $N_{D5}$ D5-branes and $N_{NS5}$ NS5-branes at distinct locations in the $x^3$ direction, with the D5s stretched along the 012456 directions and the NS5s stretched along the 012789 directions. Next, we consider $N$ semi-infinite D3-branes stretched in the 0123 directions, extending to $x^3 = \infty$, each ending on some fivebrane. Finally, we can have additional D3-branes of finite extent in $x^3$ stretched between some of the fivebranes. An example is shown in Figure \ref{fig:braneconfig}.

As explained in \cite{Gaiotto:2008ak}, the low-energy physics of such configurations does not depend on the specific positions of the fivebranes along the $x^3$ direction, and is even unchanged if we rearrange the fivebranes relative to one another, taking into account the fact that when a D5-brane is moved past an NS5-brane towards the direction of larger $x^3$, we create and additional D3-brane stretched between the D5 and NS5 \cite{Hanany:1996ie}. We consider brane configurations related by such rearrangements as being part of an equivalence class.

The distinct IR superconformal BCFTs that can arise from these brane configurations are in one-to-one correspondence with equivalence classes that obey certain additional constraints \cite{Gaiotto:2008ak}.\footnote{Configurations which do not obey the constraints may fail to have a supersymmetric vacuum or may give rise to theories which factorize into a superconformal BCFT and some other 3D SCFT.} The distinct theories satisfying the constraints may be represented by brane configurations of the type shown in Figure \ref{fig:braneconfig}, where we have $n_i$ D3-branes immediately to the right of the $i$th NS5-brane counted from the left, and $M_i$ D5-branes that intersect these, with the constraint that
\be
\label{Mconstr}
M_i \ge 2 n_i - n_{i+1} - n_{i-1} \qquad i=1 \dots N_{NS5} - 1
\ee
taking $n_0 = 0$. Additional D5-branes sit to the right of all NS5-branes, and we have a constraint that the net number of D3-branes ending on each D5-brane from the right (i.e. the number on the right minus the number on the left) increases from left to right. 

The constraints (\ref{Mconstr}) are equivalent to the requirement that by moving all D5-branes to the right of all NS5-branes (while preserving their order) as in Figure \ref{fig:braneconfig} (bottom), the net number $K_i$ of D3-branes ending from the right on the $i$th NS5-brane (starting from the left) is positive and non-decreasing with $i$. By construction, the net number $\tilde{L}_i$ of D3-branes ending from the right on the $i$th D5-brane (starting from the left) is also non-decreasing with $i$, and satisfies $\tilde{L}_i > - N_{NS5}$. The quantities $L_i = \tilde{L}_i +  N_{NS5}$ are then positive and increasing with $i$; the action of S-duality simply exchanges $\{K_i\} \leftrightarrow \{L_i\}$. The parameters $K_i$ and $\tilde{L}_i$ (or alternatively $L_i$), known as ``linking numbers,'' are closely related to the parameters appearing in the dual supergravity solutions.\footnote{Here, the parameters $(L_i,K_i)$ were introduced in \cite{Gaiotto:2008ak} while the alternative $(\tilde{L}_i,K_i)$ were used in \cite{Aharony:2011yc}.}

We can read off the linking numbers without reordering the branes by defining $K_i$ in general to be the net number of D3-branes ending on the $i$th NS5-brane from the right plus the total number of D5-branes to the left of this NS5, and defining $\tilde{L}_i$ to be the net number of D3-branes ending on the $i$th D5-brane from the right minus the total number of NS5-branes to the right of this D5. With this definition, we can check that the linking numbers do not change as we move a D5-brane past an NS5-brane. It follows that the NS5-brane linking numbers $K_i$ can be expressed in terms of $M_i$ and $n_i$ as
\be
K_i = n_i - n_{i-1} + \sum_{j=1}^{i-1} M_j \; .
\ee
Conversely, we have that $M_i$ is the number of D5-branes with linking number $\tilde{L} = i - N_{NS5}$ while 
\be
\label{defN}
n_j = \sum_{i=1}^j (K_i  + (i - j) M_i) \; ,
\ee
so the requirement that $n_j$ should be positive may be expressed as a constraint on the linking numbers.

It will also be useful to note that the rank of the gauge group for our ${\cal N} =4$ SYM theory is related to the linking numbers by
\be
N = \sum_i K_i + \sum_i \tilde{L}_i \: .
\ee

We can understand the field theory corresponding to such brane configurations as follows \cite{Gaiotto:2008ak}. The semi-infinite D3-branes give rise to the bulk ${\cal N}=4$ theory. Some subset of these end on D5-branes, so we have D5-brane boundary conditions as above for a subset of fields. These break the gauge symmetry from $U(N)$ to some subgroup $U(n)$ where $n \equiv n_{N_{NS5}}$ corresponds to the number of D3-branes intersecting the rightmost NS5-brane. The simplest situation is where these $n$ D3-branes simply end on a single NS5-brane with no additional branes to the left. This defines some particular BCFT with unbroken $U(n)$ gauge symmetry. The more general theories can be understood as coupling this theory to a 3D SCFT with global $U(n)$ symmetry, arising from the low energy dynamics of the brane configuration between the leftmost and rightmost NS5-brane.  

The 3D superconformal theories that are coupled at the boundary arise from the IR limit of certain 3-dimensional linear quiver gauge theories \cite{Gaiotto:2008ak, Assel:2011xz}, where we have gauge group $U(n_1) \times \cdots \times U(n_{N_{NS5}-1})$, and
\begin{itemize}
    \item One 3d $\mathcal{N}=4$ gauge multiplet for each gauge group factor $U(n_{i})$ (coming from strings that start and end on D3-branes stretched between NS5s)
    \item One 3d $\mathcal{N}=4$ bifundamental hypermultiplet for each neighboring pair of gauge group factors $U(n_{i}) \times U(n_{i+1})$ (coming from strings that begin and end on D3-branes on either side of an NS5-brane)
    \item One 3d $\mathcal{N}=4$ fundamental hypermultiplet for each D5-brane between NS5-branes (coming from 3-5 strings)
    \item An additional $n$ 3d $\mathcal{N}=4$ hypermultiplets in the fundamental of $U(n_{N_{NS5}-1})$.
\end{itemize}
We have a global symmetry $\left\{ \prod_i U(M_i) \right\} \times U(n)$ under which the various fundamental hypermultiplets transform. See \cite{Gaiotto:2008ak} for additional details.

\section{Dual Gravity Solutions} \label{sec:solutions}

Through the AdS/CFT correspondence, the vacuum states of the $OSp(2, 2|4)$-symmetric BCFTs descending from $U(N)$ ${\cal N} = 4$ SYM theory correspond to $OSp(2, 2|4)$-symmetric solutions of type IIB supergravity. The general local solutions with this symmetry were constructed by D'Hoker, Estes, and Gutperle in \cite{DHoker:2007zhm, DHoker:2007hhe} by solving the BPS equations. The $SO(3,2) \times SO(3) \times SO(3)$ global symmetry is reflected in the fact that the solutions are 
\begin{equation}
    \textnormal{AdS}_{4} \times S_{1}^{2} \times S_{2}^{2}  \: ,
\end{equation}
fibered over a Riemann surface $\Sigma$. Such solutions turn out to be uniquely characterized by specifying a pair of harmonic functions $h_{1}, h_{2}$ on $\Sigma$. The requirement that the solutions are non-singular imposes the extra constraint that the poles of $h_i$ lie on the boundary of $\Sigma$, and flux-quantization conditions place additional constraints on the locations of these poles. Ultimately, the harmonic functions $h_i$ and thus the entire supergravity solutions are determined by the locations of the poles. 

This set of solutions includes geometries dual to the BCFTs we are interested in, but also geometries dual to ${\cal N} = 4$ SYM theories with planar codimension-one defects or interfaces between ${\cal N} = 4$ SYM theories with different parameters. Those solutions corresponding to the BCFT case were specifically analyzed in \cite{Aharony:2011yc} (see also \cite{Assel:2011xz}).

\subsection{General Local Solution} \label{sec:localsol}

We now review explicitly the solutions of \cite{DHoker:2007zhm, DHoker:2007hhe, Aharony:2011yc}; our conventions for type IIB string theory parameters and their relation to ${\cal N}=4$ SYM theory parameters are summarized in Appendix \ref{app:conventions}. 

To describe the solutions, we take $\Sigma$ to be the first quadrant of the plane, with complex coordinate $w = r e^{i \theta} = x + i y$ and metric 
\begin{equation}
    ds_{\Sigma}^{2} = 4 \rho^{2} |dw|^{2} \; .
\end{equation}
The solutions are expressed in terms of harmonic functions $h_1,h_2$ on $\Sigma$. 

The full metric for the ten-dimensional solution takes the form
\begin{equation}
    ds^{2} = f_{4}^{2} ds_{\textnormal{AdS}_{4}}^{2} + f_{1}^{2} ds_{S_{1}^{2}}^{2} + f_{2}^{2} ds_{S_{2}^{2}}^{2} + ds_{\Sigma}^{2} \: ,
\end{equation}
where $f_{1}, f_{2}, f_{4}$ are real-valued functions on $\Sigma$, and $ds_{\textnormal{AdS}_{4}}^{2}$ and $ds_{S_{i}^{2}}^{2}$ are metrics for $AdS_4$ and two-spheres with unit radius. 

The metric functions and dilaton field can be expressed via a set of real functions
\begin{equation} \label{eq:WN1N2}
  W \equiv \partial_{w} h_{1} \partial_{\bar{w}} h_{2} + \partial_{w} h_{2} \partial_{\bar{w}} h_{1} \:  \qquad \qquad 
  N_{i} \equiv 2 h_{1} h_{2} | \partial_{w} h_{i}|^{2} - h_{i}^{2} W  \qquad (i=1,2) 
\end{equation}
in terms of which the dilaton is 
\begin{equation} \label{eq:dilaton}
    e^{2 \Phi} = e^{4 \phi} = \frac{N_{2}}{N_{1}} \: ,
\end{equation}
and the Einstein frame metric factors are
\begin{equation} \label{eq:metricfunctions}
    \rho^{2} = e^{- \frac{\Phi}{2}} \frac{\sqrt{ - N_{2} W}}{h_{1} h_{2}} \: , \quad f_{1}^{2} = 2 e^{\frac{\Phi}{2}} h_{1}^{2} \sqrt{ - \frac{W}{N_{1}}} \: , \quad f_{2}^{2} = 2 e^{- \frac{\Phi}{2}} h_{2}^{2} \sqrt{ - \frac{W}{N_{2}}} \: , \quad f_{4}^{2} = 2 e^{- \frac{\Phi}{2}} \sqrt{ - \frac{N_{2}}{W}} \: .
\end{equation}
The solutions also have a non-trivial NS/NS three-form field strengths and RR three-form and five-form field strengths. We do not need these for our analysis, but review them in Appendix \ref{app:formfields} for completeness.

\subsection{Supergravity Solutions: \texorpdfstring{$\textnormal{AdS}_{5} \times S^{5}$}{}} \label{sec:ads5}

It is useful to begin by describing the solution corresponding to AdS$_{5} \times S^5$. Making use of polar coordinates on $\Sigma$, we have
\bea
\label{h12AdS}
h_1 = \frac{L^2}{4} \frac{1}{\sqrt{g}}  \cos \theta \left( \frac{r}{r_{0}} + \frac{r_{0}}{r} \right) \: , \qquad h_2 = \frac{L^2}{4} \sqrt{g}  \sin \theta \left( \frac{r}{r_{0}} + \frac{r_{0}}{r} \right)
\eea
where $g$ is the string coupling. Using
\be
\partial_{w} \partial_{\bar{w}}f = {1 \over 4}\left[ {1 \over r} \partial_r (r \partial_r f) + {1 \over r^2} \partial_\theta^2 f\right] \; ,
\ee
we find that
\bea
W &=& -{L^4 \over 16 r^2} \sin(2 \theta) \: , \qquad
{1 \over g^2} N_2 = g^2 N_1 = {L^8 \over 1024 r_0^4} \sin(2 \theta) {(r^2 + r_0^2)^4 \over r^6} \; .
\eea
This gives a constant dilaton $e^{2 \phi} = e^{\Phi} = g$ and a metric
\be
ds^2 = L^2 \left\{ \left[d\theta^2 + \sin^2(\theta) d \Omega_2^2 + \cos^2(\theta) d \Omega_2^2\right] + \left[ {dr^2 \over r^2} + {(r^2 + r_0^2)^2 \over 4 r_0^2 r^2}({1 \over u^2}(du^2 -dt^2 + d \vec{x}^{2}))\right]  \right\} \; .
\ee
The first term in square brackets is the metric of a unit five-sphere while the second term in square brackets is the metric for AdS$_{5}$ with $\ell_{AdS} = 1$; the latter can be checked by the change of coordinates
\be
\label{coordTform}
z = u {2 r r_0 \over r^2 + r_0^2} \qquad \qquad x_{\perp} = u { r^2 - r_0^2 \over r_0^2 + r^2} \: ,
\ee
after which this factor becomes
\be
{1 \over z^2} (dz^2 + dx_{\perp}^2 - dt^2 + d \vec{x}^2) \: .
\ee
We see that the parameter $L$ corresponds to the $AdS$ radius in Einstein frame, the parameter $g$ corresponds to the string coupling, and the parameter $r_0$ is only associated with our choice of coordinates, with $r = r_0$ corresponding to the plane $x_{\perp}=0$ in Fefferman-Graham coordinates.

\subsection{Supergravity Solutions: General BCFT Solutions}

The general solution we consider may be expressed most simply using Cartesian coordinates on the first quadrant as\footnote{Here, we assume that $l_A$ and $k_B$ are distinct. Alternatively, we could omit the coefficient $c_A/\sqrt{g}$  and $d_B \sqrt{g}$ (which we will see are quantized in string theory solutions) and allow specific $l_A$s to appear with some multiplicity. The solutions described in \cite{Aharony:2011yc} have set $g=1$; we have used the symmetry $\phi \to \phi + \phi_0$, $B_2 \to e^{\phi_0} B_2$, $C_2 \to e^{- \phi_0}C_2$ to write the solution for general asymptotic string coupling $g = e^{\phi_\infty}$.}
\bea
\label{gensol}
h_1 &=&{\pi \ell_s^2 \over 2} {x \over \sqrt{g}} + {\ell_s^2 \over 4}\sum_A {c_A \over  \sqrt{g}} \ln \left( {(x + l_A)^2 + y^2 \over (x-l_A)^2 + y^2} \right) \cr
h_2 &=& {\pi \ell_s^2 \over 2} \sqrt{g} y + {\ell_s^2 \over 4}\sum_B d_B \sqrt{g} \ln \left( {x^2 + (y + k_B)^2 \over x^2 + (y-k_B)^2} \right) \; .
\eea
We see that $l_A$ give the location of poles of $h_1$ on the $x$ axis, while $k_A$ give the location of poles of $h_2$ on the $y$-axis.

Near $r=\infty$, these functions asymptote to
\begin{equation}
\begin{split}
    h_{1} & =  \frac{\ell_{s}^{2}}{\sqrt{g}} \left(  \frac{\pi}{2} r + \frac{1}{r} \sum_{A} c_{A} l_{A} \right) \cos \theta + O(r^{-2}) \\
    h_{2} & = \ell_{s}^{2} \sqrt{g} \left(  \frac{\pi}{2} r + \frac{1}{r} \sum_{B} d_{B} k_{B} \right) \sin \theta + O(r^{-2}) \: .
\end{split} \label{eq:h1h2asympt}
\end{equation}
Using these asymptotic expressions in the general equations for the metric and dilaton, we find that the asymptotic metric is AdS$_{5} \times S^{5}$, with Einstein frame AdS length
\be
L^4 = 4 \pi \ell_s^4 (\sum_A c_A l_A + \sum_B d_B k_B) 
\ee
and asymptotic dilaton $e^\phi = g$. In the asymptotic AdS$_{5} \times S^5$ region, our coordinate choice here matches with the coordinates of the previous section if we choose 
\be
\label{defr0}
r_0 = {L^2 \over 2 \pi \ell_s^2} \: .
\ee
From (\ref{LNE}), the rank of the gauge group is related to the parameters in the solution by
\be
N = \sum_A c_A l_A + \sum_B d_B k_B \; . 
\ee

\begin{figure}
\centering
\includegraphics[width=140mm]{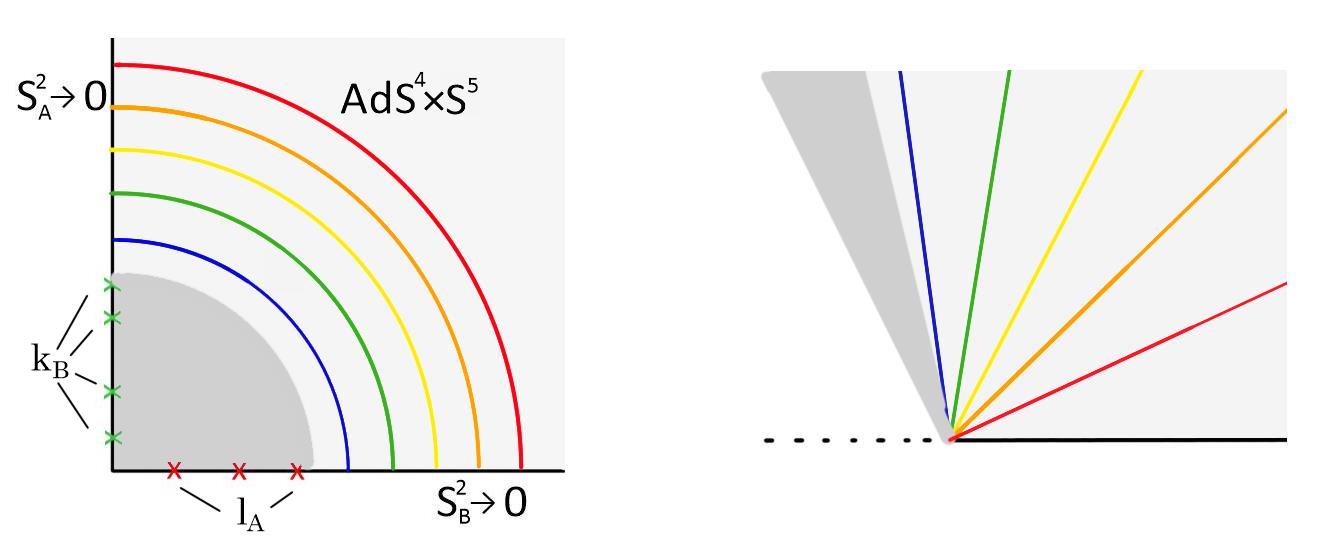}
\caption{\textbf{Left:} The dual geometries are AdS$_{4} \times S^2 \times S^2$ fibered over the quadrant shown, with the first and second  $S^2$s contracting to zero on the $y$ and $x$ axes respectively. \textbf{Right:} The geometries can be understood as corresponding to a portion of Poincar\'e AdS$_{5} \times S^5$ with Poincar\'e angle $\theta > \theta_*$, capped off by an ``end-of-the-world'' brane (shaded grey region) where the internal space degenerates smoothly. Arcs for large $r$ correspond to $AdS_4 \times S^5$ slices of the $AdS^5 \times S^5$ region.}
\label{fig:asymptotic}
\end{figure}

As shown in Figure \ref{fig:asymptotic}, the large $r$ part of the geometry (where $r$ is the radial coordinate on the quadrant) corresponds to a portion of Poincar\'e AdS$_{5} \times S^5$ with Poincar\'e angle near $\pi/2$. From (\ref{coordTform}) we have that the Poincar\'e angle is related to $r$ by
\be
\tan \theta = {1 \over 2} \left({r \over r_0} - {r_0 \over r} \right) \: .
\ee
The small $r$ region corresponds to an ``end-of-the-world'' brane in the full geometry where the internal space degenerates smoothly, apart from D5-brane throats associated with the singularities at $x = l_A, y = 0$ and NS5-brane throats associated with the singularities at $y=k_A, x=0$.

Using (\ref{Nd}) and the result (4.14) from \cite{Aharony:2011yc} for the flux integral, we find that the number of units of D5-brane flux associated to the singularity at $l_A$ is
\be
N_{D5}^{(A)} = {1 \over \sqrt{g}} c_A \: .
\ee
Similarly, from (\ref{Nn}) and the result (4.13) from \cite{Aharony:2011yc}, we have that the number of units of NS5-brane flux associated with the singularity at $k_B$ is
\be
N_{NS5}^{(B)} = \sqrt{g} d_B \: .
\ee
By analyzing the five-form fluxes in the solution, \cite{Aharony:2011yc} determined that the number of units of five-form flux (the flux associated with D3-branes) {\it per fivebrane} coming from the D5-branes in the $A$th stack and the NS5-branes in the $B$th stack are
\bea
\label{defKL}
N_{D3}^A &=& \hat{l}_A - {2 \over \pi} \sum_{B} N_{NS5}^{(B)} \arctan\left(g{\hat{k}_B \over \hat{l}_A}\right) \; .  \cr
N_{D3}^B &=& \hat{k}_B + {2 \over \pi} \sum_{A} N_{D5}^{(A)} \arctan\left(g{\hat{k}_B \over \hat{l}_A}\right)
\eea
where we have defined $\hat{k}_B = k_B/\sqrt{g}$ and $\hat{l}_A = \sqrt{g} l_A$.

In string theory, $N_{D3}^A$ and $N_{D3}^B$ should be quantized, so while we have a supergravity solution for any choice of $\left\{ l_A \right\}$ and $\left\{ k_A \right\}$, the allowed values corresponding to string theory solutions are discrete. We see that for small $g$, the parameters $\hat{l}_A$ and $\hat{k}_B$ should be integers up to small corrections.

\subsubsection{Relating Supergravity Parameters and Gauge Theory Parameters}
\label{sec:ftsugra}

As pointed out in \cite{Aharony:2011yc}, it is natural to identify the numbers on the left in (\ref{defKL}) with the linking numbers  that specify the BCFT,\footnote{Recall that these corresponded to the number of D3-branes ending on each fivebrane in the bottom picture of Figure \ref{fig:braneconfig}.} where we have
\bea
\{ \tilde{L}_i\} &=& \{N_{D3}^A \; \textnormal{with  multiplicity} \; N_{D5}^{(A)} \}\cr
\{ K_i\} &=& \{N_{D3}^B \; \textnormal{with multiplicity} \; N_{NS5}^{(B)}\} \: .
\eea
Alternatively, we can take $l_A$ with multiplicity $N_{D5}^{(A)}$ and $k_B$ with multiplicity $N_{NS5}^{(B)}$ in the original definition of $h_i$, setting $c_A/\sqrt{g} = d_B \sqrt{g}$=1. In this case, we find the original linking numbers $(L_A,K_A)$ of Gaiotto and Witten can be related simply to the supergravity parameters as
\bea
\label{defKL1}
L_A &=& \sqrt{g} l_A + {2 \over \pi} \sum_{B} \arctan{l_A \over k_B} \; . \cr
K_B &=& {k_B \over \sqrt{g}} + {2 \over \pi} \sum_{A}  \arctan{k_B \over l_A} \: .
\eea
The sum in the first expression has a geometric interpretation as the acute angle between the $x$ axis and the line segment from $(l_A,0)$ to $(0,k_B)$, summed over $k_B$, while the sum in the second expression is the acute angle between the $y$ axis and the segment from $(0,k_B)$ to $(l_A,0)$, summed over $l_A$. We note that (\ref{defKL1})  are invariant under the S-duality transformations $\{L_A\} \leftrightarrow \{K_B\}$, $\{l_A\} \leftrightarrow \{k_B\}$, $g \leftrightarrow {1/g}$.

In order to find the supergravity solution corresponding to the vacuum state of a particular BCFT defined by linking numbers $\vec{\tilde{L}}$ and $\vec{K}$, we need to use (\ref{defKL}) to solve for the parameters $\{\hat{l}_A\}$ and $\{\hat{k}_B\}$, though it is not clear how to do this explicitly in general. An interesting check is that for any linking numbers that can be expressed in terms of supergravity parameters as in \ref{defKL}, the field theory constraint that the quantities (\ref{defN}) must be positive (so that the brane configuration can be represented as in Figure \ref{fig:braneconfig} (Top)) is automatically satisfied, as we show in Appendix \ref{sec:proofineq}.  

We note that the final terms in the two equations in (\ref{defKL}) are bounded in magnitude by the total number of NS5-branes and D5-branes respectively. Thus, when the linking numbers (D3-branes per fivebrane) are all large compared with the total number of fivebranes, the solution will have $\hat{k}_B \sim K_B$ and $\hat{l}_A \sim L_A$, and we can find the corrections to these leading order expressions perturbatively in $1/K$ and/or $1/L$. Similarly, when the asymptotic string coupling $g$ is taken small with fixed linking numbers, we will have $\hat{k}_B \sim K_B + {\cal O}(g)$ and $\hat{l}_A \sim L_A + {\cal O}(g)$.

\section{Holographic Computation of Boundary \texorpdfstring{$F$}{}} \label{sec:holo}

In this section, we perform a holographic computation of boundary $F$ for the general BCFTs defined by a set of linking numbers $\{K_i, \tilde{L}_i\}$. This was done for the special case of $N$ D3-branes ending on $k$ D5-branes (linking numbers $K_i = 0, L_i = N/k$ with multiplicity $k$) in \cite{Estes:2014hka}; similar calculations of $F$ in 3D superconformal theories were performed in \cite{Assel:2012cp}. 

As we have described earlier, boundary $F$ may be computed either by evaluating the partition function for the theory on hemisphere, or by calculating the vacuum entanglement entropy for a half-ball centered on the boundary. Either of these may be computed holographically using the dual gravity solutions; the two calculations give rise to the same final expression for boundary $F$ in terms of the harmonic functions $h_1$ and $h_2$. In our presentation, we will holographically calculate the entanglement entropy, using the Ryu-Takayanagi formula \cite{Ryu:2006bv, Hubeny:2007xt}
\be
S(A) = {{\rm Area}(\tilde{A}) \over 4 G} \; ,
\ee
where $\tilde{A}$ is the minimal area codimension-two extremal surface homologous to the half-ball region on the boundary of AdS, computed using the Einstein-frame metric. The boundary $F$ is then extracted by subtracting off half of the entanglement entropy for a ball-shaped region in ${\cal N}=4$ SYM and keeping the universal piece, as in equations (\ref{defF1},\ref{defF2},\ref{defF3}).

In the ten-dimensional geometry, the extremal surface we need to consider is codimension two in the full spacetime. It wraps both of the internal $S^2$s, and the directions spanned by the Riemann surface $\Sigma$, so that the surface is specified by describing a codimension-two locus in each AdS$_{4}$ slice. It turns out that the appropriate extremal surface to compute the entanglement entropy of a half-ball region of radius $R$ centred at the BCFT boundary is just the one described by the hemisphere $\{ t = t_{0}, u^{2} + \vec{x}^{2} = R^{2}, u > 0\}$ in each AdS$_{4}$ slice, which we recall had metric
\begin{equation}
    ds_{AdS_{4}}^{2} = \frac{1}{u^{2}} \left( du^{2} - dt^{2} + d \vec{x}^{2} \right) \: , \qquad \vec{x} = (x_{1}, x_{2}) \: .
\end{equation}
Indeed, one can verify that the surface $u^{2} + \vec{x}^{2} = R^{2}$ is extremal in AdS from the Euler-Lagrange equations; this holds in any dimension, provided we let $\vec{x}$ denote the $d-2$ transverse coordinates. Moreover, in the boundary coordinates $(t, \vec{x}, x_{\perp})$ of the half-space $H\mathbb{R}^{3, 1}$, our extremal surface asymptotes to the entangling surface $\{t = t_{0}, x_{\perp}^{2} + \vec{x}^{2} = R^{2}, x_{\perp} < 0\}$. 

The area of the extremal surface diverges as usual, but we will regulate this by placing a cutoff at $Z=\epsilon$ in Fefferman-Graham coordinates. Subtracting off half the area of the RT surface for a ball of radius $R$ in ${\cal N}=4$ SYM theory with the same regulator, we will obtain a result that is finite in the limit $\epsilon \to 0$.

\subsubsection*{Regulated Areas}

Representing the $AdS_4$ metric as
\be
ds_{AdS^4}^2 = {1 \over u^2} (-dt^2 + du^2 + dx_\perp^2) = {1 \over \rho^2 \cos^2\theta_P} (-dt^2 + d\rho^2 + \rho^2 d\theta_P^2 + \rho^2 \sin^2 \theta_P d \phi^2)
\ee
we have that the extremal surface is at $\rho = R$ and fixed $t$. The eight-dimensional area of this surface is
\bea
{\rm Area} &=& 64 \pi^2 \int r dr d \theta d \phi \sin \theta_P d \theta_P \rho^2 f_1^2 f_2^2 f_4^2 {1 \over \cos^2 \theta_P}
\eea
where the regulator $Z= \epsilon$ in Fefferman-Graham coordinates corresponds to a restriction $\theta_P < \theta_P^\epsilon(r,\theta)$. The regularization procedure is described in detail in Appendix \ref{app:regularization}. After subtracting off the regulated area of the RT surface for a ball of radius $R$ in ${\cal N}=4$ SYM theory and removing the regulator, we find from the definitions (\ref{defF1},\ref{defF2},\ref{defF3}) that 
\bea
F_{\partial} &=& -\lim_{\Lambda \to \infty} {256 \pi^3 \over G} \Bigg[ \int_0^\Lambda  dr \int_0^{\pi \over 2} d \theta r h_1 h_2 \partial_w \partial_{\bar{w}}(h_1 h_2) \cr
&\qquad& \qquad \qquad \qquad - \int_{r_0}^\Lambda  dr \int_0^{\pi \over 2} d \theta r h^{AdS}_1 h^{AdS}_2 \partial_w \partial_{\bar{w}}(h^{AdS}_1 h^{AdS}_2) \Bigg]   \; .
\label{Fintegral}
\eea
where $h_i^{AdS}$ are the harmonic functions corresponding to pure $AdS^5 \times S^5$. We can easily evaluate the second term explicitly using the explicit expressions in Section \ref{sec:ads5}, to give
\bea
\label{Fintegral2}
F_{\partial} = \lim_{\Lambda \to \infty} \left\{-{256 \pi^3 \over G} \int_0^\Lambda  dr \int_0^{\pi \over 2} d \theta r h_1 h_2 \partial_w \partial_{\bar{w}}(h_1 h_2)  - {\pi \over 8} N \Lambda^2 - {1 \over 4} N^2  \ln \left({\Lambda^2 \pi \over N} \right) \right\}
\eea
where we have used (\ref{LNE}). Alternatively, we can combine the integrands to obtain a convergent integral,
\be
\label{Fintegral3}
F_{\partial} = -{ \pi^3 \over G} \int_\Sigma d^2 x \{ h_1 h_2 \partial_w \partial_{\bar{w}}(h_1 h_2) - {\cal J}_0 \}
\ee
where (recalling the definition of $r_0$ in (\ref{defr0}))
\be
{\cal J}_0 = \begin{cases} 
0 & r < r_0  \cr
h^{AdS}_1 h^{AdS}_2 \partial_w \partial_{\bar{w}}(h^{AdS}_1 h^{AdS}_2) = -{L^8 \sin^2(2 \theta) (r^2 + r_0^2)^2 \over 512 r_0^2 r^4} & r \ge r_0 \end{cases} \: .
\ee

\subsection{Boundary Free Energy: The Integral}

In this section, we will evaluate the integral (\ref{Fintegral2}) for the general solution (\ref{gensol}) in order to calculate the boundary free energy $F_{\partial}$ in the supergravity approximation. We note that the metric, expressed in terms of the parameters $(c_{A},d_{A},l_{A},k_{A})$, does not depend on the string coupling $g$, so we can work with $g=1$. However, when expressing the results in terms of the natural field theory parameters, some $g$ dependence will appear. 

In terms of the parameters $c_A, d_A, l_A$, and $k_A$, we have 
\be
F_{\partial}(c_A, d_A, l_A, k_A) = -2 \pi \lim_{\Lambda \to \infty}\left\{{\cal I}(c_A,d_A,k_A,l_A,\Lambda) + {1 \over 16} N \Lambda^2 + {1 \over 8 \pi} N^2  \ln \left({\Lambda^2 \pi \over N} \right)\right\} \: ,
\ee
where we define
\beas
{\cal I}(c_A,d_A,k_A,l_A,\Lambda) &=& \int d^2 w \left\{\hat{h}_1 \hat{h}_2 \partial_w \partial_{\bar{w}} (\hat{h}_1 \hat{h}_2) \right\} \cr
&=& \frac{1}{4} \int_0^\Lambda r dr \int_0^{\pi \over 2} d \theta \left\{ \hat{h}_1 \hat{h}_2 \left( {1 \over r} \partial_r (r \partial_r (\hat{h}_1 \hat{h}_2)) + {1 \over r^2} \partial_\theta^2 (\hat{h}_1 \hat{h}_2)\right)\right\}
\eeas
with
\beas
\hat{h}_1 &=&  r \cos(\theta) + \sum_A {c_A \over 2 \pi} \ln \left( {r^2 + 2r l_A \cos(\theta) + l_A^2 \over r^2 - 2r l_A \cos(\theta) + l_A^2} \right)\cr
\hat{h}_2 &=&  r \sin(\theta) + \sum_A {d_A \over 2 \pi} \ln \left( {r^2 + 2r k_A \sin(\theta) + k_A^2 \over r^2 - 2r k_A \sin(\theta) + k_A^2} \right) \; .
\eeas
We note that the factors of $\ell_s$ present in $h_1$ and $h_2$ have cancelled in those from the Einstein frame expression for $G$ taken from Appendix A.

There are no terms independent of $c_A$ and $d_A$, so we can express the full result as
\beas
{\cal I}(c_A,d_A,k_A,l_A,\Lambda)  &=& \sum_A c_A {\cal I}^c_A + \sum_A d_A {\cal I}^d_A + \sum_{A,B} c_A c_B {\cal I}^{cc}_{A B} \cr && + \sum_{A,B} d_A d_B {\cal I}^{dd}_{A B}  + \sum_{A,B} c_A d_B {\cal I}^{cd}_{A B} + \sum_{A,B,C} c_A c_B d_C {\cal I}^{ccd}_{A B C} 
\cr && 
+ \sum_{A,B,C} c_A d_B d_C {\cal I}^{cdd}_{A B C} + \sum_{A,B,C,D} c_A c_B d_C d_D {\cal I}^{ccdd}_{A B C D} \: .
\eeas

\subsubsection*{Integration Techniques}

There are various tricks that facilitate evaluation of the integral. First, it is helpful to use Stokes' theorem in order to rewrite the integral as a simpler integral plus a term that can be expressed as a boundary integral. We have
\beas
&&4 \,{\cal I}(c_A,d_A,k_A,l_A,\Lambda) = \int_0^\Lambda r dr \int_0^{\pi \over 2} \left\{-\hat{h}_2^2 \left(\partial_r \hat{h}_1 \partial_r \hat{h}_1 + {1 \over r^2} \partial_\theta \hat{h}_1 \partial_\theta \hat{h}_1 \right) \right\} \cr
&& \quad + \int_0^{\pi \over 2} d \theta \left\{\hat{h}_1 \hat{h}_2^2 r \partial_r \hat{h}_1 \right\}_{r = \Lambda}
 + \int_0^\Lambda dr \left\{-\hat{h}_1 \hat{h}_2^2 {1 \over r} \partial_\theta \hat{h}_1 \right\}_{\theta = 0}
 - \int_0^\Lambda dr \left\{-\hat{h}_1 \hat{h}_2^2 {1 \over r} \partial_\theta \hat{h}_1 \right\}_{\theta = \pi/2} \: .
\eeas
In evaluating the various pieces, it is helpful to differentiate with respect to the parameters $k_A$ or $l_A$ in order to convert the logarithms into rational functions of $r$. The resulting expressions can be expressed in a partial fraction expansion, with denominators that are polynomials in $r$ and $\cos(\theta)$ and numerators that are constant or linear functions of $\cos(\theta)$. After evaluating the integrals of the various parts, we can antidifferentiate with respect to $k_A$ or $l_A$ to obtain the final results. A more elegant method to obtain the results uses complex integration; see Appendix \ref{sec:complexint}.

We now present results for the various parts of the integral.

\subsubsection*{Linear Terms}

The terms linear in $c_A$ or $d_A$ are:
\be
{\cal I}^c_A = -{1 \over 16} {\it l_A}\,{\Lambda}^{2}+{\frac {{{\it l_A}}^{3}}{24}}
\ee
and
\be
{\cal I}^d_A =  -{1 \over 16} {\it k_A}\,{\Lambda}^{2}+{\frac {{{\it k_A}}^{3}}{24}} \: .
\ee

\subsubsection*{Quadratic Terms}

For the terms quadratic in $c_A$ and/or $d_A$, we find
\begin{equation}
    \begin{split}
        \pi \, {\cal I}^{cc}_{A B} & = -{1 \over 4}{\it l_A}\,{\it l_B}\,\ln  \left( \Lambda \right) -{3 \over 16} {\it
        l_A}\,{\it l_B} \\
        & \qquad+{1 \over 32} \left( {\it l_A}+{\it l_B} \right) ^{2}\ln
        \left(  \left( {\it l_A}+{\it l_B} \right) ^{2} \right) 
         - {1 \over 32}
        \left( {\it l_A}-{\it l_B} \right) ^{2}\ln  \left(  \left( {\it l_A}-{
        \it l_B} \right) ^{2} \right) 
    \end{split}
\end{equation}
\begin{equation}
    \begin{split}
        \pi \, {\cal I}^{dd}_{A B} & = -{1 \over 4}{\it k_A}\,{\it k_B}\,\ln  \left( \Lambda \right) -{3 \over 16}{\it
        k_A}\,{\it k_B} \\
        & \qquad +{1 \over 32} \left( {\it k_A}+{\it k_B} \right) ^{2}\ln
        \left(  \left( {\it k_A}+{\it k_B} \right) ^{2} \right) -{1 \over 32}
        \left( {\it k_A}-{\it k_B} \right) ^{2}\ln  \left(  \left( {\it k_A}-{
        \it k_B} \right) ^{2} \right)
    \end{split}
\end{equation}

and
\[
\pi\, {\cal I}^{cd}_{A B} = -{1 \over 2}\,{\it l_A}\,{\it k_B}\,\ln  \left( \Lambda \right) -{3 \over 8}{\it
l_A}\,{\it k_B}+ {1 \over 4}\,{\it k_B}\,{\it l_A}\,\ln  \left( {{\it k_B}}^{2}
+{{\it l_A}}^{2} \right) \: .
\]

\subsubsection*{Cubic Terms}

For the cubic terms, it is simpler to first give the derivatives with respect to one of the parameters. We have:
\beas
\pi^2 \, {d \over d k_C} {\cal I}^{ccd}_{A B C} &=& {1 \over 8} \left({\frac {{{\it l_A}}^{2}}{{{\it k_C}}^{2}+{{\it l_A}}^{2
}}}+{\frac {{{\it l_B}}^{2}}{{{\it k_C}}^{2}+{{\it l_B}}^{2}}}
 \right) \ln  \left( {\frac { \left( {\it l_A}-{\it l_B} \right) ^{2}}{
 \left( {\it l_A}+{\it l_B} \right) ^{2}}} \right)
 \cr
 &&- {1 \over 8} \left({\frac {{\it l_A}\,{\it l_B}}{{{\it k_C}}^{2}+{{\it l_A}
}^{2}}}+{\frac {{\it l_A}\,{\it l_B}}{{{\it k_C}}^{2}+{{\it l_B}
}^{2}}} \right) \ln  \left(  \left( {{\it l_A}}^{2}-{{\it l_B}}^{2}
 \right) ^{2} \right) \cr
&& +\,{1 \over 4}{\frac {{\it l_A}\,{\it l_B} }{{{\it k_C}}^{2}+{{\it l_B}}^{2}}}\,\ln  \left( {{\it k_C}}^{2}+{{
\it l_A}}^{2} \right)+{1 \over 4}{\frac {{\it l_A}\,{\it l_B}\, }{{{\it k_C}}^{2}+{{\it l_A}}^{2}}}\ln  \left( {{\it k_C}}^{2}+{{\it l_B}}^{2}
 \right) \: .
\eeas
We can integrate this with respect to $k_C$, requiring that the result vanishes at $k_C=0$. The result is conveniently written in terms of the Bloch-Wigner dilogarithm\footnote{This is Jamie Sully's favorite dilogarithm. We thank him for making us aware of it and extolling its virtues.}
\be
D(z) = {\rm Im}(\textnormal{Li}_2(z)) + \arg(1-z) \log|z| \; .
\ee
Here, $\textnormal{Li}_2$ is the dilogarithm function defined as
\be
\textnormal{Li}_2(z) = \sum_{n=1}^\infty {z^n \over n^2} = -\int_0^z {dt \over t} \log(1-t) \; .
\ee
Our result is simply
\be
{\cal I}^{ccd}_{A B C} = { l_A \over 4 \pi^2} \left\{D\left[{l_B - i k_C \over l_A + l_B} \right] + D\left[{l_B - i k_C \over l_B-l_A} \right] \right\} + \left\{l_A \leftrightarrow l_B \right\} \; .
\ee
The diagonal terms $l_B=l_A$ simplify to
\be
{\cal I}^{ccd}_{A A C} =  {l_A \over 2 \pi^2} D\left[{1 \over 2} -{i \over 2}{k_C \over l_A} \right] \; .
\ee

\subsubsection*{Quartic Terms}

We find that
\beas
{d \over d k_C} {d \over d k_D} {\cal I}^{ccdd}_{A B C D} &=&\,{1 \over 8 \pi^3} \left\{{\frac {{\it l_A}\,{\it l_B}}{ \left( {{\it k_C}}^{2}+{{\it l_B}}^{2} \right)  \left( {{\it k_D}}^{2}+{{\it l_A}}^{2} \right) }} \ln  \left( {\frac { \left( {{\it l_A}}^{2}-{{\it l_B}}^{2} \right) ^{2}
 \left( {{\it k_C}}^{2}-{{\it k_D}}^{2} \right) ^{2}}{ \left( {{\it k_C}}
^{2}+{{\it l_A}}^{2} \right) ^{2} \left( {{\it k_D}}^{2}+{{\it l_B}}^{2}
 \right) ^{2}}} \right) \right. \cr
&&+\,{\frac {{\it l_A}\,{\it l_B}}{ \left( {{\it k_D}}^{2}+{{\it l_B}
}^{2} \right)  \left( {{\it k_C}}^{2}+{{\it l_A}}^{2} \right) }} \ln  \left( {\frac { \left( {{\it l_A}}^{2}-{{\it l_B}}^{2} \right) ^{2}
 \left( {{\it k_C}}^{2}-{{\it k_D}}^{2} \right) ^{2}}{ \left( {{\it k_C}}
^{2}+{{\it l_B}}^{2} \right) ^{2} \left( {{\it k_D}}^{2}+{{\it l_A}}^{2}
 \right) ^{2}}} \right) \cr
 &&\left.+
 \left({\frac {{  {\it l_A}}^{2}}{ \left( {{\it k_D}}^{2}+{{\it l_A}}^{2} \right)
 \left( {{\it k_C}}^{2}+{{\it l_A}}^{2} \right) }}
 + {\frac {{{\it l_B}}^{2}}{ \left( {{\it k_D}}^{2}+{{\it l_B}}^{2} \right)
 \left( {{\it k_C}}^{2}+{{\it l_B}}^{2} \right) }}
 \right)\ln  \left( {\frac { \left( {\it l_A}+{\it l_B} \right) ^{2}}{
 \left( {\it l_A}-{\it l_B} \right) ^{2}}} \right) \right\} \: .
\eeas

We now need to integrate this with respect to $k_C$ and $k_D$. This time, the result involves the trilogarithm function $\textnormal{Li}_3(z)$ in addition to dilogarithms and elementary functions. Taking guidance from the cubic terms, which could be written simply in terms of the Bloch-Wigner dilogarithm, we can make the guess that the full result here may be obtained by keeping only the terms with trilogarithms, and replacing each trilogarithm with the combination
\be
{\cal L}(z) = Re(\textnormal{Li}_3(z) - \ln|z|\textnormal{Li}_2(z) + {1 \over 3} \ln^2|z| \textnormal{Li}_1(z))
\ee
which has been shown to be real analytic on $\mathbb{C} - \{0,1\}$ and continuous everywhere, and to obey various nice relations such as ${\cal L}(1/z) = {\cal L}(z)$. This turns out to be correct. The full result for the integral is
\beas
8 \pi^3 \, {\cal I}^{ccdd}_{A B C D}  &=& {\cal L} \left( {\frac { \left( {\it k_C}+i{\it l_A} \right)  \left( {\it k_D}+i
{\it l_B} \right) }{ \left( {\it k_D}+i{\it l_A} \right)  \left( {\it k_C}
+i{\it l_B} \right) }} \right) +{\cal L} \left( {\frac { \left( {\it k_C}+i{
\it l_A} \right)  \left( {\it k_D}+i{\it l_B} \right) }{ \left( {\it k_D}-
i{\it l_A} \right)  \left( {\it k_C}-i{\it l_B} \right) }} \right) \cr
&& -{\cal L}
 \left( {\frac { \left( {\it k_C}+i{\it l_A} \right)  \left( {\it k_D}-i{
\it l_B} \right) }{ \left( {\it k_D}+i{\it l_A} \right)  \left( {\it k_C}-
i{\it l_B} \right) }} \right) -{\cal L} \left( {\frac { \left( {\it k_C}+i{\it
l_A} \right)  \left( {\it k_D}-i{\it l_B} \right) }{ \left( {\it k_D}-i{
\it l_A} \right)  \left( {\it k_C}+i{\it l_B} \right) }} \right) \cr
&&
+{\cal L} \left( {\frac { \left( {\it k_D}+i{\it l_A} \right)  \left( {\it l_B}-i
{\it k_C} \right) }{ \left( {\it k_C}+{\it k_D} \right)  \left( {\it l_A}+
{\it l_B} \right) }} \right) +{\cal L} \left( {\frac { \left( {\it l_A}+i{\it
k_D} \right)  \left( {\it k_C}-i{\it l_B} \right) }{ \left( {\it k_C}-{
\it k_D} \right)  \left( {\it l_A}-{\it l_B} \right) }} \right) \cr
&&
 -{\cal L} \left( {\frac { \left( {\it l_A}+i{\it k_D} \right)  \left( {\it k_C}+i{
\it l_B} \right) }{ \left( {\it k_C}-{\it k_D} \right)  \left( {\it l_A}+{
\it l_B} \right) }} \right) -{\cal L} \left( {\frac { \left( {\it l_A}+i{\it k_D
} \right)  \left( {\it k_C}+i{\it l_B} \right) }{ \left( {\it k_C}+{\it
k_D} \right)  \left( {\it l_A}-{\it l_B} \right) }} \right) \cr
&&+\{l_A \leftrightarrow l_B \} \: .
\eeas
We note that the first two lines are already invariant under $\{l_A \leftrightarrow l_B \}$. The diagonal terms can be recovered by taking a limit in the above expression. 

\subsection{Full Result} \label{sec:result}

Combining all terms, we can now write the full result for $F_{\partial}$ (in the supergravity approximation) associated to the theory whose vacuum has supergravity dual labeled by $\mathcal{P} \equiv \{c_A,d_A,l_A,k_A\}$. The result is
\bea
\label{SugraResult}
F_{\partial}(\mathcal{P}) &=&  {3 \over 8} N^2 + {1 \over 4} N^2 \ln \left({N \over \pi}\right) \\
&&-  {\pi \over 12} \sum_A c_A l_A^3 -  {\pi \over 12} \sum_B d_B k_B^3 \cr
&&- {1 \over 16} \sum_{A,B}  c_A c_B \left\{(l_A + l_B)^2 \ln \left((l_A + l_B)^2\right) - (l_A - l_B)^2 \ln\left( (l_A - l_B)^2 \right) \right\} \cr
&&- {1 \over 16} \sum_{A,B}  d_A d_B \left\{(k_A + k_B)^2 \ln \left( (k_A + k_B)^2\right) - (k_A - k_B)^2 \ln \left(  (k_A - k_B)^2 \right) \right\} \cr
&& - {1 \over 2} \sum_{A,B} c_A d_B \left\{l_A k_B \ln \left(l_A^2 + k_B^2 \right) \right\} \cr
&& - {1 \over  \pi} \sum_{A,B,C}   c_A c_B d_C l_A \left\{ D \left[{l_B - i k_C \over l_A + l_B} \right] + D \left[{l_B - i k_C \over l_B - l_A} \right] \right\}\cr
&& - {1 \over  \pi} \sum_{A,B,C}  d_A d_B c_C k_A \left\{ D \left[{k_B - i l_C \over k_A + k_B} \right] + D \left[{k_B - i l_C \over k_B - k_A} \right] \right\} \cr
&& - {1 \over 2 \pi^2} \sum_{A,B,C,D}  c_A c_B d_C d_D \left\{
{\cal L} \left( {\frac { \left( {\it k_C}+i{\it l_A} \right)  \left( {\it k_D}+i
{\it l_B} \right) }{ \left( {\it k_D}+i{\it l_A} \right)  \left( {\it k_C}
+i{\it l_B} \right) }} \right) +{\cal L} \left( {\frac { \left( {\it k_C}+i{
\it l_A} \right)  \left( {\it k_D}+i{\it l_B} \right) }{ \left( {\it k_D}-
i{\it l_A} \right)  \left( {\it k_C}-i{\it l_B} \right) }} \right) \right. \cr
&& \qquad \qquad \qquad \qquad \quad \;
-{\cal L}
 \left( {\frac { \left( {\it k_C}+i{\it l_A} \right)  \left( {\it k_D}-i{
\it l_B} \right) }{ \left( {\it k_D}+i{\it l_A} \right)  \left( {\it k_C}-
i{\it l_B} \right) }} \right) -{\cal L} \left( {\frac { \left( {\it k_C}+i{\it
l_A} \right)  \left( {\it k_D}-i{\it l_B} \right) }{ \left( {\it k_D}-i{
\it l_A} \right)  \left( {\it k_C}+i{\it l_B} \right) }} \right) \cr
&&\qquad \qquad \qquad \qquad \quad \;
+{\cal L} \left( {\frac { \left( {\it k_D}+i{\it l_A} \right)  \left( {\it l_B}-i
{\it k_C} \right) }{ \left( {\it k_C}+{\it k_D} \right)  \left( {\it l_A}+
{\it l_B} \right) }} \right) +{\cal L} \left( {\frac { \left( {\it l_A}+i{\it
k_D} \right)  \left( {\it k_C}-i{\it l_B} \right) }{ \left( {\it k_C}-{
\it k_D} \right)  \left( {\it l_A}-{\it l_B} \right) }} \right) \cr
&&\qquad \qquad \qquad \qquad  \quad \;
\left.
 -{\cal L} \left( {\frac { \left( {\it l_A}+i{\it k_D} \right)  \left( {\it k_C}+i{
\it l_B} \right) }{ \left( {\it k_C}-{\it k_D} \right)  \left( {\it l_A}+{
\it l_B} \right) }} \right) -{\cal L} \left( {\frac { \left( {\it l_A}+i{\it k_D
} \right)  \left( {\it k_C}+i{\it l_B} \right) }{ \left( {\it k_C}+{\it
k_D} \right)  \left( {\it l_A}-{\it l_B} \right) }} \right) \right\} \: , \nonumber
\eea
where we recall that
\be
N = \sum_A c_A l_A + \sum_B d_B k_B \; .
\ee
We can express the results in terms of field theory parameters using the correspondence described in Section \ref{sec:ftsugra}.

\subsubsection*{D5-Branes Only}
We now consider various special cases. For theories descending from string theory configurations with only D3-branes and D5-branes, the result simplifies to
\beas
F_{\partial} &=&  {3 \over 8} N^2
-  \sum_A {\pi \over 12} c_A l_A^3 \cr
&& - \sum_{A,B} {1 \over 16} c_A c_B \left\{(l_A + l_B)^2 \ln \left(\pi {(l_A + l_B)^2\over N}\right) - (l_A - l_B)^2 \ln\left(\pi { (l_A - l_B)^2 \over N}\right) \right\}
\eeas
Expressed purely in terms of the linking numbers $L_A$ (which coincide with $\tilde{L}_A$ in this case), this is
\beas
F_{\partial} &=&  {N^2\over 4}\left( {3 \over 2} + \ln \left({\lambda \over 4 \pi^2}\right) \right) - {\pi^2 N \over 3 \lambda} \sum_A L_A^3 \cr
&&- {1 \over 16} \sum_{A,B}\left\{ (L_A + L_B)^2 \ln \left((L_A + L_B)^2\right) - (L_A - L_B)^2 \ln\left((L_A - L_B)^2\right) \right\} \: ,
\eeas
We recall that in the brane construction, $\{L_A\}$ represents the numbers of D3-branes ending on each individual D5-brane, such that $\sum_A L_A = N$. When we have $N$ D3-branes ending on $N_5$ D5-branes with $N/N_5$ D3-branes ending on each D5, the result simplifies further to 
\be
\label{D5simp}
F_{\partial} = {N^2 \over 8} \left[3 - {8 \pi^{2} \over 3 \lambda} {N^2 \over N_5^2}  - 2 \ln \left({16 \pi^2 \over \lambda} {N^2 \over N_5^2} \right) \right] \: .
\ee
This result corresponds to the case considered previously in \cite{Estes:2014hka}; our result agrees precisely with that computation.

\subsubsection*{NS5-Branes Only}
For boundary conditions associated with only NS5-branes, we find that
\beas
F_{\partial} &=& {3 \over 8} N^2
-  \sum_A {\pi \over 12} d_A k_A^3 \cr
&& - \sum_{A,B} {1 \over 16} d_A d_B \left\{(k_A + k_B)^2 \ln \left(\pi {(k_A + k_B)^2\over N}\right) - (k_A - k_B)^2 \ln\left(\pi { (k_A - k_B)^2 \over N}\right) \right\} \: .
\eeas
We can check that this may also be obtained from the D5-brane result by S-duality, manifested in the transformations $l_A \to k_A$, $c_A \to d_A$, $g \to 1/g$ (or $\lambda \to 16 \pi^2 N^2 /\lambda$). 
Expressed purely in terms of the linking numbers $K_A$, this gives
\beas
F_{\partial} 
&=& \frac{N^{2}}{4} \left( \frac{3}{2} + \ln \Big( \frac{4 N^{2}}{\lambda} \Big) \right)
-  { \lambda \over 48 N} \sum_A K_A^3 \cr
&& - {1 \over 16} \sum_{A,B} \left\{(K_A + K_B)^2 \ln \left( (K_A + K_B)^2\right) - (K_A - K_B)^2 \ln\left( (K_A - K_B)^2 \right) \right\} \: , \cr
\eeas
where $\{K_{A}\}$ represents the numbers of D3-branes ending on each individual NS5-brane, as for the D5-brane case above. 
In the case corresponding to $N$ D3-branes ending on $N_5$ NS5-branes with $N/N_5$ D3-branes ending on each NS5, the result simplifies to
\be
\label{NS5simp}
F_{\partial} = {N^2 \over 8} \left[3 - { \lambda \over 6 N_5^2}  - 2 \ln \left( {\lambda \over N_5^2} \right) \right] \: .
\ee

\subsubsection*{Both D5-Branes and NS5-Branes}

In the special cases with either D5-branes or NS5-branes only, we were able to write an explicit expression for $F_{\partial}$ in terms of variables in the brane constructions, i.e. the 5-brane charges and linking numbers. For the most general constructions involving both D5-branes and NS5-branes, however, we do not know how to analytically invert the relations between supergravity and field theory variables. 
In scenarios of interest, we can always choose some field theory parameters, try to solve for the SUGRA parameters numerically, and then evaluate $F_{\partial}$.

\subsection{Validity of the Supergravity Approximation}

The results of this section are based on the supergravity approximation to the dual gravity solutions and on the leading order RT formula without $\alpha'$-corrections or quantum corrections. However, we expect that the solution and the RT formula receive both string loop and $\alpha'$-corrections. These will correct our result, unless the corrections vanish, for example due to some supersymmetric non-renormalization theorem. 

Taking into account $\alpha'$ and string loop corrections, the purely gravitational sector of the effective action in string frame takes the schematic form
\begin{equation}
    S \sim \int dx \: \sqrt{g} \left[ e^{2 \phi} \left( \alpha ' R + ( \alpha ' R)^{2} + \ldots \right) + e^{4 \phi} \left( \alpha ' R + ( \alpha ' R)^{2} + \ldots \right) + \ldots \right] \: ,
\end{equation}
though certain terms vanish in type IIB supergravity due to constraints of supersymmetry.

This implies that the $\alpha'$-corrections will be suppressed if the \textit{string frame} Ricci curvature obeys
\begin{equation}
    \alpha' R \ll 1 \: ,
\end{equation}
whereas string loop corrections will be suppressed if
\begin{equation}
    e^{2 \phi} \ll 1 \: .
\end{equation}
For large $N$ and large $\lambda$, we anticipate that these expressions should hold in the asymptotically AdS region, but might break down in the vicinity of the fivebrane throats. 

In order to estimate the expected size of the corrections to the supergravity results, we can employ the following general procedure:
\begin{itemize}
    \item For an arbitrary fixed set of parameters, determine the region near a given 5-brane stack where these correction terms would naively have a similar order of magnitude to the leading supergravity results.
    \item Find the size of the supergravity contribution to $F_{\partial}$ from this region. Assuming that the corrections have a similar order of magnitude, we will take this as an estimate of size of the correction terms. Terms in the supergravity result that are parametrically larger than this will be considered reliable. 
\end{itemize}






The details of this analysis are provided in Appendix \ref{sec:corrections}. 
As a specific example of the results, we find that for the theory corresponding to $N$ D3-branes ending on a single stack of $\tilde{L} = N / N_{5}$ D5-branes, the expected contribution from the part of the NS5-brane throat where the string frame curvature is large is
\begin{equation}
    \begin{cases}
        O(N_{5}^{2}) & \tilde{L} \sim 1 \\
        O \left( ( N_{5} \ln \tilde{L})^{2} \right) & \tilde{L} \gg 1
    \end{cases} \: .
\end{equation}
Thus, we might expect corrections to the supergravity result (\ref{D5simp}) at this order.

For the case of $N$ D3-branes ending on a single stack of $K = N / N_{5}$ NS5-branes, the string frame curvature is only large in the vicinity of the NS5-brane throat provided that $N_{NS5} \sim 1$, in which case the expected contribution to $F_{\partial}$ from this region is $O(N^{2})$. Additionally, the expected contribution from the region in which the dilaton is large is
\begin{equation}
    \begin{cases}
        O \left( \left( N_{5}^{2} \ln \left( K^{2} / N_{5}^{2} \right) \right)^{2} \right) & K \gg N_{5} \\
        O(N_{5}^{4}) & K \sim N_{5} \\
        O(N^{2}) & K \ll N_{5}
    \end{cases} \: .
\end{equation}
Thus, we might expect corrections to the supergravity result (\ref{NS5simp}) at this order.

In the next section, we will be able to calculate boundary $F$ exactly using supersymmetric localization, for boundary conditions associated either with only D5-branes or only NS5-branes. We will see that the supergravity results are actually more reliable than our analysis suggests. 

\section{Localization Calculation} \label{sec:loc}

In the above analysis, we have extracted the value of $F_{\partial}$ by holographically computing the entanglement entropy for a half-ball centred at the field theory boundary. However, we recall that $F_{\partial}$ is also related to the partition function for the theory on a hemisphere; specifically, we have \cite{Gaiotto:2014gha} 
\begin{equation}
    F_{\partial} \equiv -\frac{1}{2} \lim_{r \rightarrow \infty} \ln \Big( \frac{|Z_{HS^{4}}|^{2}}{Z_{S^{4}}} \Big) \: ,
\end{equation}
where $r = R / \epsilon$ is the quotient of the radius $R$ of the (hemi)sphere and a UV regulator $\epsilon$.

Calculations of the partition function in theories with supersymmetry are often tractable using the technique of supersymmetric localization; see \cite{Pestun:2016zxk} for a review. In particular, the calculation of the partition function, in addition to generic half-BPS Wilson loop observables, for $\mathcal{N}=2$ (or $\mathcal{N}=4$) supersymmetric gauge theories on a background $S^{4}$ was first performed in \cite{Pestun:2007rz}. Localization was later applied to compute 't Hooft loop observables \cite{Gomis:2011pf} and 1/8-BPS Wilson loop observables \cite{Pestun:2009nn} in such theories, and generalizations to theories on ellipsoids appeared in \cite{Hama:2012bg}, as reviewed in \cite{Hosomichi:2016flq}. Analogous calculations were performed for $\mathcal{N}=2$ theories on $S^{3}$ in \cite{Kapustin:2009kz}, with exact evaluation of the partition function for 3-dimensional quiver gauge theories appearing in \cite{Kapustin:2010xq, Benvenuti:2011ga, Nishioka:2011dq}. Localization calculations on manifolds with boundary in two and three dimensions first appeared in \cite{Sugishita:2013jca}; in four dimensions, the first direct calculations appeared in \cite{Gava:2016oep}, which considered Neumann and Dirichlet boundary conditions only, followed by \cite{Gupta:2019qlg}, which considered more general boundary conditions for the Abelian theory. Earlier general considerations for the case with boundaries can be found in \cite{Drukker:2010jp, Gaiotto:2014gha, Dedushenko:2018tgx}. More recent results involving localization and supersymmetric boundaries and interfaces include \cite{Wang:2020seq, Komatsu:2020sup, Goto:2020per, Dedushenko:2020vgd}. 

We will therefore endeavour in this section to compare our gravity results to the calculation of $F_{\partial}$ using supersymmetric localization on the field theory side. 
In particular, we will restrict our attention to theories arising from D3-branes and NS5-branes only (i.e. with arbitrary linking numbers $\{K_i\}$, but $\{L_i \} = \emptyset$). In this case, the form of the partition function as a zero-dimensional matrix integral may be inferred by recalling the established results for the hemisphere with Neumann boundary conditions \cite{Gava:2016oep, Wang:2020seq} and 3-dimensional quiver gauge theories \cite{Kapustin:2009kz, Kapustin:2010xq,  Benvenuti:2011ga, Nishioka:2011dq}, and applying the gluing formula of \cite{Dedushenko:2018tgx}. Using S-duality, we can obtain results for general D5-like boundary conditions. 

In the following, we will denote
\begin{equation}
    \sh(x) = 2 \sinh \pi x \: , \qquad \ch(x) = 2 \cosh \pi x \: .
\end{equation}
The partition function of $U(N)$ $\mathcal{N}=4$ SYM on the hemisphere $HS^{4}$ with Neumann boundary conditions is then
\begin{equation}
    Z_{\textnormal{Neum.}}[HS_{r=1}^{4}] = \frac{1}{N!} \int \left( \prod_{i=1}^{N} d \lambda_{i} \right) e^{- \frac{4 \pi^{2}}{g_{\textnormal{YM}}^{2}} \sum_{i=1}^{N} \lambda_{i}^{2} } \prod_{i < j}^{N} (\lambda_{i} - \lambda_{j}) \sh (\lambda_{i} - \lambda_{j}) \: ,
\end{equation}
and the partition function for a 3-dimensional $\mathcal{N}=4$ $U(n_{1}) \times \ldots \times U(n_{N_{5}})$ quiver gauge theory with $l_{i}$ fundamental hypermultiplets associated to the $U(n_{i})$ factor, with hypermultiplet masses $m_{i, j}$ and FI parameters $\alpha_{i}$, is
\begin{equation}
\begin{split}
    Z_{\alpha, m}[S_{r=1}^{3}] & = \frac{1}{n_{1}! \ldots n_{N_{5}}!} \int \left( \prod_{j=1}^{N_{5}} \prod_{\ell=1}^{n_{j}} d \lambda_{j, \ell} e^{2 \pi i \alpha_{j} \lambda_{j, \ell}} \right) \prod_{j=1}^{N_{5}} \prod_{k < \ell}^{n_{j}} \sh^{2} \left( \lambda_{j, k} - \lambda_{j, \ell} \right) \\
    & \qquad \prod_{j=1}^{N_{5}-1} \prod_{k=1}^{n_{j}} \prod_{\ell=1}^{n_{j+1}} \frac{1}{\ch(\lambda_{j, k} - \lambda_{j+1, \ell})} \prod_{j=1}^{N_{5}} \prod_{\ell=1}^{n_{j}} \prod_{k=1}^{\ell_{j}} \frac{1}{\ch( \lambda_{j, \ell} - m_{j, k})} \: .
\end{split}
\end{equation}
The hemisphere partition function for the $\mathcal{N}=4$ SYM theory coupled to a quiver gauge theory at the boundary is then obtained by integrating the integrand of $Z_{\textnormal{Neum.}}[HS^{4}]$ against an appropriate ``brane factor" with respect to the bulk zero modes $(\lambda_{1}, \ldots, \lambda_{N})$; in this case, the brane factor coincides with the partition function of the boundary theory $Z_{\alpha, m}[S^{3}]$, where the masses in the terminal node of the quiver diagram are replaced by the bulk zero modes (as the restriction of the bulk vector multiplet gauges the boundary flavour symmetry). For example, in the case where the quiver gauge theory contains vanishing FI parameters and no fundamental hypermultiplets (as will arise in the present case), we recover the partition function
\begin{equation}
\begin{split}
    Z[HS^{4}] & = \frac{1}{n_{1}! \ldots n_{N_{5}}!} \int \left( \prod_{j=1}^{N_{5}} \prod_{\ell=1}^{n_{j}} d \lambda_{j, \ell}  \right) e^{- \frac{4 \pi^{2}}{g_{\textnormal{YM}}^{2}} \sum_{i=1}^{N} \lambda_{N_{5}, i}^{2}} \prod_{i < j}^{N} (\lambda_{N_{5}, i} - \lambda_{N_{5}, j}) \: \sh (\lambda_{N_{5}, i} - \lambda_{N_{5}, j}) \\
    & \qquad \times  \prod_{j=1}^{N_{5}-1} \prod_{k < \ell}^{n_{j}} \sh^{2} \left( \lambda_{j, k} - \lambda_{j, \ell} \right) \prod_{j=1}^{N_{5}-1} \prod_{k=1}^{n_{j}} \prod_{\ell=1}^{n_{j+1}} \frac{1}{\ch(\lambda_{j, k} - \lambda_{j+1, \ell})} \: ,
\end{split}
\end{equation}
where we will let $n_{N_{5}} \equiv N$ for convenience.
In the brane construction, there are $n_{j}$ D3-branes stretched between the $j^{\textnormal{th}}$ and $(j+1)^{\textnormal{th}}$ NS5-brane, so that for a configuration satisfying the Gaiotto-Witten constraints, one has $0< K_{1} \leq \ldots \leq K_{N_{5}}$ where $K_{i} \equiv n_{i} - n_{i-1}$. 

Since the calculation of $F_{\partial}$ involves a subtraction of the partition function for the theory on the full $S^{4}$, we will need to know the partition function for $U(N)$ $\mathcal{N}=4$ SYM on $S^{4}$. 
One has matrix integral partition function \cite{Pestun:2007rz}
\begin{equation}
\label{fullsphere}
    \begin{split}
        Z_{S^{4}_{r=1}} & = \frac{1}{N!} \int \left( \prod_{i=1}^{N} d \lambda_{i} \right) e^{ - \frac{8 \pi^{2}}{g_{\textnormal{YM}}^{2}} \sum_{i=1}^{N} \lambda_{i}^{2}} \prod_{i < j}^{N} (\lambda_{i} - \lambda_{j})^{2} 
    \end{split}
\end{equation}
on $S^{4}$ with unit radius $r=1$, where the measure factor $\frac{1}{N!} \prod_{i<j}(\lambda_{i} - \lambda_{j})^{2}$ arises from reducing the integration over the full Lie algebra $\mathfrak{u}(N)$ to the Cartan subalgebra, and the exponential factor is the classical contribution to the partition function, coming from evaluating the on-shell action. (The one-loop and instanton corrections vanish in this highly symmetric situation.) For $S^{4}_{r}$ with arbitrary radius, the purely gauge-theoretic measure should be invariant, but the classical contribution has
\begin{equation}
    S_{\textnormal{on-shell}}^{r} \sim r^{2} S_{\textnormal{on-shell}}^{r=1} \: .
\end{equation}
The calculation can be found in Appendix \ref{sec:locint}: it provides
\begin{equation}
    Z_{S_{r}^{4}} = (2 \pi)^{N/2} \left( \frac{g_{\textnormal{YM}}}{4 \pi r} \right)^{N^{2}} G_{2}(N+1) \: ,
\end{equation}
where 
\begin{equation}
   G_{2}(N+1) \equiv \prod_{k=1}^{N-1} k!   
\end{equation}
is the Barnes G-function.
One then has
\begin{equation}
    \ln Z_{S^{4}_{r}} = -N^{2} \ln r + \frac{N^{2}}{2} \ln \Big( \frac{\lambda}{16 \pi^{2} N} \Big) + \ln G_{2}(N+1) + \frac{N}{2} \ln 2 \pi \: .
\end{equation}
For the purposes of comparing to the gravity calculation, we will typically be interested in the large $N$ behaviour of this expression, so we require the asymptotics of
\begin{equation}
    \ln G_{2}(N+1) = \sum_{k=1}^{N-1} (N-k) \ln k = N \ln (N-1)! - \sum_{k=1}^{N-1} k \ln k \: .
\end{equation}
The asymptotics of the first term are given by the Stirling formula
\begin{equation}
    N \ln (N-1)! = N^{2} \ln N - N^{2} + \mathcal{O}(N \ln N) \: .
\end{equation}
To find an asymptotic expression for the sum $\sum_{k=1}^{N-1} k \ln k$, we will use the Euler-Maclaurin formula
\begin{equation}
        \sum_{k=a}^{b} f(k) \sim \int_{a}^{b} f(x) \: dx + \frac{f(b)+f(a)}{2} + \sum_{k=1}^{\infty} \frac{B_{2k}}{(2k)!} \big( f^{2k-1}(b) - f^{(2k-1)}(a) \big) \: ,
\end{equation}
whence
\begin{equation}
    \sum_{k=1}^{N-1} k \ln k = \frac{N^{2} \ln N}{2} - \frac{N^{2}}{4} + \mathcal{O}(N \ln N) \: .
\end{equation}
It is straightforward to determine the higher order terms if needed. 
All together, we have
\begin{equation}
    \ln G_{2}(N+1) = \frac{N^{2}}{2} \ln N - \frac{3}{4} N^{2} + \mathcal{O}(N \ln N) \: ,
\end{equation}
and so
\begin{equation}
    \ln Z_{S^{4}_{r}} = - N^{2} \ln r + \frac{N^{2}}{2} \ln \Big( \frac{\lambda}{16 \pi^{2}} \Big) - \frac{3}{4} N^{2} + \frac{N}{2} \ln 2 \pi + \mathcal{O}(N \ln N) \: .
\end{equation}

It is worth noting that, from the general theory of the structure of UV divergences in the partition function, we anticipate 
\begin{equation}
    \ln Z_{S^{4}_{r}} = A_{1} r^{4} + A_{2} r^{2} + A \ln r + F_{4} \: ;
\end{equation}
here, $A_{1}, A_{2}$ can be tuned through the addition of local counterterms, as can $F_{4}$ (the local counterterm corresponds to the Euler density). Although these quantities are scheme-dependent, they will cancel out in the calculation of $F_{\partial}$ as long as we are consistent. The coefficient $A$ of the logarithmic divergence, however, is physically meaningful: it is proportional to the A-type anomaly $a$ for the $\mathcal{N}=4$ SYM theory on $S^{4}$, with
\begin{equation}
    \frac{\partial}{\partial \ln r} \ln Z_{S_{r}^{4}} = - 64 \pi^{2} a \: .
\end{equation}
The general Weyl anomaly in four dimensions is
\begin{equation}
    \langle T^{\mu}_{\: \mu} \rangle =  a E - c W^{2} \: ,
\end{equation}
with $E$ the Euler density and $W^{2}$ shorthand for a contraction of the Weyl tensor, and in the super-Yang-Mills theory,
\begin{equation}
    \langle T^{\mu}_{\: \mu} \rangle =  \frac{N^{2}}{64 \pi^{2}} \left( E - W^{2} \right) \: .
\end{equation}
We thus indeed recover $a = \frac{N^{2}}{64 \pi^{2}}$, and therefore $A = - N^{2}$, which confirms the $r$-dependence. 

\subsection{Neumann Boundary Condition}

As a warm-up to the case of general NS5-like boundary conditions, we can consider a pure Neumann boundary condition. This corresponds to $N$ D3-branes ending on a single NS5-brane, associated with parameter values
\begin{equation}
    N_{NS5} = \sqrt{g} d = 1 \: , \qquad \hat{k} = \frac{1}{\sqrt{g}} k = N \: ,
\end{equation}
that is,
\begin{equation}
    d = \sqrt{\frac{4 \pi N}{\lambda}} \: , \qquad k = \sqrt{\frac{\lambda N}{4 \pi}} \: .
\end{equation}
The partition function for this theory  (expressed as a matrix integral in \cite{Gava:2016oep}) on the unit hemisphere is
\begin{equation}
    Z_{HS^{4}_{r=1}}^{\textnormal{Neum.}} = \frac{1}{N! } \int \left( \prod_{i=1}^{N} d \lambda_{i} \right) e^{- \frac{4 \pi^{2}}{g_{\textnormal{YM}}^{2}} \sum_{i=1}^{N} \lambda_{i}^{2} } \prod_{i < j}^{N} (\lambda_{i} - \lambda_{j}) \sh  (\lambda_{i} - \lambda_{j}) \: .
\end{equation}
This is similar to the $S^{4}$ partition function (\ref{fullsphere}), except one now has one-loop determinant
\begin{equation}
    Z_{1-\textnormal{loop}}^{\textnormal{Neum.}} = \prod_{i<j} \frac{\sh(\lambda_{i} - \lambda_{j})}{\lambda_{i} - \lambda_{j}} \: ,
\end{equation}
where we have combined 1-loop factors from a $\mathcal{N}=2$ vector multiplet and an adjoint $\mathcal{N}=2$ hypermultiplet to recover the full 1-loop determinant for the $\mathcal{N}=4$ vector multiplet theory. 
Using the results of Appendix \ref{sec:locint}, this yields
\begin{equation}
\begin{split}
    Z_{HS^{4}_{r=1}}^{\textnormal{Neum.}} & = (2 \pi)^{\frac{N^{2}}{2}} \Big( \frac{ g_{\textnormal{YM}}^{2}}{8 \pi^{2}} \Big)^{\frac{N^{2}}{2}} e^{ \frac{g_{\textnormal{YM}}^{2} N(N+1)(N-1)}{48}} G_{2}(N+1) \: .
\end{split}
\end{equation}
We thus have
\begin{equation}
    \begin{split}
        \ln \big| Z_{HS^{4}_{r=1}}^{\textnormal{Neum.}} \big| 
        & =   \frac{\lambda (N+1)(N-1)}{48} + \frac{N^{2}}{2} \ln \left( \frac{\lambda}{4 \pi N} \right)  + \ln G_{2}(N+1) \: .
    \end{split}
\end{equation}
We therefore find that $F_{\partial}$ is given by
\begin{equation}
\begin{split}
    F_{\partial}^{\textnormal{Neum.}} & = - \left( \ln |Z_{HS_{r}^{4}}^{\textnormal{Neum.}}|^{2} - \ln Z_{S^{4}_{r}} \right) \\
    & =-\frac{\lambda(N^{2} - 1)}{48} - \frac{N^{2}}{4} \ln \left( \frac{\lambda}{N} \right) + \frac{ N}{4} \ln 2 \pi - \frac{1}{2 } \ln G_{2}(N+1) \: .
\end{split}
\end{equation}
Using the results above, we we can expand this for large $N$ as
\begin{equation}
    F_{\partial}^{\textnormal{Neum.}} = \frac{N^{2}}{8} \Big( -\frac{\lambda}{6} - 2 \ln ( \lambda) + 3 \Big)  + \mathcal{O}(N \ln N) \: .
\end{equation}
This may be compared to the gravity result
\begin{equation}
    \begin{split}
        F_{\partial}^{\textnormal{SUGRA}}
        & = \frac{N^{2}}{8} \Big(- \frac{\lambda}{6} - 2 \ln ( \lambda) + 3 \Big) \: .
    \end{split}
\end{equation}
Remarkably, at leading order in $N$, the exact expression for $F_{\partial}$ agrees exactly with the supergravity result as a function of $\lambda$.

\subsection{General NS5-Like Boundary Conditions}

We would like to evaluate the integral
\begin{equation}
\begin{split}
    Z[HS^{4}] & = \lim_{\alpha_{1}, \ldots, \alpha_{N_{5}-1} \rightarrow 0} \frac{1}{n_{1}! \ldots n_{N_{5}}!} \int \left( \prod_{j=1}^{N_{5}} \prod_{\ell=1}^{n_{j}} d \lambda_{j, \ell}  \right) \left( \prod_{j=1}^{N_{5}-1} \prod_{\ell=1}^{n_{j}} e^{2 \pi i \alpha_{j} \lambda_{j, \ell}} \right) e^{- \frac{4 \pi^{2}}{g_{\textnormal{YM}}^{2}} \sum_{i=1}^{N} \lambda_{N_{5}, i}^{2}} \\
    & \qquad \times \prod_{i < j}^{N} (\lambda_{N_{5}, i} - \lambda_{N_{5}, j}) \: \sh (\lambda_{N_{5}, i} - \lambda_{N_{5}, j}) \prod_{j=1}^{N_{5}-1} \prod_{k < \ell}^{n_{j}} \sh^{2} \left( \lambda_{j, k} - \lambda_{j, \ell} \right)  \\
    & \qquad \times \prod_{j=1}^{N_{5}-1} \prod_{k=1}^{n_{j}} \prod_{\ell=1}^{n_{j+1}} \frac{1}{\cosh(\lambda_{j, k} - \lambda_{j+1, \ell})} \: .
\end{split}
\end{equation}
As detailed in Appendix \ref{sec:locint}, this integral yields
\begin{equation}
\begin{split}
    Z[HS^{4}] & =  (2 \pi)^{-\sum_{i=1}^{N_{5}-1} n_{i}} \left( \frac{\gym^{2}}{4 \pi} \right)^{ \frac{N^{2}}{2}} e^{ \frac{\gym^{2} }{48} \sum_{c=1}^{N_{5}} K_{c} (K_{c}-1)(K_{c}+1)}  \left( \prod_{c=1}^{N_{5}} G_{2}(K_{c}+1) \right) \\
    & \qquad \times  \prod_{c<d}^{N_{5}} \left[ 2^{-(K_{d} - K_{c}) K_{c}} \left( \frac{\pi}{2} \right)^{\epsilon_{cd} K_{c}} \left( \left( (K_{d} - K_{c})!! \right)^{K_{c}} \prod_{k=1}^{K_{c}-1} \left( \frac{K_{d} - K_{c}}{2} + k \right)^{K_{c} - k} \right)^{2}  \right] \: ,
    \end{split}
\end{equation}
where $K_{i} \equiv n_{i} - n_{i-1}$ is the $i^{\textnormal{th}}$ linking number (satisfying $0 < K_{1} \leq \ldots \leq K_{N_{5}}$), and 
\begin{equation}
    \epsilon_{cd} \equiv \frac{1 - (-1)^{K_{c} - K_{d}}}{2}  \: .
\end{equation} 
We thus find 
\begin{equation}
    \begin{split}
        F_{\partial} & = - \ln \big| Z[HS^{4}] \big| + \frac{1}{2} \ln Z[S^{4}] \\
        & =
    - \left( \frac{N(2N-1)}{4}  - \sum_{p=1}^{N_{5}} (N_{5} - p) K_{p} \right) \ln (2 \pi) - \frac{N^{2}}{4} \ln \left( \frac{\lambda}{4 \pi^{2}N}  \right) -  \frac{\lambda}{48 N}\left( \sum_{p=1}^{N_{5}} K_{p}^3 -N \right) \\
        & \qquad - \sum_{p=1}^{N_{5}} \ln G_{2}(K_{p}+1) + \frac{1}{2} \ln G_{2}(N+1)  + \ln 2 \sum_{p<q}^{N_{5}} K_{p}(K_{q} - K_{p})  -  \ln \left( \frac{\pi}{2} \right) \sum_{p<q}^{N_{5}} \epsilon_{pq} K_{p} \\
        & \qquad -2 \sum_{p<q}^{N_{5}} K_{p} \ln  \left( \left( K_{q} - K_{p} \right) !! \right) -2 \sum_{p<q}^{N_{5}} \sum_{k=1}^{K_{p}-1} (K_{p} - k) \ln  \Big( \frac{K_{q} - K_{p}}{2} + k\Big)  \: .
    \end{split}
\label{ExactLocalization}    
\end{equation}
Equation (\ref{ExactLocalization}) is our exact expression for the boundary free energy, in the case with exclusively NS5-branes.

One particular case of interest is when we have $N$ D3-branes ending on $N_{5}$ NS5-branes of equal linking number $K = N / N_{5}$. In this case,
\begin{equation}
\label{NS5simpExact}
    \begin{split}
        F_{\partial} 
        & = - \left( \frac{N(N-N_{5})}{2}  + \frac{N}{4} \right) \ln (2 \pi) - \frac{N^{2}}{4} \ln \left( \frac{\lambda}{4 \pi^{2}N}  \right) -  \frac{\lambda}{48} \left({N^2 \over N_5^2}-1\right) \\
        & \qquad -N_{5}^{2} \ln G_{2}\left( {N \over N_5} +1 \right) + \frac{1}{2} \ln G_{2}(N+1)  \: .
    \end{split}
\end{equation}
This is the exact version of the supergravity expression (\ref{NS5simp}).

\subsection{General D5-Like Boundary Conditions}

We can obtain $F_{\partial}$ for a general D5-like boundary condition by applying an S-duality transformation to the above result, which simply amounts to replacing the NS5-brane linking numbers with D5-brane linking numbers, and performing an S-transformation to the gauge coupling $\frac{\lambda}{4 \pi N} \rightarrow \frac{4 \pi N}{\lambda}$. We thus obtain 
\begin{equation}
    \begin{split}
         F_{\partial} & = -\ln |Z[HS^{4}]| + \frac{1}{2} \ln Z[S^{4}] \\
        & = - \left( \frac{N(2N-1)}{4}  - \sum_{p=1}^{N_{5}} (N_{5} - p) L_{p} \right) \ln (2 \pi) - \frac{N^{2}}{4} \ln \left( \frac{4 N}{ \lambda}  \right) -  \frac{\pi^2 N}{3 \lambda} \left( \sum_{p=1}^{N_{5}} L_{p}^3 - N \right) \\
        & \qquad - \sum_{p=1}^{N_{5}} \ln G_{2}(L_{p}+1) + \frac{1}{2} \ln G_{2}(N+1)  + \ln 2 \sum_{p<q}^{N_{5}} L_{p}(L_{q} - L_{p})  - \ln \left( \frac{\pi}{2} \right) \sum_{p<q}^{N_{5}} \epsilon_{pq} L_{p}\\
        & \qquad - 2 \sum_{p<q}^{N_{5}} L_{p} \ln  \left( \left( L_{q} - L_{p} \right) !! \right) -2\sum_{p<q}^{N_{5}} \sum_{k=1}^{L_{p}-1} (L_{p} - k) \ln  \Big( \frac{L_{q} - L_{p}}{2} + k\Big) \: .
    \end{split}
    \label{ExactLocalizationD}  
\end{equation}

For $N$ D3-branes ending on $N_{5}$ D5-branes of equal linking number $L = N / N_{5}$, we obtain
\begin{equation}
\label{D5simpExact}
    \begin{split}
        F_{\partial} 
        & = - \left( \frac{N(N-N_{5})}{2}  + \frac{N}{4} \right) \ln (2 \pi) - \frac{N^{2}}{4} \ln \left( \frac{4 N}{\lambda}  \right) -  \frac{\pi^2 N^2}{3 \lambda} \left( {N^2 \over N_5^2}-1 \right) \\
        & \qquad -N_{5}^{2} \ln G_{2}\left(  {N \over N_5} +1 \right) + \frac{1}{2} \ln G_{2}(N+1)  \: .
    \end{split}
\end{equation}
This is the exact version of the supergravity expression (\ref{D5simp}).

\subsubsection*{Comparison with Supergravity Results}

We now compare the localization result (\ref{ExactLocalization}) with our supergravity results.

When the $K_{k}$ (and their differences) are taken to be large in (\ref{ExactLocalization}), then we can use the Euler-Maclaurin approximation for the last term to find
\begin{equation}
    \begin{split}
        -2  \sum_{p<q}^{N_{5}} \sum_{k=0}^{K_{p}-1} k \ln \left( \frac{K_{q} + K_{p}}{2} - k \right) & \approx -2  \sum_{p<q}^{N_{5}} \int_{k=0}^{K_{p}-1} dx \: x \ln \left( \frac{K_{q} + K_{p}}{2} - x \right) \\
        & = - \frac{1}{4} \sum_{p<q}^{N_{5}} \left( (K_{q} + K_{p})^{2} \ln (K_{q} + K_{p}) - (K_{q} - K_{p})^{2} \ln (K_{q} - K_{p}) \right) \\
        & \qquad - \sum_{p<q}^{N_{5}} K_{p}(K_{p} - K_{q}) \ln (K_{q} - K_{p}) \\
        & \qquad + \frac{1}{2} \sum_{p<q}^{N_{5}} K_{p} (K_{q} + 2(1 + \ln 2) K_{p}) + O(N_{5}^{2} K \ln K) \: .
    \end{split}
\end{equation}
Meanwhile, using the Stirling approximation, we find
\begin{equation}
\begin{split}
    \ln (M!!) & = \frac{M}{2} \ln M - \frac{M}{2} + O(\ln M) \: .
\end{split}
\end{equation}
Thus,
\begin{equation}
    -2  \sum_{p<q}^{N_{5}} K_{p} \ln ((K_{q} - K_{p})!!) = -2  \sum_{p<q}^{N_{5}} K_{p} \left( \frac{(K_{q} - K_{p})}{2} \ln (K_{q}  - K_{p}) - \frac{(K_{q}  - K_{p})}{2} \right) + O(N_{5}^{2} K \ln K) \: .
\end{equation}
We thus find 
\begin{equation}
    \begin{split}
        F_{\partial} & = -  \left( \frac{N^{2}}{2}  - \sum_{p<q}^{N_{5}} K_{p} \right) \ln (2 \pi)  - \frac{N^{2}}{4} \ln \left( \frac{\lambda}{4 \pi^{2}N}  \right) - \frac{\lambda}{48 N} \sum_{p=1}^{N_{5}} K_{p}^{3} \\
        & \qquad - \sum_{p=1}^{N_{5}} \left( \frac{K_{p}^{2}}{2} \ln K_{p} - \frac{3 K_{p}^{2}}{4} \right) + \frac{1}{2} \left( \frac{N^{2}}{2} \ln N - \frac{3 N^{2}}{4} \right)  + \ln 2 \sum_{p<q}^{N_{5}} K_{p}(K_{q} - K_{p})  \\
        & \qquad - \ln \left( \frac{\pi}{2} \right) \sum_{p<q}^{N_{5}} \epsilon_{pq} K_{p} - 2 \sum_{p<q}^{N_{5}} K_{p} \left( \frac{(K_{q} - K_{p})}{2} \ln (K_{q}  - K_{p}) - \frac{(K_{q}  - K_{p})}{2} \right) \\
        & \qquad - \frac{1}{4} \sum_{p<q}^{N_{5}} \left( (K_{q} + K_{p})^{2} \ln (K_{q} + K_{p}) - (K_{q} - K_{p})^{2} \ln (K_{q} - K_{p}) \right) \\
        & \qquad - \sum_{p<q}^{N_{5}} K_{p}(K_{p} - K_{q}) \ln (K_{q} - K_{p}) + \frac{1}{2} \sum_{p<q}^{N_{5}} K_{p} (K_{q} + 2(1 + \ln 2) K_{p}) + O(N_{5}^{2} K \ln K) \: .
    \end{split}
\end{equation}
Massaging this expression, we arrive at
\begin{equation}
    \begin{split}
        F_{\partial}
        &  =  \frac{N^{2}}{4} \left( \frac{3}{2} +  \ln \left( \frac{4 N^{2}}{\lambda}  \right) \right) - \frac{\lambda}{48 N} \sum_{p=1}^{N_{5}} K_{p}^{3}  \\
        & \qquad - \frac{1}{16} \sum_{p, q}^{N_{5}} \left( (K_{q} + K_{p})^{2} \ln \left( (K_{q} + K_{p})^{2} \right) - (K_{q} - K_{p})^{2} \ln \left( (K_{q} - K_{p})^{2} \right) \right) \\
        & \qquad + O(N_{5}^{2} K \ln K) \: .
    \end{split}
\end{equation}
This limit exactly reproduces our result from the supergravity calculation. We can similarly check that the exact expression for general D5-brane boundary conditions reproduces the supergravity answer when the linking numbers and their differences are large.

\subsubsection*{Comparison for finite $N$}

We can also compare the exact results with the supergravity results for finite $N$. We note that the $\log \lambda$ term agrees exactly between the supergravity and localization calculations, while the term of order $\lambda$ in the NS5-brane supergravity expression (or $1/\lambda$ in the D5-brane expression) becomes exact under the replacement
\be
\sum_A K_A^3 \to \sum_A (K_A^3 - K_A) \; 
\ee
(or the same replacement with $L_A$ for the D5-brane expression).

\begin{figure}
\centering
\includegraphics[width=70mm]{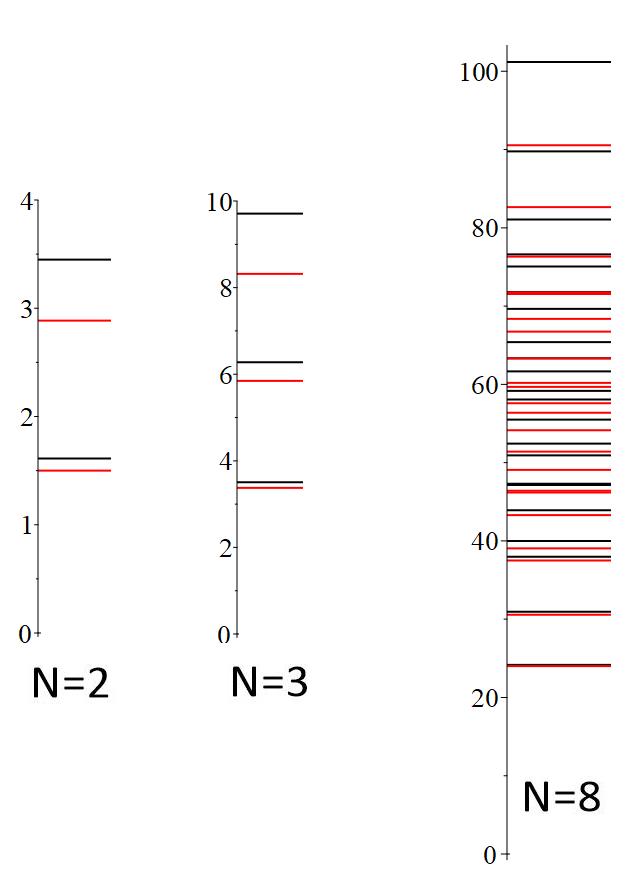}
\caption{The $\lambda$-independent part of the spectrum of possible boundary $F$ values for $U(N)$ ${\cal N}=4$ SYM theory with $N=2,3,8$. Black lines represent the exact values while red lines give the supergravity approximation.}
\label{fig:Spectrum}
\end{figure}

The remaining terms are $\lambda$-independent. It is straightforward to calculate these for all possible boundary conditions for small fixed values of the gauge group rank $N$ and compare supergravity results with the exact results. For $N=2$,$N=3$, and $N=8$, this $\lambda$-independent part of the spectrum of boundary $F$ values is shown in Figure \ref{fig:Spectrum}. 

We see that the results agree reasonably well even for small values of $N$. As an example, for the  $N=8$ case, the $\lambda$-independent parts of the boundary $F$ values for linking numbers
\bea
&([1, 1, 1, 1, 1, 1, 1, 1], [1, 1, 1, 1, 1, 1, 2], [1, 1, 1, 1, 2, 2], [1, 1, 2, 2, 2],  [2, 2, 2, 2],  [1, 1, 1, 1, 1, 3],\cr & [1, 1, 1, 2, 3], [1, 2, 2, 3], [1, 1, 3, 3], [2, 3, 3], [1, 1, 1, 1, 4],  [1, 1, 2, 4], [2, 2, 4], [1, 3, 4], [4, 4], \cr & [1, 1, 1, 5], [1, 2, 5], [3, 5],   [1, 1, 6], [2, 6], [1, 7], [8])
\eea
are (rounded to the nearest integer)
\be
(101, 89, 81, 75, 71, 76, 69, 65, 61, 59, 63, 58, 55, 52, 47, 50, 47, 43, 39, 37, 30, 24)
\ee
using the exact results and
\be
(90, 82, 76, 71, 68, 71, 66, 63, 59, 57, 60, 56, 54, 51, 46, 49, 46, 43, 39, 37, 30, 24)
\ee
with the supergravity expressions. 

\section{Statistics of Boundary \texorpdfstring{$F$}{}} \label{sec:stats}

In this section, we will use our results above to investigate the distribution of possible values for $F_{\partial}$ for a given $N$, for various types of boundary conditions.

For fixed $\lambda$ and $N$, there are infinitely many superconformal boundary conditions that one can impose, since we can couple in an arbitrarily complicated 3D SCFT. We expect that there is a lower bound, but no upper bound on the allowed value of $F_{\partial}$, which can be thought of as a measure of the number of local boundary degrees of freedom.

For the class of theories corresponding to D3-branes ending on D5-branes only or NS5-branes only, we have only a finite set of possibilities, enumerated by partitions of $N$, the rank of the gauge group. In this case, we have upper and lower bounds for $F_{\partial}$ that depend on $N$ and $\lambda$, and we can investigate the distribution of $F_{\partial}$ values for a given $N$ and $\lambda$ either using the supergravity expressions or the exact results from localization.

\subsubsection*{D5-Brane Boundary Conditions}

Defining $p_{A} = L_{A} / N$, our supergravity expression for $F_{\partial}$ for the theories associated with D3-branes ending on D5-branes is
\begin{equation}
\begin{split}
    F^{\textnormal{SUGRA}}_{\partial} & =  \frac{N^{2}}{4} \Big( \frac{3}{2} + \ln \Big( \frac{\lambda}{4 \pi^{2} N^{2}} \Big) \Big) - \frac{\pi^2 N^{4}}{3} \sum_{A} p_{A}^{3}  \\
    & \qquad  - \frac{N^{2}}{16} \sum_{A, B} \Big[ (p_{A} + p_{B})^{2} \ln \big( (p_{A} + p_{B})^{2} \big) - (p_{A} - p_{B})^{2} \ln \big( (p_{A} - p_{B})^{2} \big) \Big] \: ,
\end{split}
\end{equation}
where the positivity of $L_i$ and the relation $\sum_{A} L_{A} = N$ give $p_{A} \geq 0$ and $\sum_{A} p_{A} = 1$. Thus $\{ p_A\}$ satisfies the constraints of a probability distribution.

In Appendix \ref{sec:StatDetails}, we show that the minimum and maximum values of $F^{\textnormal{SUGRA}}_{\partial}$ are obtained by considering the distribution $\{ p_A\}$ with the minimum and maximum entropy respectively, i.e. where $\{ p_A\} = \{ 1\}$ and $\{ p_A\} = \{ 1/N,\dots, 1/N\}$. This yields
\begin{equation}
\label{rangeS}
N^{2} \left(- \frac{ \pi^2 N^{2}}{3 \lambda}  - {1 \over 4}\ln \Big( \frac{16 \pi^{2} N^{2}}{\lambda} \Big) +  \frac{3}{8} \right) \leq
     F^{\textnormal{SUGRA}}_{\partial}\leq N^{2} \left(  {1 \over 4} \ln \Big( \frac{\lambda}{16 \pi^{2}} \Big) + \frac{3}{8} - \frac{ \pi^{2}}{3 \lambda} \right)  \: .
\end{equation}
Assuming that the same sets of linking numbers lead to the minimum and maximum values for $F_{\partial}$ with the exact expression, we find a range of allowed values 
\be 
F_{\partial}^{-} \le F_{\partial} \le F_{\partial}^{+}
\ee
where $F_{\partial}^{+}$ corresponds to the maximum entropy configuration and is given by (setting $L=1$ in (\ref{D5simpExact}))
\begin{equation}
    \begin{split}
        F_{\partial}^+ & = -\frac{N}{4} \ln (2 \pi) - \frac{N^{2}}{4} \ln \left( \frac{4 N}{\lambda}  \right)  + \frac{1}{2} \ln G_{2}(N+1) \: ,
    \end{split}
\end{equation}
and $F_{\partial}^{-}$ corresponds to the minimum entropy configuration and is given by (setting $L = N$ in (\ref{D5simpExact}))
\begin{equation}
    \begin{split}
        F_{\partial}^-
        & = - \left( \frac{N^{2}}{2}  - \frac{N}{4} \right) \ln (2 \pi) - \frac{N^{2}}{4} \ln \left( \frac{4 N}{\lambda}  \right) -  \frac{\pi^2 N^2} {3 \lambda} (N-1)(N+1) - \frac{1}{2} \ln G_{2} \left( N +1 \right)   \: .
    \end{split}
\end{equation}
Using the large $N$ approximation to the Barnes G-function, we then find that up to ${\cal O}(N \ln N)$ corrections we have a range of allowed values
\begin{equation}
   N^{2} \left( -\frac{\pi^2 N^2}{3\lambda} - {1 \over 4} \ln \left( {16 \pi^2 N^2 \over \lambda} \right)  + \frac{3}{8} \right)   \leq F_{\partial}^{D5} \leq N^{2} \left( {1 \over 4} \ln \left( \frac{ \lambda}{4} \right)  - {3 \over 8} \right)   \: .
\end{equation}
We note that the upper bound is modified here compared to the supergravity result (\ref{NSrangeS}). We emphasize that we have not proven that the left and right sides here are actually the upper and lower bounds on $F_{\partial}$; this will be true assuming that the same boundary conditions giving rise to the minimum and maximum for $F^{\textnormal{SUGRA}}_{\partial}$ also give rise to the minimum and maximum for $F_{\partial}$.

We see that this allowed range covers primarily negative values, with the upper end of the range positive only for sufficiently large $\lambda$. We can understand the large negative values of boundary $F$ that arise for boundary conditions associated with D3-branes ending on few D5-branes by the fact that the scalars are developing an expectation value, and this results in a large fraction of the $N^2$ fields becoming massive, with mass increasing as we approach the boundary. Thus, we lose degrees of freedom compared with the situation where the scalar vevs are vanishing. The quantity boundary $F$ is in some sense a measure of the number of boundary degrees of freedom, but in this case, the negative value indicates that it is taking away from the bulk degrees of freedom.

\subsubsection*{NS5-Brane Boundary Conditions}

A similar analysis applies to the NS5-brane boundary conditions. Defining $p_{A} = K_{A} / N$, we have  
\begin{equation}
\begin{split}
     F^{\textnormal{SUGRA}}_{\partial} & =  \frac{N^{2}}{4} \left( \frac{3}{2} + \ln \Big( \frac{4}{\lambda} \Big) \right) - \frac{\lambda N^{2}}{48} \sum_{A} p_{A}^{3}  - \frac{N^{2}}{16}\\
    & \qquad  - \frac{N^{2}}{16} \sum_{A, B} \Big[ (p_{A} + p_{B})^{2} \ln \big( (p_{A} + p_{B})^{2} \big) - (p_{A} - p_{B})^{2} \ln \big( (p_{A} - p_{B})^{2} \big) \Big] \: ,
\end{split}
\end{equation}
A similar argument to that for the D5-brane boundary conditions shows that $F^{\textnormal{SUGRA}}_{\partial}$ is again minimized/maximized on the minimum/maximum entropy distribution, yielding
\begin{equation}
\label{NSrangeS}
N^{2} \left( - \frac{\lambda}{48} - {1 \over 4}\ln \lambda + \frac{3}{8} \right) 
      \leq   F^{\textnormal{SUGRA}}_{\partial} \leq   N^{2} \left(- \frac{\lambda}{48 N^{2}} + {1 \over 4} \ln \Big( \frac{N^{2}}{\lambda} \Big) + \frac{3}{8}   \right) \: .
\end{equation}

Assuming that the same sets of linking numbers lead to the minimum and maximum values for $F_{\partial}$ with the exact expression, we find a range of allowed values 
\be 
F_{\partial}^{-} \le F_{\partial} \le F_{\partial}^{+}
\ee
where $F_{\partial}^{-}$ corresponds to the ``minimum entropy" configuration and is given by (setting $K = N$ in (\ref{NS5simpExact}))
\begin{equation}
    \begin{split}
        F_{\partial} ^+
        & = - \left( \frac{N^{2}}{2}  - \frac{N}{4} \right) \ln (2 \pi) - \frac{N^{2}}{4} \ln \left( \frac{\lambda}{4 \pi^{2}N}  \right) -  \frac{\lambda}{48} (N-1)(N+1) - \frac{1}{2} \ln G_{2} \left( N +1 \right)   \: ,
    \end{split}
\end{equation}
and $F_{\partial}^{+}$ corresponds to the ``maximum entropy" configuration and is given by (setting $K=1$ in (\ref{NS5simpExact}))
\begin{equation}
    \begin{split}
       F_{\partial}^- & = -\frac{N}{4} \ln (2 \pi) - \frac{N^{2}}{4} \ln \left( \frac{\lambda}{4 \pi^{2}N}  \right)  + \frac{1}{2} \ln G_{2}(N+1) \: ,
    \end{split}
\end{equation}
Using the large $N$ approximation to the Barnes G-function, we then find that up to ${\cal O}(N \ln N)$ corrections, we have a range of allowed values
\begin{equation}
N^{2} \left( -  \frac{\lambda}{48} - \frac{1}{4} \ln \lambda    + \frac{3}{8}  \right) \leq F_{\partial}^{NS5} \leq N^{2} \left(  \frac{1}{4} \ln \left( \frac{4 \pi^{2}N^{2}}{\lambda}  \right) - \frac{3}{8} \right)    \: .
\end{equation}
As above, the upper bound is modified here compared to the supergravity result (\ref{NSrangeS}), which is expected since the linking numbers are not large in this case. We emphasize that we have not proven that the left and right sides here are actually the upper and lower bounds on $F_{\partial}$; this will be true assuming that the same boundary conditions giving rise to the minimum and maximum for $F^{\textnormal{SUGRA}}_{\partial}$ also give rise to the minimum and maximum for $F_{\partial}$.

We see that at least for small values of $\lambda$, the range of allowed boundary $F$ values for these boundary conditions is positive, consistent with the fact that the scalar vevs are zero for these boundary conditions and the full set of massless bulk degrees of freedom remain.

\begin{figure}[h]
\centering
\includegraphics[width=75mm]{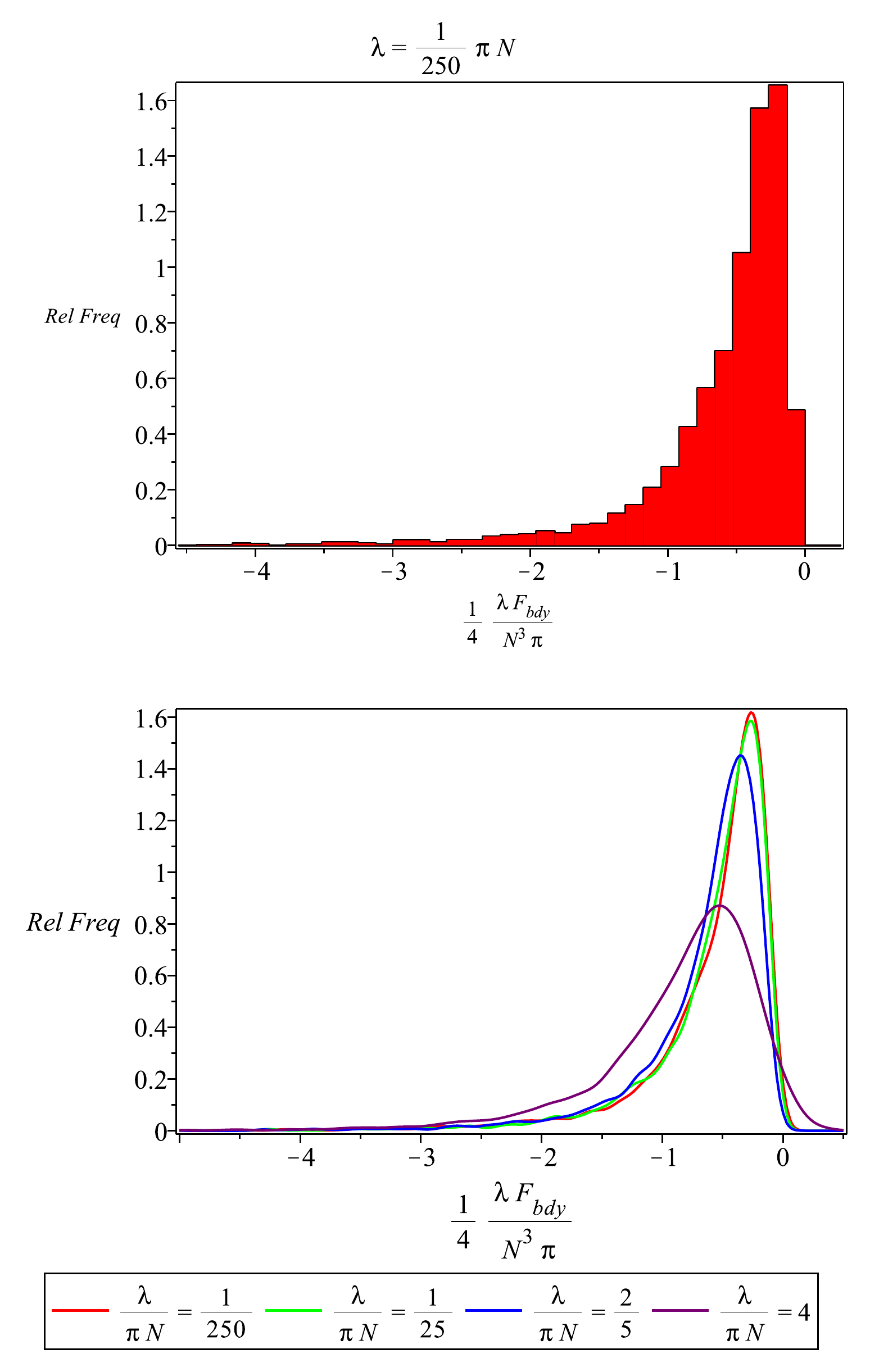}
\includegraphics[width=75mm]{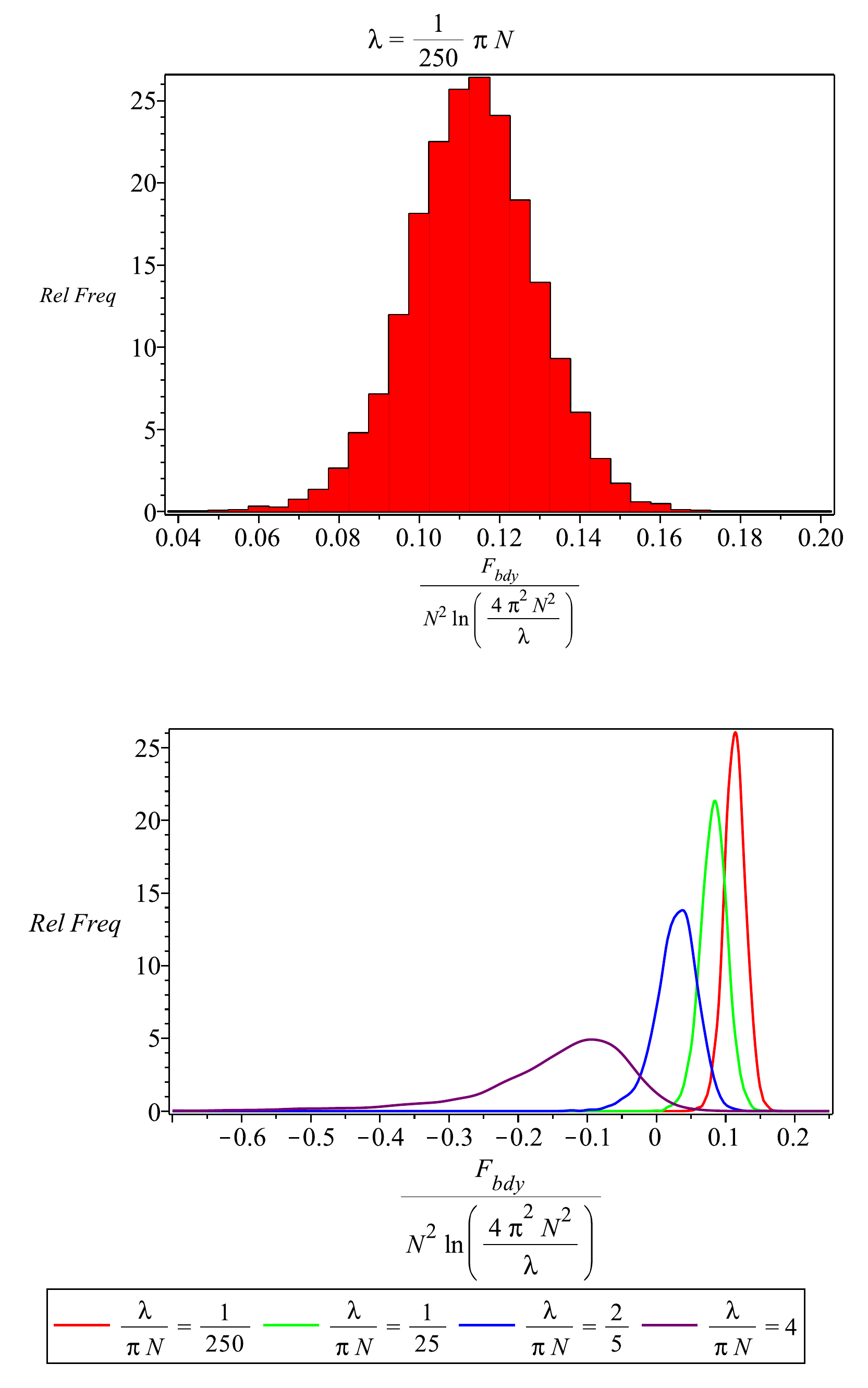}
\caption{\textbf{Top Left:} Histogram of values of $\frac{\lambda}{4 \pi N} \frac{F_{\partial}}{N^{2}}$ for D5-like boundary conditions, with $N=100$ and $\frac{\lambda}{4 \pi N} = 10^{-3}$. \textbf{Bottom Left:} Contours of histograms of $\frac{\lambda}{4 \pi N} \frac{F_{\partial}}{N^{2}}$ (bins removed for clarity) for D5-like boundary conditions, with $N=100$ and various values of $\lambda$ up to the self-dual value $\lambda = 4 \pi N$. \textbf{Top Right:} Histogram of values of $\frac{1}{\ln \left( 4 \pi^{2} N^{2} /\lambda \right)} \frac{F_{\partial}}{N^{2}}$ for NS5-like boundary conditions, with $N=100$ and $\frac{\lambda}{4 \pi N} = 10^{-3}$. \textbf{Bottom Right:} Contours of histograms of $\frac{1}{\ln \left( 4 \pi^{2} N^{2} /\lambda \right)} \frac{F_{\partial}}{N^{2}}$ (bins removed for clarity) for NS5-like boundary conditions, with $N=100$ and various values of $\lambda$ up to the self-dual value $\lambda = 4 \pi N$. For each histogram, we uniformly sample 5000 partitions of the integer $N$, and compute $F_{\partial}$ for the associated boundary conditions.}
\label{fig:FhistsN100}
\end{figure} 

\subsubsection*{Distribution of Boundary $F$ Values}

It is also of interest to ask about the distribution of allowed $F_{\partial}$ values for a given $N$ and $\lambda$. In Figure \ref{fig:FhistsN100}, we display contour plots for histograms of allowed values (scaled by positive factors involving $\lambda$ and $N$ for convenience) for the case $N=100$ with various values of $\lambda$. We display the results for D5-brane and NS5-brane boundary conditions with up to the self-dual value $\lambda = 4 \pi N$ for the 't Hooft coupling; these confirm that, for $\lambda$ below the self-dual value, $F_{\partial}$ is predominantly negative for D5-brane boundary conditions, and predominantly positive for NS5-brane boundary conditions. These plots also implicitly reveal the behaviour of $F_{\partial}$ for $\lambda$ above the self-dual value; the distribution of $F_{\partial}$ for D5-brane boundary conditions with such $\lambda$ is identical to that for NS5-brane boundary conditions with the dual value of the 't Hooft coupling, and vice versa.  

\begin{figure}[h]
\centering
\includegraphics[width=75mm]{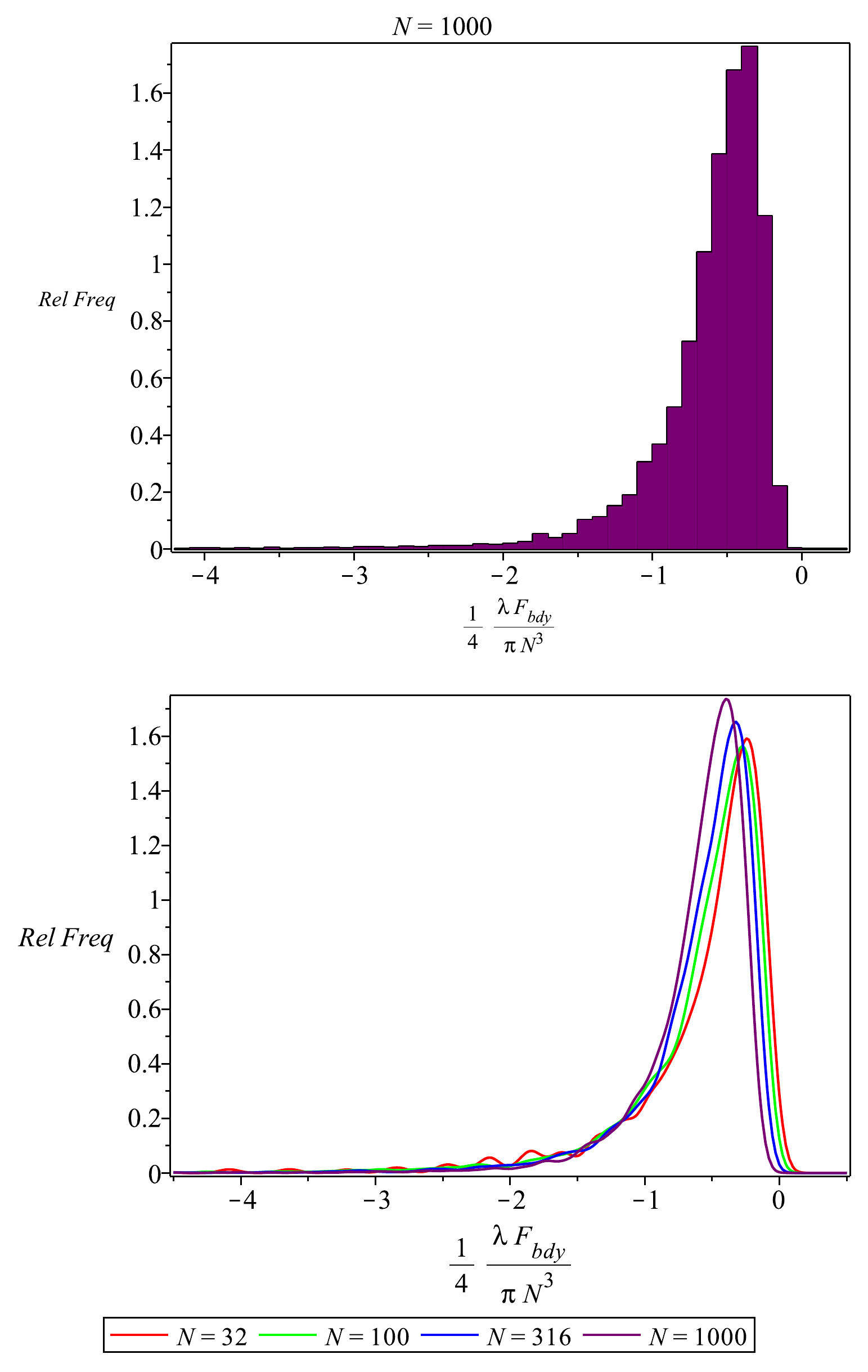}
\includegraphics[width=75mm]{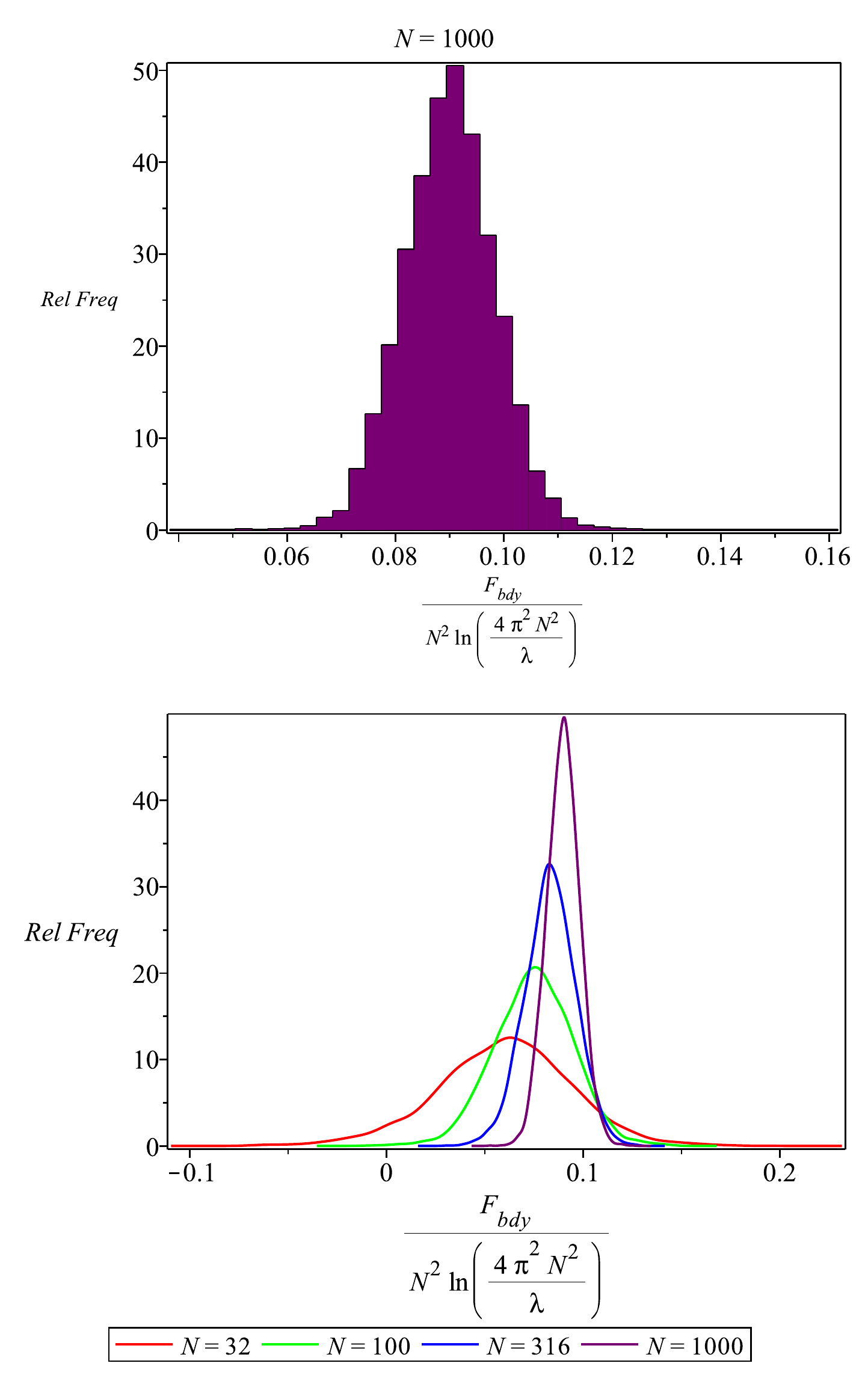}
\caption{\textbf{Top Left:} Histogram of values of $\frac{\lambda}{4 \pi N} \frac{F_{\partial}}{N^{2}}$ for D5-like boundary conditions, with $\lambda=20$ and $N=10^{3}$. \textbf{Bottom Left:} Contours of histograms of $\frac{\lambda}{4 \pi N} \frac{F_{\partial}}{N^{2}}$ (bins removed for clarity) for D5-like boundary conditions, with $\lambda=20$ and various values of $N$. \textbf{Top Right:} Histogram of values of $\frac{1}{\ln \left( 4 \pi^{2} N^{2} /\lambda \right)} \frac{F_{\partial}}{N^{2}}$ for NS5-like boundary conditions, with $\lambda = 20$ and $N = 10^{3}$. \textbf{Bottom Right:} Contours of histograms of $\frac{1}{\ln \left( 4 \pi^{2} N^{2} /\lambda \right)} \frac{F_{\partial}}{N^{2}}$ (bins removed for clarity) for NS5-like boundary conditions, with $\lambda = 20$ and various values of $N$. For each histogram, we uniformly sample 5000 partitions of the integer $N$, and compute $F_{\partial}$ for the associated boundary conditions. }\label{fig:Fhistsl20}
\end{figure}

\begin{figure}[h]
\centering
\includegraphics[width=155mm]{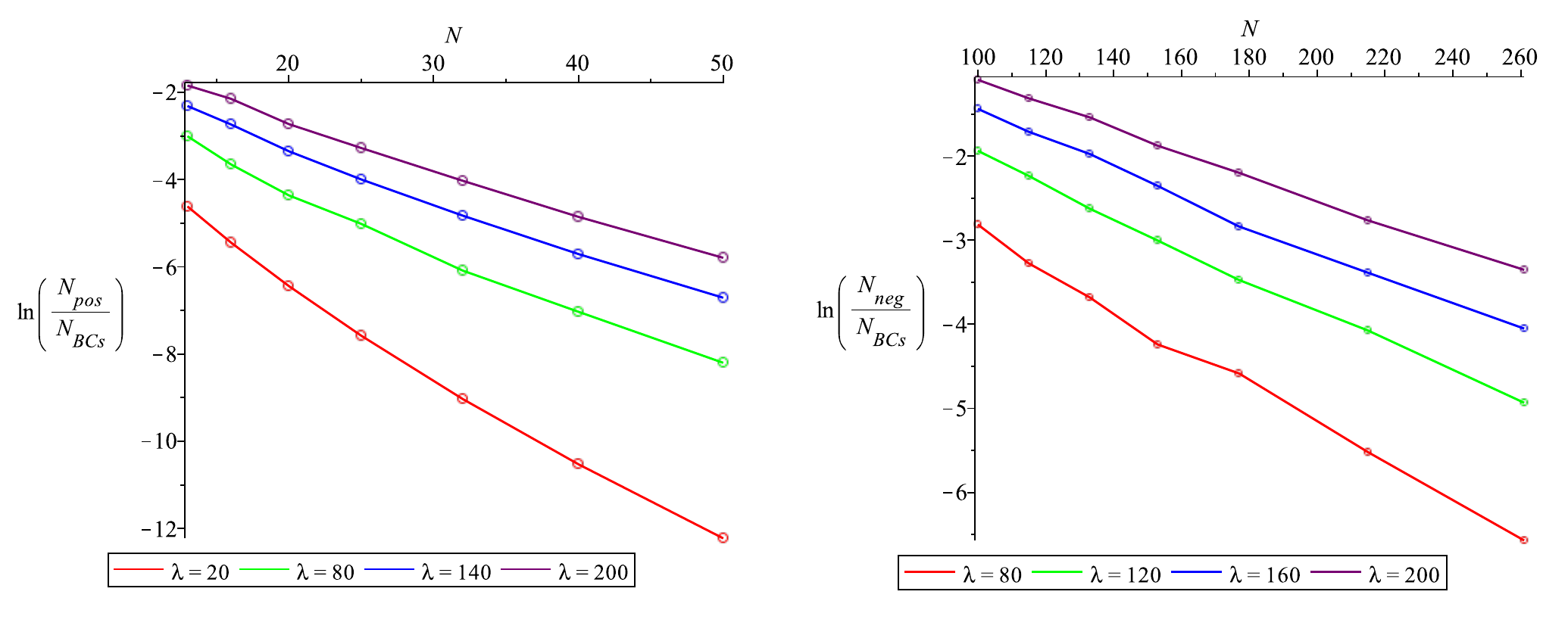}
\caption{\textbf{Left:} Logarithm of the proportion of D5-like boundary conditions giving rise to positive $F_{\partial}$, for various values of $\lambda$ and increasing $N$. Values are exact, as we include every possible such boundary condition. \textbf{Right:} Logarithm of the proportion of NS5-like boundary conditions giving rise to negative $F_{\partial}$, for various values of $\lambda$ and increasing $N$. Each point is based on 5000 uniformly sampled partitions of the integer $N$.}
\label{fig:prop_negs}
\end{figure}

We also display similar plots for the case of fixed $\lambda$ and increasing $N$ in Figure \ref{fig:Fhistsl20}. One notable feature of these plots is that, for fixed $\lambda$, the proportion of D5-like/NS5-like boundary conditions for which $F_{\partial}$ is positive/negative appears to asymptote to zero for increasing $N$; this is illustrated further in Figure \ref{fig:prop_negs}. 

\subsubsection*{Arbitrarily Large Boundary $F$ for General Boundary Conditions}

To conclude this section, we verify the claim that by considering general boundary conditions involving D5-branes and NS5-branes, we can make $F_{\partial}$ arbitrarily large. This is expected, since the general boundary conditions can be understood as coupling in a SCFT to one of the theories with D5-branes or NS5-branes only, and we can take this SCFT to have arbitrarily many degrees of freedom. We are therefore motivated to verify this claim by considering such a boundary condition with a large number of boundary degrees of freedom; for simplicity, we consider the case of a single stack of many D5-branes and a single stack of many NS5-branes, with linking numbers
\begin{equation}
    \tilde{L} = - 1 \: , \qquad K = 1 \: ,
\end{equation}
and with $N, N_{NS5}$ taken to be large independent parameters, with $N \ll N_{NS5}$. We then have
\begin{equation}
    N_{D5} =  N_{NS5} - N \: .
\end{equation}
The supergravity parameters $\hat{l}, \hat{k}$ are given by
\begin{equation}
    -1 = \hat{l} - \frac{2}{\pi} N_{NS5} \arctan(g \hat{k} / \hat{l}) \: , \qquad 1 = \hat{k} + \frac{2}{\pi}(N_{NS5} - N) \arctan(g \hat{k} / \hat{l} ) \: ,
\end{equation}
which has perturbative solution
\begin{equation}
    \begin{split}
        \hat{k} & = \frac{2 \pi^{2}}{\lambda} \frac{N^{2}}{N_{NS5}^{2}} + O \left( \frac{N^{3}}{N_{NS5}^{3}} \right) \\
        \hat{l} & = \frac{N}{N_{NS5}} + O \left( \frac{N^{2}}{N_{NS5}^{2}} \right) \: .
    \end{split}
\end{equation}
Most of the terms appearing in the uncorrected $F_{\partial}$ in the case of this boundary condition are suppressed by $\frac{N}{N_{NS5}}$, and will vanish in the limit $\frac{N}{N_{NS5}} \rightarrow 0$ with fixed $\lambda$ and $N$; the terms which are not suppressed in this limit are the constant contribution $\frac{3}{8}N^{2}$, the ``cubic terms"
\begin{equation}
        - \frac{1}{\pi} c^{2} d l D \left[ \frac{1}{2} - \frac{i k}{2l} \right] \sim N N_{NS5} \ln 2 \: , \qquad 
        - \frac{1}{\pi} d^{2} c k D \left[ \frac{1}{2} - \frac{i l}{2k} \right] \sim \frac{2 \pi^{2}}{\lambda}  N^{2} \ln(N_{NS5})  \: ,
\end{equation}
and the ``quartic term"
\begin{equation}
    \begin{split}
        - {1 \over 2 \pi^2} c^{2} d^{2} \left\{
{\cal L} \left(  \frac{(k+il)^{2}}{(k-il)^{2}} \right) + {\cal L} \left( \frac{ (k+il)^{2}}{4 i k l} \right) - \zeta(3) \right\} \sim  N_{NS5}^{2} \ln (N_{NS5}) \: .
    \end{split}
\end{equation}
Meanwhile, the anticipated corrections from the vicinity of the D5-branes and NS5-branes are $O(N_{NS5}^{2})$ (see Appendix \ref{sec:corrections}). Consequently, the leading term in the uncorrected $F_{\partial}$, which is $N_{NS5}^{2} \ln N_{NS5}$, should provide a good approximation to $F_{\partial}$ when $N_{NS5} \gg N$. Since $N_{NS5}$ can take arbitrarily large values, we see that $F_{\partial}$ is unbounded from above.

\section{Discussion}

In this final section, we mention a few possible applications of our results. 

\subsubsection*{RG Ordering of BCFTs}

We recall that $F_{\partial}$ has been conjectured to decrease under boundary renormalization group flows. Assuming that this is true, our results provide very detailed information about which boundary RG flows are possible between the various BCFTs we consider. For cases where the endpoints of an RG flow are known, for example where we add supersymmetric mass terms or Fayet-Illiopoulos parameters to a UV theory, it would be interesting to verify the decrease of boundary $F$ to provide support for the conjecture; this was done in \cite{Estes:2014hka} for the simple case considered there.

We note that for $N > 7$, the ordering of boundary $F$ for different theories depends on the bulk 't Hooft coupling parameter $\lambda$. Thus, if the boundary $F$ monotonicity conjecture is correct, we could have the interesting situation where some relevant perturbation of theory $A$ flows to theory $B$ for small values of $\lambda$ while some relevant perturbation of theory $B$ flows to theory $A$ for large values of $\lambda$. Of course, it may also be the case that no RG flows are possible between theories whose boundary $F$ values switch orderings as a function of $\lambda$.

\subsubsection*{Holographic Interpretation}

As we discussed in Section \ref{sec:solutions}, the addition of a boundary to the ${\cal N}=4$ theory corresponds to the addition of a certain type of ``end of the world'' brane in the five-dimensional gravity picture. This corresponds in the higher-dimensional picture to a region where the internal space smoothly degenerates. In many holographic applications of BCFTs, the gravity side is described using a bottom-up approach, in which such an ETW brane is simply described by adding a boundary action with certain parameters to the bulk gravitational theory \cite{Karch:2000ct,Takayanagi:2011zk}. The simplest such parameter is the tension of the ETW brane. An interesting question, one of the questions that motivated this work, is to understand the range of tension parameters in bottom-up models for which the qualitative physics can be reproduced in microscopic constructions. 

As discussed in \cite{Takayanagi:2011zk}, there is a direct relationship between the tension parameter of a bottom up model and the boundary entropy, obtained by performing a holographic calculation of boundary $F$ as a function of this tension. We provide this calculation in the four-dimensional case in Appendix \ref{BottomUp}, with the result that
\be
\label{FvsT}
F = c_{bulk} \left( {T \over 1 - T^2} + {1 \over 2} \ln{1 + T \over 1 - T}\right) \; .
\ee
where we define $c_{bulk} = (L^3 \pi / 4G)$ and the tension is $3L_{AdS}T/(8 \pi G)$. This provides a guide to choosing the tension parameter if one wishes to model the physics of our more detailed microscopic theories using a bottom-up model.

\subsubsection*{Generalizations}

There is a significantly larger class of theories with the same symmetry as the theories considered in this paper. The more general theories correspond to ${\cal N}=4$ SYM theory with a supersymmetric planar defect, or to supersymmetric interfaces between ${\cal N}=4$ SYM theories with different parameters. Type IIB supergravity solutions for these theories are also known, so it should be straightforward to use the methods of this paper to calculate the defect/interface entropy for these theories. 

\section*{Acknowledgments}

We would like to thank David Gaiotto, Jaume Gomis, Michael Gutperle, and Kristan Jensen for helpful comments. This work is supported in part by the Natural Sciences and Engineering Research Council of Canada and the Simons Foundation via the It From Qubit Collaboration and a Simons Investigator Award. 

\appendix

\section{AdS/CFT Correspondence: Conventions}
\label{app:conventions}

We here establish various formulae relevant to type IIB string theory. The Planck scale and string scale are related by
\be
\ell_p = g^{1 \over 4} \ell_s \; ,
\ee
where $g$ is the string coupling and $\ell_s$ is defined in terms of the string tension $\frac{1}{2 \pi \alpha'}$ by
\be
\alpha' = \ell_s^2 \; .
\ee
The ten-dimensional Newton constant is defined as
\be
G = 8 \pi^6 g^2 \ell_s^8 \; .
\ee
In the AdS/CFT correspondence relating $U(N)$ ${\cal N} = 4$ SYM theory to type IIB string theory on AdS$_{5} \times S^5$, we have that the AdS radius in string frame is related to the rank of the gauge group by
\be
\label{LN}
L_{SF}^4 = 4 \pi g N \ell_s^4 \; .
\ee
If we make the transformation $g_{\mu \nu} \to e^{- \phi} g_{\mu \nu}$ to Einstein frame, including the asymptotic value of the dilaton $g = e^\phi_{\infty}$ in $\phi$, this becomes
\be
\label{LNE}
L^4 = 4 \pi N \ell_s^4 \; .
\ee
In this case, we should use $G = 8 \pi^6 \ell_s^8$ for the Newton constant. 
The string coupling (equal to the asymptotic value of $e^\phi$ where $\phi$ is the dilaton) is related to the Yang-Mills coupling by
\be
4 \pi g = g^2_{YM} \; .
\ee
The 't Hooft coupling is
\be
\lambda = g_{YM}^2 N \; .
\ee

To evaluate the number of units of quantized 3-form flux through a sphere, we use
\be
\label{Nd}
N_{D5} = {1 \over 4 \pi^2 \ell_s^2} \int_{S^3} F_3
\ee
and
\be
\label{Nn}
N_{NS5} = {1 \over 4 \pi^2 \ell_s^2} \int_{S^3} H_3
\ee
where $F_3$ and $H_3$ are the R-R and NS-NS three-form field strengths. In the absence of three-form fields, the number of units of five-form flux through a five-sphere is given by
\be
N_{D3} = {1 \over 16 \pi^4 \ell_s^4} \int_{S^5} F_5 \: .
\ee
The analysis of five-form fluxes and their relation to D3-brane charges is more subtle when three-form fields are present (as they are in the solutions we consider). See \cite{Marolf:2000cb, Aharony:2011yc} or \cite{Assel:2011xz} for a detailed discussion.

\section{Supergravity Solutions: Form Fields}
\label{app:formfields}

In this appendix, we review for completeness the gauge fields in the supergravity solutions, following the conventions of \cite{Assel:2011xz}.

The form fields are again expressed in terms of the harmonic functions $h_i$ together with the harmonic duals $h_i^D$ defined so that  
\beas
{\cal A}_1 &=& {1 \over 2} (h_1^D + i h_1) \cr
{\cal A}_2 &=& {1 \over 2} (h_2 - i h_2^D)
\eeas
are holomorphic. The ambiguity in choosing $h_i^D$ corresponds to gauge freedom in defining the potentials for the form fields. 

The NS-NS 3-form field strength $H_3$ and the R-R 3-form field strength $F_3$ take the form
\be
H_3 = \omega^{45} \wedge db_1 \qquad \qquad F_3 = \omega^{67} \wedge db_2
\ee
where $\omega^{45}$ and $\omega^{67}$ are volume forms on the first and second unit-radius $S^2$s. The real functions $b_i$ are defined in terms of the harmonic functions by
\begin{equation}
    b_{1} = 2 h^D_{2} + 2 h_{1}^{2} h_{2} \frac{X}{N_{1}} \: , \qquad b_{2} = - 2 h^D_{1} + 2 h_{1} h_{2}^{2} \frac{X}{N_{2}} \: ,
\end{equation}
where
\be
X \equiv i \left( \partial_{w} h_{1} \partial_{\bar{w}} h_{2} - \partial_{w} h_{2} \partial_{\bar{w}} h_{1} \right) \: .
\ee
The fiveform field strength can be expressed as
\be
F_{5} = -4 f_4^4 \omega^{0123} \wedge {\cal F} + 4 f_1^2 f_2^2 \omega^{45} \wedge \omega^{67} \wedge (*_2 {\cal F}) \; .
\ee
Here, $\omega^{0123}$ is the volume form on the unit-radius $AdS_4$, ${\cal F}$ is a one-form on $\Sigma$, and $*_2$ denotes Poincar\/'e duality with respect to the metric on $\Sigma$.

We have that 
\be
f_4^4 {\cal F} = d j_1
\ee
where
\beas
j_1 &=& 3 {\cal C} + 3 \bar{\cal C} - 3 {\cal D} + {h_1 h_2 X \over W} \cr
\partial_w {\cal C} &=& {\cal A}_1 \partial_w {\cal A}_2 - {\cal A}_2 \partial_w {\cal A}_1 \cr
{\cal D} &=& \bar{{\cal A}}_1 {\cal A}_2 + {\cal A}_1 \bar{{\cal A}}_2 \: .
\eeas
So far, we have assumed that the R-R zero-form potential vanishes, but more general solutions with non-vanishing axion can be obtained using the $SL(2, \mathbb{R})$ symmetry of type IIB supergravity. 

\section{Regularization of the Area Integrals}
\label{app:regularization}

In this appendix we explain in detail the regularization procedure used in computing boundary $F$ via the RT formula. Given the metric dual to one of the BCFTs, we can redefine coordinates to place the metric in Fefferman-Graham form
\be
ds^2 = {L^2 \over Z^2} (dZ^2 + dY^2 - dt^2 + d \vec{x}^2) + d \Omega_5^2 + {\cal O}(Z^2) 
\ee
where the correction terms do not involve $dZ$. We then compute the area of the $Z >  \epsilon$ portion of the RT surface for a half-ball region of radius $R$ centered on the BCFT boundary, and subtract half the area of the RT-surface for a ball of radius $R$ in ${\cal N}=4$ SYM theory.

\subsubsection*{Regulated Area in the BCFT Duals}

To calculate the regulated area of the extremal surface  corresponding to a half ball in one of our BCFT duals, we need to understand where the cutoff surface $Z=\epsilon$ lies in the coordinates we are using. Representing the AdS$_{4}$ metric as
\be
ds_{AdS^4}^2 = {1 \over u^2} (-dt^2 + du^2 + dx_\perp^2) = {1 \over \rho^2 \cos^2\theta} (-dt^2 + d\rho^2 + \rho^2 d\theta^2 + \rho^2 \sin^2 \theta d \phi^2) \: ,
\ee
we will have that the cutoff surface lies at some $u_{min}(r,\theta)$. In the full metric, this AdS$_{4}$ slice enters as
\be
ds^2 = f_4^2 ({1 \over u^2} (-dt^2 + du^2 + dx_\perp^2)) + \dots
\ee
Converting to Fefferman-Graham coordinates, this will become asymptotically
\be
ds^2 = {L^2 \over Z^2} (dZ^2 + dY^2 - dt^2 + dx_\perp^2)  + \dots
\ee
where $Z$ and $Y$ are determined in terms of the other coordinates, and $L$ is the asymptotic AdS curvature scale that will be determined in terms of the parameters appearing in the metric. Thus, asymptotically, we must have that
\be
{f_4^2 \over u^2} = {L^2 \over Z^2} \: .
\ee
This allows us to fix the cutoff surface as
\be
u_{min}(r,\theta) = {\epsilon \over L} f_4(r,\theta) \; .
\ee

The locus of the extremal surface in each AdS$_{4}$ slice is $\rho^2 = u^2 + x^2 + y^2 = R^2$, and the two-dimensional area of the portion of this surface inside the cutoff is
\be
\label{AdS4area}
\int_0^{\cos^{-1}{u_{min} \over R}} d \theta {2 \pi \sin \theta \over \cos^2 \theta}
= 2 \pi  \left( {R \over u_{min}} - 1\right) \: .
\ee
Using this, we find that the regulated eight-dimensional area of the extremal surface is given by 
\bea
\label{regA}
{\rm Area} &=& 128 \pi^3 \int_0^{u_{min}(r,\theta) = R} r dr d \theta \rho^2 f_1^2 f_2^2 f_4^2 \left({R \over u_{min}(r,\theta) }   - 1 \right) \cr
&=& - 1024 \pi^3 \int_0^{f_4(r,\theta) = RL/\epsilon} r dr d \theta h_1 h_2 \partial_w \partial_{\bar{w}}(h_1 h_2) \left({R L \over \epsilon f_4(r, \theta)}   - 1 \right) \: .
\eea

From this expression, we need to subtract off half the area of the extremal surface corresponding to a ball in the parent ${\cal N}=4$ SYM theory. The area to be subtracted off can be expressed in a similar way to (\ref{regA}) by taking $h_1$ and $h_2$ to be the expressions (\ref{h12AdS}) relevant to pure AdS. Since we would like to subtract off half of the regulated area of the extremal surface in pure AdS, we can keep only the part for $x \le 0$ in Fefferman-Graham coordinates, which translates to the part with $r \ge r_0$ in the coordinates we are using. Thus, the regulated half-hemisphere area is
\beas
{1 \over 2} {\rm Area}_{AdS} = - 1024 \pi^3 \int_{r_0}^{f^{AdS}_4(r,\theta) \le RL/\epsilon} r dr d \theta h^{AdS}_1 h^{AdS}_2 \partial_w \partial_{\bar{w}}(h^{AdS}_1 h^{AdS}_2) \left({R L \over \epsilon f^{AdS}_4(r, \theta)}   - 1 \right) \: .
\eeas

\subsubsection*{Details of the Subtraction}

In order to evaluate the integrals, it is convenient to split the integration domain into a part with $r \in [0,\Lambda]$ and an asymptotic part $\{r \ge \Lambda, f_4(r,\theta) \le RL/\epsilon \}$, for some large $\Lambda$ that we will take to infinity as $\epsilon \to 0$.

For the first part,
\be
- 1024 \pi^3 \int_0^\Lambda  dr \int_0^{\pi \over 2} d \theta r h_1 h_2 \partial_w \partial_{\bar{w}}(h_1 h_2) \left({R L \over \epsilon f_4(r, \theta)}   - 1 \right) \; ,
\ee
the first term does not contribute to the final result since it gives an $R/ \epsilon$ term that is eliminated by the derivative in the definition (\ref{defF1},\ref{defF2},\ref{defF3}) of boundary $F$. Thus, this part of the integral gives a contribution to boundary $F$ of 
\be
\label{F1}
F_1 = - {256 \pi^3 \over G} \int_0^\Lambda  dr \int_0^{\pi \over 2} d \theta r h_1 h_2 \partial_w \partial_{\bar{w}}(h_1 h_2) \; .
\ee

From this, we subtract off the corresponding integral for pure AdS, so we have a contribution 
\be
\label{F2}
F_2 = {256 \pi^3 \over G} \int_{r_0}^\Lambda  dr \int_0^{\pi \over 2} d \theta r h^{AdS}_1 h^{AdS}_2 \partial_w \partial_{\bar{w}}(h^{AdS}_1 h^{AdS}_2 )\; .
\ee

To evaluate the asymptotic part of the integral (i.e the region with $r > \Lambda$), we use that the asymptotic form of $f_4$ in the general solution is
\be
f_4(r,\theta) = A r + B(\theta){1 \over r} + {\cal O}(1 / r^2) \; .
\ee
while the asymptotic form of the integrand is
\be
I(r,\theta) = I_1(\theta) r + I_2(\theta){1 \over r} + {\cal O}(1 / r^2)
\ee
Then the integral in the asymptotic region takes the form
\be
\int_0^{\pi \over 2} d \theta \int_\Lambda^{{RL \over \epsilon A} - {B(\theta) \epsilon \over RL} + \dots} dr \left[I_1(\theta) r + I_2(\theta){1 \over r} + \dots \right] \left[{RL \over \epsilon A r} - 1 + \dots \right]
\ee
where the omitted terms give contributions that vanish in the limit $\epsilon \to 0$ and $\Lambda \to \infty$. Evaluating the integral for the remaining terms gives
\be
\int_0^{\pi \over 2}d \theta \left[ {R^2 L^2 \over 2 \epsilon^2 A^2} I_1(\theta) - {R L \Lambda \over \epsilon} I_1 (\theta) + {1 \over 2} I_1(\theta) \Lambda^2 - I_2(\theta) \ln \left( {RL \over \epsilon A \Lambda} \right) + \dots \right] \: .
\ee
Now, we can check that $A$, $I_1(\theta)$ and $\int d \theta I_2(\theta)$ all give the same results for the general solution and for the pure AdS case with the corresponding $L$ and $r_0$. Thus, when we perform the subtraction, there are no terms that contribute from this $r > \Lambda$ region in the limits $\epsilon \to 0$ and $\Lambda \to \infty$.

To summarize, our final result is that boundary $F$ is given by the $\Lambda \to \infty$ limit of the sum of the two contributions (\ref{F1}) and (\ref{F2}),
\be
\label{FintegralA}
F_{\partial} = - \lim_{\Lambda \to \infty} {256 \pi^3 \over G} \left\{ \int_0^\Lambda  dr \int_0^{\pi \over 2} d \theta r h_1 h_2 \partial_w \partial_{\bar{w}}(h_1 h_2) - \int_{r_0}^\Lambda  dr \int_0^{\pi \over 2} d \theta r h^{AdS}_1 h^{AdS}_2 \partial_w \partial_{\bar{w}}(h^{AdS}_1 h^{AdS}_2) \right\}
\ee

\section{Evaluation of Boundary \texorpdfstring{$F$}{} Using Complex Variables} \label{sec:complexint}

Our goal in this section is to demonstrate how to evaluate the integral
\begin{equation}
\begin{split}
    \mathcal{I}(c_{A}, d_{A}, k_{A}, l_{A}, \Lambda) & \equiv \int_{\Lambda} d^{2} w \Big( \hat{h}_{1} \hat{h}_{2} \partial_{w} \partial_{\bar{w}} (\hat{h}_{1} \hat{h}_{2} ) \Big) \: ,
\end{split}
\end{equation}
where we have coordinates $(w, \bar{w}) = (r e^{i \theta}, r e^{- i \theta})$ in the first quadrant of the complex plane, and
\begin{equation}
    \begin{split}
        \hat{h}_{1} & = r \cos \theta + \sum_{A} \frac{c_{A}}{2 \pi} \ln \Big( \frac{r^{2} + 2 r l_{A} \cos \theta + l_{A}^{2}}{r^{2} - 2 r l_{A} \cos \theta + l_{A}^{2}} \Big) \\
        & = \Big( \frac{w + \bar{w}}{2} \Big) + \sum_{A} \frac{c_{A}}{2 \pi} \ln \Big( \frac{(w + l_{A})(\bar{w} + l_{A})}{(w - l_{A})(\bar{w}-l_{A})} \Big) \\
        \hat{h}_{2} & = r \sin \theta + \sum_{A} \frac{d_{A}}{2 \pi} \ln \Big( \frac{r^{2} + 2 r k_{A} \sin \theta + k_{A}^{2}}{r^{2} - 2 r k_{A} \sin \theta + k_{A}^{2}} \Big) \\
        & = \Big( \frac{w - \bar{w}}{2i} \Big) + \sum_{A} \frac{d_{A}}{2 \pi} \ln \Big( \frac{(w + i k_{A})(\bar{w} - i k_{A})}{(w - i k_{A})(\bar{w} + i k_{A})} \Big) \: .
    \end{split}
\end{equation}
For the sake of brevity, we will actually compute only the quartic term $\mathcal{I}_{ACBD}^{ccdd}$; the remaining terms can be approached similarly. Given that
\begin{equation}
\begin{split}
    \partial_{w} \partial_{\bar{w}} (\hat{h}_{1} \hat{h}_{2}) & = \frac{1}{2 \pi i} \sum_{A} \frac{c_{A}}{l_{A}} F_{A}(r, \theta) - \frac{1}{2 \pi i} \sum_{A} \frac{d_{A}}{k_{A}} G_{A}(r, \theta) \\
    & \qquad + \frac{1}{\pi^{2} i} \sum_{A, B} \frac{c_{A} d_{B}}{l_{A} k_{B}} \frac{(l_{A}^{2} + k_{B}^{2})}{r^{2}} H_{A, B}(r, \theta)
\end{split}
\end{equation}
with
\begin{equation}
    \begin{split}
        F_{A}(r, \theta) & \equiv \frac{e^{4 i \theta} - 1}{(e^{2 i \theta} - l_{A}^{2} / r^{2})(e^{2 i \theta} - r^{2} / l_{A}^{2})} \: , \qquad G_{A}(r, \theta) \equiv \frac{e^{4 i \theta} - 1}{(e^{2 i \theta} + k_{A}^{2} / r^{2})(e^{2 i \theta} + r^{2} / k_{A}^{2})} \: , \\
        H_{A, B}(r, \theta) & \equiv \frac{e^{2 i \theta} (e^{4 i \theta} - 1)}{(e^{2 i \theta} - l_{A}^{2} / r^{2})(e^{2 i \theta} - r^{2} / l_{A}^{2})(e^{2 i \theta} + k_{B}^{2} / r^{2})(e^{2 i \theta} + r^{2} / k_{B}^{2})} \: ,
    \end{split}
\end{equation}
we see that
\begin{equation}
\begin{split}
    \mathcal{I}_{ACBD}^{ccdd} & \equiv \frac{1}{(2 \pi)^{2}} \Big( \frac{(l_{C}^{2} + k_{D}^{2})}{ i \pi^{2} l_{C} k_{D}} \Big) \int_{0}^{\Lambda} r dr \: \int_{0}^{\pi / 2} d \theta \: \frac{H_{C, D}(r, \theta)}{r^{2}} \\
    & \qquad \times \ln \Big( \frac{r^{2} + 2 r l_{A} \cos \theta + l_{A}^{2}}{r^{2} - 2 r l_{A} \cos \theta + l_{A}^{2}} \Big)  \ln \Big( \frac{r^{2} + 2 r k_{B} \sin \theta + k_{B}^{2}}{r^{2} - 2 r k_{B} \sin \theta + k_{B}^{2}} \Big) \: .
\end{split}
\end{equation}
We notice that
\begin{equation}
    \begin{split}
        \ln \Big( \frac{(w + l_{A})(\bar{w} + l_{A})}{(w - l_{A})(\bar{w}-l_{A})} \Big) & = 2 r \int_{0}^{l_{A}} ds \: \Big( \frac{e^{i \theta}}{r^{2} e^{2 i \theta} - s^{2}} + \frac{e^{- i \theta}}{r^{2} e^{- 2 i \theta} - s^{2}} \Big) \\
        \ln \Big( \frac{(w + i k_{B})(\bar{w} - i k_{B})}{(w - i k_{B})(\bar{w} + i k_{B})} \Big) & = 2 i r \int_{0}^{k_{B}} dt \: \Big( \frac{e^{i \theta}}{r^{2} e^{2 i \theta} + t^{2}} - \frac{e^{- i \theta}}{r^{2} e^{- 2 i \theta} + t^{2}}  \Big) \: ,
    \end{split}
\end{equation}
so
\begin{equation}
    \frac{\partial}{\partial l_{A}} \frac{\partial}{\partial k_{B}} \mathcal{I}_{ACBD}^{ccdd} \equiv  \frac{1}{\pi^{4}} \frac{(l_{C}^{2} + k_{D}^{2})}{l_{C} k_{D}} \int_{0}^{\Lambda} r dr \int_{0}^{\pi / 2} d \theta \: H_{C, D}(r, \theta) A_{l_{A}}(r, \theta) B_{k_{B}}(r, \theta) \: ,
\end{equation}
with
\begin{equation}
\begin{split}
    A_{s}(r, \theta) & = \Big( \frac{e^{i \theta}}{r^{2} e^{2 i \theta} - s^{2}} + \frac{e^{- i \theta}}{r^{2} e^{- 2 i \theta} - s^{2}} \Big) = \frac{e^{i \theta}(e^{2 i \theta} + 1)(1/r^{2} - 1/s^{2})  }{(e^{2 i \theta} - s^{2} / r^{2})(e^{2 i \theta} - r^{2} / s^{2})} \\
    B_{t}(r, \theta) & = \Big( \frac{e^{i \theta}}{r^{2} e^{2 i \theta} + t^{2}} - \frac{e^{- i \theta}}{r^{2} e^{- 2 i \theta} + t^{2}}  \Big) = \Big( \frac{e^{i \theta} (e^{2 i \theta} - 1)(1/r^{2} - 1/t^{2})}{(e^{2 i \theta} + t^{2} / r^{2})(e^{2 i \theta} + r^{2} / t^{2})}  \Big) \: .
\end{split}
\end{equation}
We observe that
\begin{center}
\begin{tabular}{ c c c }
$H_{C, D}(r, - \theta) = - H_{C, D}(r, \theta)$ & $A_{s}(r, - \theta) = A_{s}(r, \theta)$ & $B_{t}(r, - \theta) = - B_{t}(r, \theta)$ \\ 
 $H_{C, D}(r, \theta + \pi) = H_{C, D}(r, \theta)$ & $A_{s}(r, \theta + \pi) = - A_{s}(r, \theta)$ & $B_{t}(r, \theta + \pi) = - B_{t}(r, \theta)$  
\end{tabular}
\end{center}
and thus the integrand is invariant under both $\theta \rightarrow - \theta$ and $\theta \rightarrow \theta + \pi$. We may therefore write
\begin{equation}
\begin{split}
    \int_{0}^{\pi/2} d \theta H_{C, D}(r, \theta) A_{s}(r, \theta) B_{t}(r, \theta) & = \frac{1}{4} \int _{0}^{2 \pi} d \theta H_{C, D}(r, \theta) A_{s}(r, \theta) B_{t}(r, \theta) \\
    & = \frac{1}{4i} \oint_{|z| = 1} \frac{dz}{z} H_{C, D}(r, z) A_{s}(r, z) B_{t}(r, z) \: ,
\end{split}
\end{equation}
where we have introduced the complex coordinate $z = e^{i \theta}$, and have slightly abused notation in defining
\begin{equation}
    \begin{split}
        & H_{C, D}(r, z) \equiv \frac{z^{2} (z^{4} - 1)}{(z^{2} - l_{C}^{2} / r^{2})(z^{2} - r^{2} / l_{C}^{2})(z^{2} + k_{D}^{2}/r^{2})(z^{2} + r^{2} / k_{D}^{2})} \: , \\
        & A_{s}(r, z) \equiv \frac{z(z^{2} + 1)(1/r^{2} - 1/s^{2})}{(z^{2} - s^{2} / r^{2})(z^{2} - r^{2} / s^{2})} \: , \: B_{t}(r, z) \equiv \frac{z(z^{2} - 1)(1/r^{2} - 1/t^{2})}{(z^{2} + t^{2} / r^{2})(z^{2} + r^{2} / t^{2})} \: .
    \end{split}
\end{equation}
We emphasize that the complex variable $z$ is independent of the real radial variable $r$; the angular integration is over a contour with $|z| = 1$. Invoking the residue theorem, this becomes
\begin{equation}
    \int_{0}^{\pi/2} d \theta H_{C, D}(r, \theta) A_{s}(r, \theta) B_{t}(r, \theta) = \frac{\pi}{2} \sum_{\textnormal{poles}} \textnormal{Res} \Big( \frac{1}{z} H_{C, D}(r, z) A_{s}(r, z) B_{t}(r, z) \Big) \: ,
\end{equation}
where the sum is over poles inside the contour $|z| = 1$. It transpires that the ordering of parameters $l_{C}, k_{D}, s, t$ is irrelevant (as it must be in order to evaluate the integral by partial fractions), so we will assume for the purposes of computation that $0 < s < t < l_{C} < k_{D} < \Lambda$. Consequently, depending on the size of the variable $r$, we have five inequivalent collections of eight poles:
\begin{enumerate}
    \item $0 < r < s$: poles at $z = \pm \frac{r}{s}, \pm \frac{ir}{t}, \pm \frac{r}{l_{C}} , \pm \frac{ir}{k_{D}}$ 
    \begin{equation}
        \frac{\pi}{2} \sum_{\textnormal{poles}} \textnormal{Res} =  \frac{\pi l_{C}^{2} k_{D}^{2} r^{4} (r^{8} - s^{2} t^{2} l_{C}^{2} k_{D}^{2})(r^{2} - s^{2})(r^{2} - t^{2})}{(r^{4} - s^{2} l_{C}^{2})(r^{4} - t^{2}k_{D}^{2})(r^{4} + s^{2} k_{D}^{2})(r^{4} + t^{2} l_{C}^{2})(r^{4} + s^{2} t^{2})(r^{4} + l_{C}^{2} k_{D}^{2})}
    \end{equation}
    \item $s < r < t$: poles at $z = \pm \frac{s}{r}, \pm \frac{ir}{t}, \pm \frac{r}{l_{C}} , \pm \frac{ir}{k_{D}}$
    \begin{equation}
        \frac{\pi}{2} \sum_{\textnormal{poles}} \textnormal{Res} =  \frac{\pi l_{C}^{2} k_{D}^{2} (r^{4} s^{2} - t^{2} l_{C}^{2} k_{D}^{2})(r^{2} - s^{2})(r^{2} - t^{2})}{
        (s^{2} + k_{D}^{2})(s^{2} - l_{C}^{2})(s^{2} + t^{2})(r^{4} - t^{2} k_{D}^{2})(r^{4} + t^{2} l_{C}^{2})(r^{4} + l_{C}^{2} k_{D}^{2})}
    \end{equation}
    \item $t < r < l_{C}$: poles at $z = \pm \frac{s}{r}, \pm \frac{it}{r}, \pm \frac{r}{l_{C}} , \pm \frac{ir}{k_{D}}$
    \begin{equation}
        \frac{\pi}{2} \sum_{\textnormal{poles}} \textnormal{Res} =  \frac{\pi l_{C}^{2} k_{D}^{2} (s^{2} t^{2} - l_{C}^{2} k_{D}^{2})(r^{2} - s^{2})(r^{2} - t^{2})}{(s^{2} - l_{C}^{2})(t^{2} - k_{D}^{2})(s^{2} + k_{D}^{2})(t^{2} + l_{C}^{2})(r^{4} + s^{2} t^{2})(r^{4} + l_{C}^{2} k_{D}^{2})}
    \end{equation}
    \item $l_{C} < r < k_{D}$: poles at $z = \pm \frac{s}{r}, \pm \frac{it}{r}, \pm \frac{l_{C}}{r} , \pm \frac{ir}{k_{D}}$
    \begin{equation}
        \frac{\pi}{2} \sum_{\textnormal{poles}} \textnormal{Res} =  \frac{\pi l_{C}^{2} k_{D}^{2} (r^{4} k_{D}^{2} - s^{2} t^{2} l_{C}^{2})(r^{2} - s^{2})(r^{2} - t^{2})}{
        (t^{2} - k_{D}^{2})(s^{2} + k_{D}^{2})(l_{C}^{2} + k_{D}^{2})(r^{4} + s^{2} t^{2})(r^{4} - s^{2} l_{C}^{2})(r^{4} + t^{2} l_{C}^{2})}
    \end{equation}
    \item $k_{D} < r < \Lambda$: poles at $z = \pm \frac{s}{r}, \pm \frac{it}{r}, \pm \frac{l_{C}}{r} , \pm \frac{ik_{D}}{r}$
    \begin{equation}
        \frac{\pi}{2} \sum_{\textnormal{poles}} \textnormal{Res} =  - \frac{\pi l_{C}^{2} k_{D}^{2} r^{4} (r^{8} - s^{2} t^{2} l_{C}^{2} k_{D}^{2})(r^{2} - s^{2})(r^{2} - t^{2})}{(r^{4} - t^{2} k_{D}^{2})(r^{4} + s^{2} k_{D}^{2})(r^{4} + l_{C}^{2} k_{D}^{2})(r^{4} + t^{2} l_{C}^{2})(r^{4} - s^{2} l_{C}^{2})(r^{4} + s^{2} t^{2})} \: .
    \end{equation}
\end{enumerate}
Consequently, performing the integral over $r$, and keeping only terms that do not vanish in the limit $\Lambda\rightarrow \infty$, we find
\begin{equation}
    \begin{split}
        & \int_{0}^{\Lambda} r dr \int_{0}^{\pi/2} d \theta \: H_{C, D}(r, \theta) A_{s}(r, \theta) B_{t}(r, \theta) \\
        & \qquad = \frac{\pi k l}{2} \Big[ k l \Big( \frac{ \ln \big( (s^{2} + k^{2})(t^{2} + l^{2}) \big)}{(l^{2} + k^{2})(s^{2} - l^{2})(t^{2} - k^{2})}  - \frac{(s^{2} + t^{2})\ln \big( (s^{2} + t^{2})(k^{2} + l^{2}) \big) }{(s^{2} - l^{2})(t^{2} - k^{2})(s^{2} + k^{2})(t^{2} + l^{2})}  \Big) \\
        & \qquad \qquad  + \frac{l \Big( (k-t)(k t-s^{2}) \ln \big( (k - t)^{2} \big) - (k+t)(kt+s^{2})\ln \big( (k + t )^{2} \big) \Big)}{2 (s^{2} + t^{2})(l^{2} + k^{2})(s^{2} + k^{2})(t^{2} + l^{2})}  \\
        & \qquad \qquad + \frac{k \Big( (l-s)(ls-t^{2}) \ln \big( ( l - s )^{2} \big) - (l+s)(ls+t^{2}) \ln \big( (l + s)^{2} \big) \Big)}{2 (s^{2} + t^{2})(l^{2} + k^{2})(s^{2} + k^{2})(t^{2} + l^{2})} \Big] \\
        & \equiv \frac{\pi k l}{2} \: \tilde{\mathcal{I}}_{s, l, t, k} \: ,
    \end{split}
\end{equation}
where we have momentarily suppressed indices on parameters $l_{C}, k_{D}$. We therefore find
\begin{equation}
    \frac{\partial}{\partial l_{A}} \frac{\partial}{\partial k_{B}} \mathcal{I}^{ccdd}_{ACBD} = \frac{(l_{C}^{2} + k_{D}^{2})}{2 \pi^{3}} \tilde{\mathcal{I}}_{l_{A}, l_{C}, k_{B}, k_{D}} \: .
\end{equation}
One can integrate this with respect to $l_{A}, k_{D}$ to recover the result in the main text.

\section{Verification of Field Theory Constraints for Linking Numbers Defined in Terms of Supergravity Parameters } \label{sec:proofineq}

In this appendix, we show that for any linking numbers defined in terms of supergravity parameters as in (\ref{defKL}), the field theory constraints on linking numbers that (\ref{defN}) are positive are automatically satisfied.

For this appendix, we define $M_n$ to be the number of D5-branes with linking number $\tilde{L} = n - N_{NS5}$, where $1 \leq n < N_{NS5}$. We will also let the indices on the linking numbers $\{\tilde{L}_{i}, K_{i}\}$ refer to the $i^{\textnormal{th}}$ 5-brane, rather than the $i^{\textnormal{th}}$ 5-brane stack.

We will prove that  for linking numbers violating the inequalities
\begin{equation} \label{eq:appineqs}
        \sum_{n=1}^{j-1} (j-n) M_{n} < \sum_{i=1}^{j} K_i \: , \qquad j \in \{1, 2, \ldots, N_{NS5} \} \: ,
\end{equation}
i.e. for which not all of the quantities (\ref{defN}) are positive, 
there is no set of supergravity parameters that can give rise to these linking numbers via (\ref{defKL}). 

We see immediately that if we define index subset
\begin{equation}
   \mathcal{I} \equiv \{ i \: : \: \tilde{L}_{i} > 0 \} \: ,
\end{equation}
then violating the final inequality 
\begin{equation}
    \sum_{n=1}^{N_{NS5}-1} (N_{NS5} - n) M_{n} < \sum_{i=1}^{N_{NS5}} K_{i} \: ,
\end{equation}
which can be written as
\begin{equation}
    - \sum_{i \notin \mathcal{I}}  \tilde{L}_{i} < \sum_{i} K_{i} \: ,
\end{equation}
implies
\begin{equation}
    N = \sum_{i \in \mathcal{I}} \tilde{L}_{i} + \sum_{i \notin \mathcal{I}} \tilde{L}_{i} + \sum_{i}  K_{i}  \leq \sum_{i \in \mathcal{I}} \tilde{L}_{i} \: , 
\end{equation}
and therefore
\begin{equation}
        N = \left(\sum_{i \notin \mathcal{I}} \hat{l}_{i} + \sum_{i} \hat{k}_{i} \right) + \sum_{i \in \mathcal{I}} \hat{l}_{i}  > \sum_{i \in \mathcal{I}} \hat{l}_{i} \geq \sum_{i \in \mathcal{I}} \tilde{L}_{i} \geq N \: , 
\end{equation}
a contradiction, implying that the system of equations has no solution. Here, we have used that $\hat{l}_{i}, \hat{k}_{i} > 0$ and
\begin{equation}
        \tilde{L}_{i} = \hat{l}_{i} - \frac{2}{\pi} \sum_{j} \arctan \left( \frac{g \hat{k}_{j}}{\hat{l}_{i}} \right) \leq \hat{l}_{i}  \: .
\end{equation}

We would like to check that violating the other inequalities similarly leads to a system with no solutions. We restrict to the case that $K_{1} > 0$, i.e. the first of the inequalities in (\ref{eq:appineqs}) is always satisfied; this is because we are interested in configurations which will correspond to theories with boundaries rather than interfaces. Moreover, we may restrict to the case that the last of the inequalities is satisfied, since we have already shown that violating this inequality leads to an insoluble system. 
To this end, let us fix arbitrary $N_{NS5} \geq 3$; our task is to show that violating the inequality in (\ref{eq:appineqs}) indexed by $j \in \{2, \ldots, N_{NS5}-1\}$ leads to a contradiction in our system of equations defining the supergravity parameters. This system is constituted by the relations
\begin{equation} \label{eq:SUGRArelations}
\begin{split}
    \hat{k}_{i} & = K_{i} - \frac{2}{\pi} \sum_{j} \arctan \left( \frac{g \hat{k}_{i}}{\hat{l}_{j}} \right) \: , \\
    \hat{l}_{i} & = \tilde{L}_{i} + \frac{2}{\pi} \sum_{j}  \arctan \left( \frac{g \hat{k}_{j}}{\hat{l}_{i}} \right) \: ,
\end{split}
\end{equation}
which in particular furnish inequalities
\begin{equation}
    K_{i} > \frac{2}{\pi} \sum_{j} \arctan \left( \frac{g \hat{k}_{i}}{\hat{l}_{j}} \right) \: , \qquad
    \tilde{L}_{i} > - \frac{2}{\pi} \sum_{j}  \arctan \left( \frac{g \hat{k}_{j}}{\hat{l}_{i}} \right) \: .
\end{equation}

First, suppose that we violate the inequality indexed by $j=2$; that is, suppose
\begin{equation}
    K_{1} + K_{2} \leq M_{1} \: .
\end{equation}
We may assume $M_{1} > 0$ without loss of generality, so that $\tilde{L}_{1} = \ldots = \tilde{L}_{M_{1}} = - (N_{NS5}-1)$, and in particular $N_{D5} > 0$; otherwise, $M_{1} \leq 0$ and $K_{1} + K_{2} \leq M_{1}$ would imply $K_{2} < 0$, which is incompatible with  (\ref{eq:SUGRArelations}) and the assumption that $\hat{k}_{A}, \hat{l}_{A}$ are positive. But since
\begin{equation}
        \frac{\pi}{2} K_{i} > \sum_{j} \arctan \left( \frac{g \hat{k}_{i}}{\hat{l}_{j}} \right) \: , \qquad \frac{\pi}{2} \tilde{L}_{1} > - \sum_{B} \arctan  \left( \frac{g \hat{k}_{j}}{\hat{l}_{i}} \right)
\end{equation}
by (\ref{eq:SUGRArelations}), we find
\begin{equation}
    \begin{split}
        \sum_{j=1}^{M_{1}} \sum_{i \geq 3} \arctan  \left( \frac{g \hat{k}_{i}}{\hat{l}_{j}} \right) & > \sum_{j=1}^{M_{1}} \left( - \frac{\pi}{2} \tilde{L}_{j} - \arctan \left( \frac{g \hat{k}_{1}}{\hat{l}_{j}} \right) - \arctan \left( \frac{g \hat{k}_{2}}{\hat{l}_{j}} \right) \right) \\
        & > \frac{\pi}{2} \left( (N_{NS5}-1) M_{1} - K_{1} - K_{2} \right) \\
        & \geq \frac{\pi}{2} (N_{NS5}-2) M_{1} \: ,
    \end{split}
\end{equation}
contradicting the bound $\arctan(x) < \frac{\pi}{2}$. 

More generally, suppose that we violate the inequality indexed by $j \in \{2, \ldots, N_{NS5}-1\}$; that is, suppose that we have
\begin{equation}
    \sum_{i=1}^{j} K_{i} \leq \sum_{n=1}^{j-1} (j-n) M_{n} \: .
\end{equation}
We may assume that at least one of $M_{1}, \ldots, M_{j-1}$ is positive (since otherwise at least one of the $K_{i}$ would be negative). 
Then, letting $M \equiv M_{1} + \ldots + M_{j-1}$,
\begin{equation}
    \begin{split}
        \sum_{i > j} \sum_{m=1}^{M} \arctan \left( \frac{g \hat{k}_{i}}{\hat{l}_{m}} \right) & > - \frac{\pi}{2} \sum_{m=1}^{M} \tilde{L}_{m} - \sum_{i = 1}^{j} \sum_{m=1}^{M} \arctan \left( \frac{g \hat{k}_{i}}{\hat{l}_{m}} \right) \\
        & > \frac{\pi}{2} \left( \sum_{n=1}^{j-1} (N_{NS5}-n) M_{n} - \sum_{i=1}^{j} K_{i} \right) \\
        & \geq \frac{\pi}{2} (N_{NS5}-j) \sum_{n=1}^{j-1} M_{n} = \frac{\pi}{2} (N_{NS5}-j) M \: ,
    \end{split}
\end{equation}
again contradicting $\arctan(x) < \frac{\pi}{2}$. This demonstrates our original claim. 

\section{Corrections to the Supergravity Approximation} \label{sec:corrections}

In this appendix, we estimate the size of the corrections to the supergravity result, following the procedure outlined at the end of Section \ref{sec:holo}.

\subsection{Estimating the Corrections}

Recall that our solutions are generated by the harmonic functions in (\ref{gensol}), determined by positive real constants $l_{A}, k_{B}$. These can be combined to define \textit{Einstein frame} metric functions and dilaton field, using (\ref{eq:WN1N2}), (\ref{eq:dilaton}), (\ref{eq:metricfunctions}) in Section \ref{sec:localsol}; to transform to the \textit{string frame}, we should multiply all of the metric functions by $e^{\phi} \equiv e^{\Phi / 2}$. We begin by determining the string frame Ricci curvature and dilaton field in the vicinity of a D5-brane or NS5-brane stack. It will be useful to define
\begin{equation}
\begin{split}
    \gamma_{C} & \equiv \pi + 2 \sum_{B} \frac{d_{B} k_{B}}{l_{C}^{2} + k_{B}^{2}} = \frac{\pi}{c_{C}} \frac{d}{d l_{C}} N_{D3}^{(C)} \\
    \delta_{D} & \equiv \pi + 2 \sum_{A} \frac{c_{A} l_{A}}{k_{D}^{2} + l_{A}^{2}} = \frac{\pi}{d_{D}} \frac{d}{d k_{D}} N_{D3}^{(D)} \: ;
\end{split}
\end{equation}
note that in the case with only D5-branes one has $\gamma_{C} = \pi$, and in the case with only NS5-branes one has $\delta_{D} = \pi$.

First, we consider the vicinity of the D5-stack at $(x, y) = (l_{C}, 0)$, and let $L_{0}$ denote the distance in the first quadrant $\Sigma$ from $l_{C}$ to the nearest 5-brane stack (or the origin), namely
\begin{equation}
    L_{0} \equiv \min_{A, B \neq C} \{ |l_{A} - l_{C}|, \sqrt{k_{B}^{2} + l_{C}^{2}}, l_{C} \} \: .
\end{equation}
Using polar coordinates $(x, y) = (l_{C} + r \cos \theta, r \sin \theta)$, we therefore have the expansion
\begin{equation}
    \begin{split}
        h_{1} & = \frac{\pi \ell_{s}^{2}}{2} \frac{1}{\sqrt{g}} \Big[  - \frac{c_C}{2 \pi} \ln (r^2 / 4 l_{C}^{2})  + \left( l_{C} + \sum_{A \neq C} {c_A \over 2 \pi} \ln \left( {(l_{C} + l_A)^2  \over (l_C-l_A)^2 } \right) \right) \\
        & \qquad + r \cos \theta \left( 1 + \frac{c_C}{2 \pi l_C} - \frac{2}{\pi} \sum_A {c_A l_A \over (l_C^2 - l_A^2)} \right) + \mathcal{O}(r^2/L_{0}^{2}) \Big] \\
        h_{2} & = {\pi \ell_s^2 \over 2} \sqrt{g} \Bigg[ r \sin \theta \left( 1 + \frac{2}{\pi} \sum_B { d_B k_B \over (l_C^2 + k_B^2) } \right) \\
        & \qquad + r^2 \sin \theta \cos \theta \left( - \frac{4 l_C}{\pi} \sum_B {d_B k_B \over (l_C^2 + k_B^2)^2}  \right) + \mathcal{O}(r^3/L_{0}^{3}) \Bigg] \: .
    \end{split}
\end{equation}
We therefore have string frame metric functions given at leading order in $r/L_{0}$ by
\begin{equation}
    \begin{split}
        \rho^{2} &  = \frac{\sqrt{2g} \gamma_{C} \ell_{\textnormal{s}}^{2}}{4 } \frac{1}{r \ln(4 l_{C}^{2} / r^{2})^{1/2}}  \: , \qquad f_{4}^{2} = \frac{\sqrt{2g} \gamma_{C} \ell_{\textnormal{s}}^{2}}{2} \: r \ln(4 l_{C}^{2} / r^{2})^{1/2} \: , \\
        f_{1}^{2} & = \frac{\sqrt{2g} \gamma_{C} \ell_{\textnormal{s}}^{2}}{2} \: r \ln(4 l_{C}^{2} / r^{2})^{1/2} \: , \qquad 
        f_{2}^{2} = \sqrt{2g} \gamma_{C} \ell_{\textnormal{s}}^{2} \frac{r \sin^{2} \theta}{\ln(4 l_{C}^{2} / r^{2})^{1/2}} \: ,
    \end{split}
\end{equation}
and dilaton
\begin{equation}
    e^{ 2 \phi} = \frac{\sqrt{2} g \gamma_{C}}{c_{C}} \frac{r}{\sqrt{\ln(4 l_{C}^{2} / r^{2})}}
    \: .
\end{equation}
We thereby deduce string frame Ricci scalar at leading order
\begin{equation}
    \alpha ' R = - \frac{6}{\gamma_{C} r} \sqrt{\frac{2}{g} \ln( 4 l_{C}^{2} / r^{2})} \: .
\end{equation}
We can perform a similar analysis near an NS5-brane stack at $(x, y) = (0, k_{D})$, for which we find dilaton and Ricci scalar
\begin{equation}
    e^{2 \phi} = \frac{g d_{D}}{\sqrt{2} \delta_{D}} \frac{\ln(4 k_{D}^{2} / r^{2})^{1/2}}{r}  \: , \qquad
    \alpha ' R = \frac{6}{\sqrt{g} d_{D}} 
\end{equation}
at leading order. 

The above expressions tell us the minimum radius $ r_{\textnormal{max}}$ past which the correction terms appearing in the effective action should be suppressed. Evidently, for the D5-brane stacks, the divergence of the string frame curvature implies that we are only justified in ignoring corrections in the region $r \gg r_{\textnormal{max}}$ with
\begin{equation}
    r_{\textnormal{max}} \sim \frac{1}{\sqrt{g} \gamma_{C}} \sqrt{ W ( g l_{C}^{2} \gamma_{C}^{2} )} \: ,
\end{equation}
where $W( \cdot )$ denotes the Lambert W-function, and we suppress order one numerical factors. 
The contribution to $F_{\partial}$ from the complementary region is
\begin{equation}
    \int_{0 < r < r_{\textnormal{max}}} r dr d\theta \: \hat{h}_{1} \hat{h}_{2} \partial \bar{\partial} (\hat{h}_{1} \hat{h}_{2}) \sim c_{C}^{2} \gamma_{C}^{2} r_{\textnormal{max}}^{2} \ln(4 l_{C}^{2} / r_{\textnormal{max}}^{2}) \: .
\end{equation}
For the NS5-brane stacks, we see that the curvature corrections will be suppressed provided we take $N_{NS5}^{(D)} \gg 1$, but will be large \textit{throughout} the region $r \ll L_{0}$ otherwise; evaluating the contribution to $F_{\partial}$ from a region within $r_{\textnormal{max}} \sim L_{0}$ gives
\begin{equation}
    \int_{0 < r < r_{\textnormal{max}}} r dr d\theta \: \hat{h}_{1} \hat{h}_{2} \partial \bar{\partial} (\hat{h}_{1} \hat{h}_{2}) \sim d_{D}^{2} \delta_{D}^{2} L_{0}^{2} \ln(4 k_{D}^{2} / L_{0}^{2}) \: .
\end{equation}
Meanwhile, the string loop corrections are small outside the region
\begin{equation}
    r_{\textnormal{max}} \sim \frac{g d_{D}}{\delta_{D}} \sqrt{W \left( \frac{\delta_{D}^{2} k_{D}^{2}}{g^{2} d_{D}^{2}} \right) } \: ,
\end{equation}
and the contribution to boundary $F$ from the complementary region is
\begin{equation}
    \int_{0 < r < r_{\textnormal{max}}} r dr d\theta \: \hat{h}_{1} \hat{h}_{2} \partial \bar{\partial} (\hat{h}_{1} \hat{h}_{2}) \sim d_{D}^{2} \delta_{D}^{2} r_{\textnormal{max}}^{2} \ln(4 k_{D}^{2} / r_{\textnormal{max}}^{2}) \: .
\end{equation}
In cases of interest, we can compare these contributions for each stack to those appearing in our classical SUGRA calculation of $F_{\partial}$; if there are terms in $F_{\partial}$ which dominate all of the naive estimates of the corrections from near the fivebrane stacks, then these terms should provide a reliable approximation to $F_{\partial}$.

\subsection{Examples}

Here we will consider some examples to illustrate the procedure of comparing the anticipated corrections to the terms appearing in the uncorrected expression for $F_{\partial}$. To recover a classical supergravity dual in the asymptotic region, we should always consider the limit $N \rightarrow \infty$ and $\lambda \gg 1$. 

\subsection*{Single Stack of D5-Branes}

Suppose we have a single stack of $N_{5}$ D5-branes, each with linking number $\tilde{L} = N / N_{5}$; here $N_{5}$ is $\Omega(N^{0})$ and $O(N)$. 
The anticipated correction in the vicinity of this stack is of order
\begin{equation}
    \mathcal{O} \left( N_{5}^{2} \cdot W \big( \pi^{2} \tilde{L}^{2} \big) \cdot \ln \Big( \frac{4 \pi^{2} \tilde{L}^{2}}{W \big( \pi^{2} \tilde{L}^{2} \big)} \Big) \right) = \begin{cases}
    O(N_{5}^{2}) & \tilde{L} \sim 1 \\
    O \left( (N_{5} \ln \tilde{L})^{2} \right) & \tilde{L} \gg 1
    \end{cases} \: ,
\end{equation}
while our uncorrected expression for $F_{\partial}$ is
\be
F_{\partial} = {N^2 \over 8} \left[3 - {8 \pi^{2} \over 3 \lambda} {N^2 \over N_5^2}  - 2 \ln \left({16 \pi^2 \over \lambda} {N^2 \over N_5^2} \right) \right] \: .
\ee
When $N$ is taken to be large, we see that the $\alpha'$-corrections are expected to be suppressed relative to all terms appearing in the uncorrected $F_{\partial}$ unless we have $\tilde{L} \sim 1$, in which case the corrections become comparable. Note that $\tilde{L}=1$ corresponds to the Dirichlet boundary condition for the gauge theory, which we refer to as a ``maximum entropy" boundary condition in Section \ref{sec:stats}; in that section we will see that the exact evaluation of $F_{\partial}$ for this boundary condition does indeed demonstrate that $F_{\partial}$ receives corrections at leading order.

\subsection*{Single Stack of NS5-Branes}

We now consider the case with a single stack of $N_{5}$ NS5-branes, each with linking number $K = N / N_{5}$; again, $N_{5}$ is $\Omega(N^{0})$ and $O(N)$. The expected $\alpha'$-correction is of order $\mathcal{O}(N^{2})$ if $N_{5} \sim 1$, and should be subleading if $N_{5} \gg 1$.
The expected string loop correction is of order
\begin{equation}
    \mathcal{O} \left( N_{5}^{4} \cdot W( \pi^{2} K^{2} / N_{5}^{2}) \cdot \ln \Big( \frac{4 \pi^{2} K^{2}}{N_{5}^{2} \cdot W \big( \pi^{2} K^{2} / N_{5}^{2} \big)} \Big) \right) = \begin{cases}
        O \left( \left( N_{5}^{2} \ln \left( K^{2} / N_{5}^{2} \right) \right)^{2} \right) & K \gg N_{5} \\
        O(N_{5}^{4}) & K \sim N_{5} \\
        O(N^{2}) & K \ll N_{5}
    \end{cases} \: .
\end{equation}
Meanwhile, the uncorrected expression is
\be
F = {N^2 \over 8} \left[3 - { \lambda \over 6 N_5^2}  - 2 \ln \left( {\lambda \over N_5^2} \right) \right] \: .
\ee
When $N$ is taken to be large, we see that the $\alpha'$-corrections and string loop corrections are both expected to be suppressed relative to the leading term in $F_{\partial}$, which is order $O(N^{2} \ln N_{5})$, provided that we take $N_{5} \gg 1$. Moreover, they will also be suppressed relative to the second leading term, which is $O(N^{2})$, provided that we take $1 \ll N_{5} \ll K$. However, they will not be suppressed relative to the third term, which is order $O(K^{2})$, unless $N_{5} = o(\sqrt{K})$. Note that $N_{5} = N$ is referred to as a ``maximum entropy" boundary condition in Section \ref{sec:stats}; in that section, we see that the exact evaluation of $F_{\partial}$ for this boundary condition demonstrates that the leading $O(N^{2} \ln N)$ term is uncorrected while the next-to-leading $O(N^{2})$ term is corrected, as we have predicted here.


\subsection*{Single Stack of D5-Branes and Single Stack of NS5-Branes}

We will focus here on a specific choice of boundary configuration involving one stack of D5-branes and one stack of NS5-branes, where we fix 
\begin{equation}
    \tilde{L} = - 1 \: , \qquad K = 1 \: ,
\end{equation}
and take $N, N_{NS5}$ to be large independent parameters, with $N \ll N_{NS5}$. We have
\begin{equation}
    N_{D5} =  N_{NS5} - N \: .
\end{equation}
This is the situation considered in Section \ref{sec:stats} to illustrate the unboundedness of $F_{\partial}$; it is a natural boundary condition to consider in order to understand a situation where the number of boundary degrees of freedom is taken to be much larger than the number of bulk degrees of freedom. 
Given that the supergravity parameters are given at leading order by
\begin{equation}
    \hat{l} = \frac{N}{N_{NS5}} + O \left( \frac{N^{2}}{N_{NS5}^{2}} \right) \: , \qquad \hat{k} = \frac{2 \pi^{2}}{\lambda} \frac{N^{2}}{N_{NS5}^{2}} + O \left( \frac{N^{3}}{N_{NS5}^{3}} \right) \: ,
\end{equation}
we find
\begin{equation}
    \gamma = \frac{ \pi N_{NS5}}{N} + O(1) \: , \qquad \delta = \frac{\lambda}{2 \pi} \frac{N_{NS5}^{2}}{N^{2}} + O \left( \frac{N_{NS5}}{N} \right) \: ,
\end{equation}
and thus an $\alpha'$-correction of order 
\begin{equation}
    (N_{NS5} - N)^{2} W( \gamma^{2} \hat{l}^{2}) \ln \left( \frac{4 \gamma^{2} \hat{l}^{2}}{W( \gamma^{2} \hat{l}^{2})} \right) = O \left( N_{NS5}^{2} \right) 
\end{equation}
from the vicinity of the D5-brane stack, and a string loop correction of order
\begin{equation}
    N_{NS5}^{4} W \left( \frac{\delta^{2} \hat{k}^{2}}{N_{NS5}^{2}} \right) \ln \left( \frac{4 \delta^{2} \hat{k}^{2}}{N_{NS5}^{2} W \left( \frac{\delta^{2} \hat{k}^{2}}{N_{NS5}^{2}} \right)} \right) = O \left( N_{NS5}^{2} \right)
\end{equation}
from the vicinity of the NS5-brane stack. We stated in Section \ref{sec:stats} that the leading contribution to the uncorrected $F_{\partial}$ was $O(N_{NS5}^{2} \ln N_{NS5})$ while the next largest contribution is of order $O(N_{NS5})$; consequently, we expect only the leading large $N_{NS5}$ term in the uncorrected expression to be reliable.

\section{Localization Integrals} \label{sec:locint}

In Section \ref{sec:loc}, we need to evaluate integrals of the form
\begin{equation}
\begin{split}
    I_{1}(a, b, s, N) & = \frac{1}{N!} \int \prod_{i=1}^{N} \frac{d \lambda_{i}}{\sqrt{2 \pi}} e^{- \frac{1}{2s} \sum_{i=1}^{N} \lambda_{i}^{2}} \prod_{i<j}^{N} 2 \sinh \left( \frac{a}{2} (\lambda_{i} - \lambda_{j}) \right) \: 2 \sinh \left( \frac{b}{2} (\lambda_{i} - \lambda_{j}) \right) \\
    I_{2}(b, s, N) & = \frac{1}{N!} \int \prod_{i=1}^{N} \frac{d \lambda_{i}}{\sqrt{2 \pi}} e^{- \frac{1}{2s} \sum_{i=1}^{N} \lambda_{i}^{2}} \prod_{i<j}^{N}  2 (\lambda_{i} - \lambda_{j})  \sinh \left( \frac{b}{2} (\lambda_{i} - \lambda_{j}) \right) \\
    I_{3}(s, N) & = \frac{1}{N!} \int \prod_{i=1}^{N} \frac{d \lambda_{i}}{\sqrt{2 \pi}} e^{- \frac{1}{2s} \sum_{i=1}^{N} \lambda_{i}^{2}} \prod_{i<j}^{N} (\lambda_{i} - \lambda_{j})^{2} \: .
\end{split}
\end{equation}
Noting that
\begin{equation}
    I_{2}(b, s, N) = \lim_{a \rightarrow 0} a^{-\frac{N(N-1)}{2}} I_{1}(a, b, s, N) \: , \qquad I_{3}(s, N) = \lim_{b \rightarrow 0} b^{-\frac{N(N-1)}{2}} I_{2}(b, s, N) \: ,
\end{equation}
we see it is sufficient to calculate $I_{1}(a, b, s, N)$, and take the appropriate limits to recover $I_{2}(b, s, N)$ and $I_{3}(s, N)$. 
Using the identity
\begin{equation} \label{eq:sinhid}
    \prod_{i < j}^{N} 2 \sinh \left( \frac{\lambda_{i} - \lambda_{j}}{2} \right) = \sum_{\sigma \in S_{N}} (-1)^{\sigma} \prod_{j=1}^{N} \exp \left( \Big( \frac{N+1}{2} - \sigma_{j} \Big) \lambda_{j} \right) \: ,
\end{equation}
we may write
\begin{equation}
\begin{split}
    I_{1}(a, b, s, N) & = \frac{1}{N!} \int \prod_{i=1}^{N} \frac{d \lambda_{i}}{\sqrt{2 \pi}} e^{- \frac{1}{2s} \sum_{i=1}^{N} \lambda_{i}^{2}} \\
    & \qquad \times \sum_{\sigma, \hat{\sigma} \in S_{N}} (-1)^{\sigma + \hat{\sigma}} \exp \left( \sum_{j=1}^{N} \Big[ a \Big( \frac{N+1}{2} - \sigma_{j} \Big) + b \Big( \frac{N+1}{2} - \hat{\sigma}_{j} \Big) \Big] \right) \: .
\end{split}
\end{equation}
Recalling the Gaussian integration
\begin{equation}
    \int \frac{d \lambda}{\sqrt{2 \pi}} e^{- \frac{1}{2s} \lambda^{2}} e^{b \lambda} = \sqrt{s} e^{s b^{2} / 2} \: ,
\end{equation}
we obtain
\begin{equation}
\begin{split}
    I_{1}(a, b, s, N) & = \frac{s^{\frac{N}{2}}}{N!} \sum_{\sigma, \hat{\sigma} \in S_{N}} (-1)^{\sigma + \hat{\sigma}} \exp \left( \frac{s}{2} \sum_{j=1}^{N} \Big[ a \Big( \frac{N+1}{2} - \sigma_{j} \Big) + b \Big( \frac{N+1}{2} - \hat{\sigma}_{j} \Big) \Big]^{2} \right) \\
    & = s^{\frac{N}{2}} \sum_{\sigma ' \in S_{N}} (-1)^{\sigma '} \exp \left( \frac{s}{2} \sum_{j=1}^{N} \Big[ a \Big( \frac{N+1}{2} - \sigma_{j} ' \Big) + b \Big( \frac{N+1}{2} - j \Big) \Big]^{2} \right) \: ,
\end{split}
\end{equation}
where $\sigma '$ denotes the relative permutation between $\sigma$ and $\hat{\sigma}$. 
We therefore find
\begin{equation}
    I_{1}(a, b, s, N)
    = s^{\frac{N}{2}} e^{- \frac{s (a+b)^{2} N (N+1)^{2}}{8}} e^{ \frac{s(a^{2} + b^{2})N (N+1)(2N+1)}{12}} \sum_{\sigma ' \in S_{N}} (-1)^{\sigma '} \prod_{j=1}^{N} e^{s a b j \sigma_{j} '} \: .
\end{equation}
Noting
\begin{equation}
        \sum_{\sigma} (-1)^{\sigma} \prod_{j=1}^{N} e^{s a b j \sigma_{j}} =
        e^{\frac{sabN(N+1)^{2}}{4}} \prod_{i < j}^{N} 2 \sinh \left( \frac{sab(j-i)}{2} \right) \: ,
\end{equation}
where we have used our earlier identity (\ref{eq:sinhid}) with $\lambda_{j} \rightarrow - s a b j$, we find
\begin{equation}
\begin{split}
    I_{1}(a, b, s, N)
    & = s^{\frac{N}{2}} e^{ \frac{s (a^{2} + b^{2}) N (N+1)(N-1)}{24}} \prod_{i < j} 2 \sinh \left( \frac{sab(j-i)}{2} \right) \\
    & = s^{\frac{N}{2}} e^{ \frac{s (a^{2} + b^{2}) N (N+1)(N-1)}{24}} \prod_{j=1}^{N-1} \left( 2 \sinh \Big( \frac{sabj}{2} \Big) \right)^{N-j} \: .
\end{split}
\end{equation} 
We therefore also deduce
\begin{equation}
    I_{2}(b, s, N)  = s^{\frac{N^{2}}{2}} b^{\frac{N(N-1)}{2}} e^{ \frac{s b^{2} N (N+1)(N-1)}{24}} G_{2}(N+1) \: ,
\end{equation}
and
\begin{equation}
    I_{3}(s, N)  = s^{\frac{N^{2}}{2}} G_{2}(N+1) \: ,
\end{equation}
where one recalls the definition of the Barnes G-function
\begin{equation}
    \prod_{k=1}^{N-1} k! \equiv G_{2}(N+1) \: .
\end{equation}
We can extract the partition function of $\mathcal{N}=4$ $U(N)$ SYM on $S^{4}$ from
\begin{equation}
    Z[S^{4}] = (2 \pi)^{\frac{N}{2}} I_{3} \left( \frac{\gym^{2}}{16 \pi^{2}}, N \right) \: ,
\end{equation}
and the partition function of $\mathcal{N}=4$ $U(N)$ SYM on $HS^{4}$ with Neumann boundary conditions from
\begin{equation}
    Z_{\textnormal{Neum.}}[HS^{4}] = (2 \pi)^{\frac{N}{2}} \lim_{b \rightarrow 2 \pi} I_{2} \left( b, \frac{\gym^{2}}{8 \pi^{2}}, N \right) \: .
\end{equation}

\subsection{General NS5-Like Localization Integrals}

We also need to evaluate integrals of the form
\begin{equation}
\begin{split}
    Z[HS^{4}] & = \frac{1}{n_{1}! \ldots n_{N_{5}}!} \int \left( \prod_{j=1}^{N_{5}} \prod_{\ell=1}^{n_{j}} d \lambda_{j, \ell}  \right) \left( \prod_{j=1}^{N_{5}-1} \prod_{\ell=1}^{n_{j}} e^{2 \pi i \alpha_{j} \lambda_{j, \ell}} \right) e^{- \frac{4 \pi^{2}}{g_{\textnormal{YM}}^{2}} \sum_{i=1}^{N} \lambda_{N_{5}, i}^{2}} \\
    & \qquad \times \prod_{i < j}^{N} (\lambda_{N_{5}, i} - \lambda_{N_{5}, j}) \: \sh (\lambda_{N_{5}, i} - \lambda_{N_{5}, j}) \prod_{j=1}^{N_{5}-1} \prod_{k < \ell}^{n_{j}} \sh^{2} \left( \lambda_{j, k} - \lambda_{j, \ell} \right)  \\
    & \qquad \times \prod_{j=1}^{N_{5}-1} \prod_{k=1}^{n_{j}} \prod_{\ell=1}^{n_{j+1}} \frac{1}{\ch(\lambda_{j, k} - \lambda_{j+1, \ell})} \: ,
\end{split}
\end{equation}
where we recall the notation
\begin{equation}
    \textnormal{sh}(x) \equiv 2 \sinh \pi x \: , \qquad \textnormal{ch}(x) \equiv 2 \cosh \pi x \: .
\end{equation}
We are ultimately interested in taking the limit $\alpha_{i} \rightarrow 0$. 
It will be convenient to introduce the function 
\begin{equation}
    h_{s}(\alpha) \equiv \sum_{n=0}^{\infty} (-1)^{ns} e^{- (2n + 1) \pi \alpha} = \begin{cases}
        \frac{1}{\ch(\alpha)} & 2 \nmid s \\
        \frac{1}{\sh(\alpha)} & 2 \mid s
    \end{cases} \: .
\end{equation}
We let $n_{0} \equiv 0, n_{m+1} \equiv N$ for notational ease. Additionally, we denote the index sets 
\begin{equation}
    S_{a} \equiv \{n_{a-1} + 1, n_{a-1}+2, \ldots, n_{a} \} \: .
\end{equation}

We will begin by showing that the integral
\begin{equation}
    \begin{split}
        \mathcal{I}_{n_{1}, \ldots, n_{s+1}}(\alpha_{1}, \ldots, \alpha_{s}) & \equiv \int \left( \prod_{j=1}^{n_{s}} d \lambda_{s, j} \right) e^{2 \pi i \sum_{j=1}^{s} \alpha_{j} \sum_{k \in S_{j}} \lambda_{s, k}}  \prod_{a=1}^{s} \prod_{\substack{k, \ell \in S_{a} \\ k < \ell}} \sh^{2}(\lambda_{s, k} - \lambda_{s, \ell}) \\
        & \qquad \prod_{a<b}^{s} \prod_{k \in S_{a}} \prod_{\ell \in S_{b}} \sh(\lambda_{s, k} - \lambda_{s, \ell}) \prod_{k=1}^{n_{s}} \prod_{\ell=1}^{n_{s+1}} \frac{1}{\ch(\lambda_{s, k} - \lambda_{s+1, \ell})} 
    \end{split}
\end{equation}
is given by
\begin{equation}
    \begin{split}
        \mathcal{I}_{n_{1}, \ldots, n_{s+1}}(\alpha_{1}, \ldots, \alpha_{s}) & = \frac{i^{-n_{s} (n_{s+1} - n_{s})}}{(n_{s+1}-n_{s})!} \tilde{H}^{s}_{n_{1}, \ldots, n_{s+1}}(\alpha_{1}, \ldots, \alpha_{s}) \\
        & \qquad \sum_{\sigma \in S_{n_{s+1}}} \frac{\prod_{a=1}^{s} e^{2 \pi i \alpha_{a} \sum_{\ell=1}^{n_{a}} \lambda_{s+1, \sigma(\ell)}}}{\prod_{a<b}^{s+1} \prod_{k \in S_{a}} \prod_{\ell \in S_{b}} \sh(\lambda_{s+1, \sigma(k)} - \lambda_{s+1, \sigma(\ell)})} \: ,
    \end{split}
\end{equation}
where we have the recursive relation
\begin{equation}
    \tilde{H}^{s}_{n_{1}, \ldots, n_{s+1}}(\alpha_{1}, \ldots, \alpha_{s}) = h_{n_{1} + n_{s} + n_{s+1}}(\alpha_{1} + \ldots + \alpha_{s})^{n_{1}} \times \tilde{H}^{s-1}_{n_{2}-n_{1}, \ldots, n_{s+1}-n_{1}}(\alpha_{2}, \ldots, \alpha_{s}) \: ,  
\end{equation}
with $\tilde{H}^{1}_{n_{1}, n_{2}}(\alpha_{1}) = h_{n_{2}}(\alpha_{1})^{n_{1}}$. 
\begin{proof}
We can verify this claim inductively. To begin, we determine 
\begin{equation}
    \mathcal{I}_{n_{1}, n_{2}}(\alpha_{1}) \equiv \int \left( \prod_{\ell=1}^{n_{1}} d \lambda_{1, \ell} e^{2 \pi i \alpha_{1} \lambda_{1, \ell}} \right) \prod_{k<\ell}^{n_{1}} \textnormal{sh}^{2}(\lambda_{1, k} - \lambda_{1, \ell}) \prod_{k=1}^{n_{1}} \prod_{\ell=1}^{n_{2}} \frac{1}{\textnormal{ch}(\lambda_{1, k} - \lambda_{2, \ell})} \: .
\end{equation}
First, we integrate out the variable $\lambda_{1, 1}$. Specifically, we would like to evaluate
\begin{equation}
    \int d \lambda_{1, 1} e^{2 \pi i \alpha_{1} \lambda_{1, 1}} \prod_{i=2}^{n_{1}} \sh^{2} (\lambda_{1, 1} - \lambda_{1, i}) \prod_{j=1}^{n_{2}} \frac{1}{\ch (\lambda_{1, 1} - \lambda_{2, j})} \: .
\end{equation}
Noting that the integrand is suppressed in the upper half plane for large $|\lambda_{1}|$ (when $\alpha_{1} > 0$), we may close the integration contour in the upper half plane and apply the residue theorem. The poles occur at $\lambda_{1, 1} = \lambda_{2, j} + (n+1/2)i$ for $j \in \{1, \ldots, n_{2}\}, n \in \mathbb{N}$, and the contribution to the integral from such a pole is given by
\begin{equation}
    i^{-(n_{2}-1)} (-1)^{(n_{1}-1)-n n_{2}} \times \frac{e^{2 \pi i \alpha_{1} (\lambda_{2, j} + (n+1/2)i)} \prod_{i=2}^{n_{1}} \ch^{2} (\lambda_{2, j} - \lambda_{1, i})}{ \prod_{\ell \neq j}^{n_{2}} \sh (\lambda_{2, j} - \lambda_{2, \ell})} \: .
\end{equation}
Consequently, summing over all of the poles, we find the full integral
\begin{equation}
    i^{-(n_{2}-1) + 2(n_{1} - 1)} h_{n_{2}}(\alpha_{1}) \sum_{j_{1}=1}^{n_{2}} \frac{e^{2 \pi i \alpha_{1} \lambda_{2, j_{1}}} \prod_{i=2}^{n_{1}} \ch^{2} ( \lambda_{1, i} - \lambda_{2, j_{1}})}{\prod_{\ell \neq j_{1}}^{n_{2}} \sh (\lambda_{2, j_{1}} - \lambda_{2, \ell})} \: ,
\end{equation}
and substituting this into $\mathcal{I}_{1}$ gives
\begin{equation}
\begin{split}
    & \frac{i^{-(n_{2} - 1) + 2 (n_{1} - 1)}}{n_{1}!} h_{n_{2}}(\alpha_{1}) \sum_{j_{1} = 1}^{n_{2}} \frac{e^{2 \pi i \alpha_{1} \lambda_{2, j_{1}}}}{\prod_{\ell \neq j_{1}} \sh(\lambda_{2, j_{1}} - \lambda_{2, \ell})} \int \left( \prod_{\ell=2}^{n_{1}} d \lambda_{1, \ell} e^{2 \pi i \alpha_{1} \lambda_{1, \ell}} \right) \\
    & \qquad \prod_{1 < k<\ell}^{n_{1}} \textnormal{sh}^{2}(\lambda_{1, k} - \lambda_{1, \ell}) \prod_{k=2}^{n_{1}} \left( \ch(\lambda_{1, k} - \lambda_{2, j_{1}}) \prod_{\ell \neq j_{1}}^{n_{2}} \frac{1}{\textnormal{ch}(\lambda_{1, k} - \lambda_{2, \ell})} \right) \: .
\end{split}
\end{equation}
Applying this approach to each successive integration, we find after integrating out all of the $\lambda_{1, i}$ variables that $\mathcal{I}_{n_{1}, n_{2}}(\alpha_{1})$ is given by
\begin{equation}
    \begin{split}
        & \frac{i^{- n_{1}(n_{2} - n_{1})} (-1)^{n_{1}(n_{1}-1)/2}}{n_{1}!} [h_{n_{2}}(\alpha_{1})]^{n_{1}} \sum_{j_{1} \neq \ldots \neq j_{n_{1}}}^{n_{2}} \frac{e^{2 \pi i \alpha_{1} \sum_{\ell=1}^{n_{1}} \lambda_{2, j_{\ell}}} \prod_{a<b}^{n_{1}} \sh(\lambda_{2, j_{b}} - \lambda_{2, j_{a}})}{\prod_{\ell \neq j_{1}}^{n_{2}} \sh(\lambda_{2, j_{1}} - \lambda_{2, \ell}) \ldots \prod_{\ell \neq j_{1}, \ldots, j_{n_{1}}}^{n_{2}} \sh(\lambda_{2, j_{n_{1}}} - \lambda_{2, \ell})} \\
        & = \frac{i^{- n_{1}(n_{2} - n_{1})}}{n_{1}!(n_{2} - n_{1})!} [h_{n_{2}}(\alpha_{1})]^{n_{1}} \sum_{\sigma \in S_{n_{2}}} \frac{e^{2 \pi i \alpha_{1} \sum_{\ell=1}^{n_{1}} \lambda_{2, \sigma(\ell)}} }{\prod_{k = 1}^{n_{1}} \prod_{\ell=n_{1} + 1}^{n_{2}} \sh(\lambda_{2, \sigma(k)} - \lambda_{2, \sigma(\ell)})} \: , 
    \end{split}
\end{equation}
which indeed is of the desired form.

Now suppose that the claim holds for $\mathcal{I}_{n_{1}, \ldots, n_{s}}(\alpha_{1}, \ldots, \alpha_{s-1})$. To determine $\mathcal{I}_{n_{1}, \ldots, n_{s+1}}(\alpha_{1}, \ldots, \alpha_{s})$, we may again apply the residue theorem to perform the first $n_{1}$ integrals
\begin{equation}
\begin{split}
    & \int \left(  \prod_{\ell=1}^{n_{1}} d \lambda_{s, \ell} \right)  e^{2 \pi i (\alpha_{1} + \ldots + \alpha_{s}) \sum_{\ell=1}^{n_{1}} \lambda_{s, \ell}}   \prod_{k < \ell}^{n_{1}} \textnormal{sh}^{2} \left( \lambda_{s, k} - \lambda_{s, \ell} \right) \\
    & \qquad \prod_{j=1}^{s} \prod_{a \in S_{0}} \prod_{b \in S_{j}} \textnormal{sh} \left( \lambda_{s, a} - \lambda_{s, b} \right)  \prod_{k \in S_{0}} \prod_{\ell =1}^{n_{s+1}} \frac{1}{\textnormal{ch}(\lambda_{s, k} - \lambda_{s+1, \ell})} \: ,
\end{split}
\end{equation}
which gives
\begin{equation}
\begin{split}
    & i^{-n_{1}(n_{s+1} - n_{s}) + n_{1}(n_{1} - 1)} [h_{n_{1} + n_{s} + n_{s+1}}(\alpha_{1}+ \ldots + \alpha_{s})]^{n_{1}} \\
    & \qquad \sum_{j_{1} \neq \ldots \neq j_{n_{1}}}^{n_{s+1}} \frac{e^{2 \pi i (\alpha_{1} + \ldots + \alpha_{s}) \sum_{k=1}^{n_{1}} \lambda_{s+1, j_{k}}} \prod_{k<\ell}^{n_{1}} \sh(\lambda_{s+1, j_{\ell}} - \lambda_{s+1, j_{k}}) \prod_{k=1}^{n_{1}} \prod_{b=n_{1}+1}^{n_{s}} \ch(\lambda_{s, b} - \lambda_{s+1, j_{k}})}{\prod_{\ell \neq j_{1}}^{n_{s+1}} \sh (\lambda_{s+1, j_{1}} - \lambda_{s+1, \ell}) \ldots \prod_{\ell \neq j_{1}, \ldots, j_{n_{1}}}^{n_{s+1}} \sh (\lambda_{s+1, j_{n_{1}}} - \lambda_{s+1, \ell}) } \: .
\end{split}
\end{equation}
Substituting this into $\mathcal{I}_{n_{1}, \ldots, n_{s+1}}(\alpha_{1}, \ldots, \alpha_{s})$ gives
\begin{equation}
    \begin{split}
        & i^{-n_{1}(n_{s+1} - n_{s})} [h_{n_{1} + n_{s} + n_{s+1}}(\alpha_{1}+ \ldots + \alpha_{s})]^{n_{1}} \\
        & \qquad \sum_{j_{1} \neq \ldots \neq j_{n_{1}}}^{n_{s+1}} \frac{e^{2 \pi i (\alpha_{1} + \ldots + \alpha_{s}) \sum_{k=1}^{n_{1}} \lambda_{s+1, j_{k}}} \prod_{k<\ell}^{n_{1}} \sh(\lambda_{s+1, j_{k}} - \lambda_{s+1, j_{\ell}})}{\prod_{\ell \neq j_{1}}^{n_{s+1}} \sh (\lambda_{s+1, j_{1}} - \lambda_{s+1, \ell}) \ldots \prod_{\ell \neq j_{1}, \ldots, j_{n_{1}}}^{n_{s+1}} \sh (\lambda_{s+1, j_{n_{1}}} - \lambda_{s+1, \ell}) } \\
        & \qquad \int \left(  \prod_{\ell=n_{1}+1}^{n_{s}} d \lambda_{s, \ell}   \right)  e^{2 \pi i \alpha_{2} \sum_{\ell=n_{1}+1}^{n_{2}} \lambda_{s, \ell}} \ldots e^{2 \pi i \alpha_{s} \sum_{\ell=n_{1}+1}^{n_{s}} \lambda_{s, \ell}}  
        \prod_{a=1}^{s-1} \prod_{n_{a}<k < \ell}^{n_{a+1}} \textnormal{sh}^{2} \left( \lambda_{s, k} - \lambda_{s, \ell} \right) \\
        & \qquad  \prod_{1 < a < b}^{s} \prod_{i \in S_{a}} \prod_{j \in S_{b}} \textnormal{sh} \left( \lambda_{s, i} - \lambda_{s, j} \right)  \prod_{k=n_{1}+1}^{n_{s}} \prod_{\ell\neq j_{1}, \ldots, j_{n_{1}}}^{n_{s+1}} \frac{1}{\textnormal{ch}(\lambda_{s, k} - \lambda_{s+1, \ell})} \: .
    \end{split}
\end{equation}
Evidently, the integral appearing in this expression is simply
\begin{equation}
    \mathcal{I}_{n_{2}-n_{1}, n_{3}-n_{1}, \ldots, n_{s+1}-n_{1}}(\alpha_{2}, \ldots, \alpha_{s}) \: ,
\end{equation}
which by inductive hypothesis is
\begin{equation}
    \begin{split}
         & i^{-(n_{s}-n_{1})(n_{s+1}-n_{s}) } \tilde{H}^{s-1}_{n_{2} - n_{1}, \ldots, n_{s+1}-n_{1}}(\alpha_{2}, \ldots, \alpha_{s}) \\
         & \qquad \sum_{j_{n_{1}+1} \neq \ldots \neq j_{n_{s}}} \frac{\prod_{a=2}^{s} e^{2 \pi i \alpha_{a} \sum_{\ell=n_{1}+1}^{n_{a}} \lambda_{s+1, j_{\ell}}}}{\prod_{1<a<b}^{s+1} \prod_{k \in S_{a}} \prod_{\ell \in S_{b}} \sh(\lambda_{s+1, j_{k}} - \lambda_{s+1, j_{\ell}})} \: .
    \end{split}
\end{equation}
Thus, we have
\begin{equation}
    \begin{split}
        \mathcal{I}_{n_{1}, \ldots, n_{s+1}}(\alpha_{1}, \ldots, \alpha_{s}) & =  i^{-n_{s}(n_{s+1} - n_{s})} [h_{n_{1} + n_{s} + n_{s+1}}(\alpha_{1}+ \ldots + \alpha_{s})]^{n_{1}} \tilde{H}^{s-1}_{n_{2} - n_{1}, \ldots, n_{s+1}-n_{1}}(\alpha_{2}, \ldots, \alpha_{s}) \\
        & \qquad \sum_{j_{1} \neq \ldots \neq j_{n_{s}}}^{n_{s+1}} \frac{ \prod_{a=1}^{s} e^{2 \pi i \alpha_{a} \sum_{k=1}^{n_{a}} \lambda_{s+1, j_{k}}} \prod_{k<\ell}^{n_{1}} \sh(\lambda_{s+1, j_{k}} - \lambda_{s+1, j_{\ell}})}{\prod_{\ell \neq j_{1}}^{n_{s+1}} \sh (\lambda_{s+1, j_{1}} - \lambda_{s+1, \ell}) \ldots \prod_{\ell \neq j_{1}, \ldots, j_{n_{1}}}^{n_{s+1}} \sh (\lambda_{s+1, j_{n_{1}}} - \lambda_{s+1, \ell}) } \\
        & \qquad \times \frac{1}{\prod_{1<a<b}^{s+1} \prod_{k \in S_{a}} \prod_{\ell \in S_{b}} \sh(\lambda_{s+1, j_{k}} - \lambda_{s+1, j_{\ell}})} \\
        & = \frac{i^{-n_{s}(n_{s+1} - n_{s})}}{(n_{s+1}-n_{s})!} \tilde{H}^{s}_{n_{1}, \ldots, n_{s}}(\alpha_{1}, \ldots, \alpha_{s}) \\
        & \qquad \sum_{\sigma \in S_{n_{s+1}}} \frac{ \prod_{a=1}^{s} e^{2 \pi i \alpha_{a} \sum_{k=1}^{n_{a}} \lambda_{s+1, \sigma(k)}}}{\prod_{a<b}^{s+1} \prod_{k \in S_{a}} \prod_{\ell \in S_{b}} \sh(\lambda_{s+1, \sigma(k)} - \lambda_{s+1, \sigma(\ell)})}  \: ,
    \end{split}
\end{equation}
which verifies the claim. 
\end{proof}

We now claim that performing the integral
\begin{equation}
    \begin{split}
        \mathcal{I}_{s} & \equiv \frac{1}{n_{1}! \ldots n_{s}!} \int \left( \prod_{a=1}^{s} \prod_{\ell=1}^{n_{a}} d \lambda_{a, \ell}  e^{2 \pi i \alpha_{a} \lambda_{a, \ell}} \right) \prod_{a=1}^{s} \prod_{k < \ell}^{n_{a}} \textnormal{sh}^{2} \left( \lambda_{a, k} - \lambda_{a, \ell} \right) \prod_{a=1}^{s} \prod_{k=1}^{n_{a}} \prod_{\ell=1}^{n_{a+1}} \frac{1}{\textnormal{ch}(\lambda_{a, k} - \lambda_{a+1, \ell})}
    \end{split}
\end{equation}
yields
\begin{equation}
    \begin{split}
        \mathcal{I}_{s} & = \frac{i^{ - \sum_{\ell=0}^{s-1} (n_{\ell+1} - n_{\ell})(n_{s+1} - n_{\ell + 1}) }}{\prod_{\ell=0}^{s} (n_{\ell+1} - n_{\ell})!} H_{(n_{1}, \ldots, n_{s+1})}(\alpha_{1}, \ldots, \alpha_{s})
        \\
        & \qquad \sum_{\sigma \in S_{n_{s+1}}} \frac{ \prod_{a=1}^{s} e^{2 \pi i \alpha_{a} \sum_{\ell=1}^{n_{a}} \lambda_{s+1, \sigma(\ell)}}}{ \prod_{a<b}^{s+1} \prod_{i \in S_{a}} \prod_{j \in S_{b}} \sh (\lambda_{s+1, \sigma(i)} - \lambda_{s+1, \sigma(j)})} \: ,
    \end{split}
\end{equation}
where $H_{(n_{1}, \ldots, n_{s+1})}(\alpha_{1}, \ldots, \alpha_{s})$ is an expression involving the $h_{n}(\alpha)$, given recursively by
\begin{equation}
    H_{(n_{1}, \ldots, n_{s+1})}(\alpha_{1}, \ldots, \alpha_{s}) = H_{(n_{1}, \ldots, n_{s})}(\alpha_{1}, \ldots, \alpha_{s-1}) \tilde{H}^{s}_{n_{1}, \ldots, n_{s+1}}(\alpha_{1}, \ldots, \alpha_{s}) \: ,
\end{equation}
and $H_{(n_{1}, n_{2})}(\alpha_{1}) = h_{n_{2}}(\alpha_{1})^{n_{1}}$. 
\begin{proof}
We can inductively verify our expression for $\mathcal{I}_{s}$, using our previous inductive result. We have already checked the base case above. 
Now suppose that the claim holds for $\mathcal{I}_{s-1}$. We have by induction hypothesis
\begin{equation}
    \begin{split}
        \mathcal{I}_{s} & = \frac{1}{n_{s}!} \int \left(  \prod_{\ell=1}^{n_{s}} d \lambda_{s, \ell}  e^{2 \pi i \alpha_{s} \lambda_{s, \ell}} \right)  \prod_{k < \ell}^{n_{s}} \textnormal{sh}^{2} \left( \lambda_{s, k} - \lambda_{s, \ell} \right)  \prod_{k=1}^{n_{s}} \prod_{\ell=1}^{n_{s+1}} \frac{1}{\textnormal{ch}(\lambda_{s, k} - \lambda_{s+1, \ell})} \\
        & \qquad \times  \frac{i^{ - \sum_{\ell=0}^{s-2} (n_{\ell+1} - n_{\ell})(n_{s} - n_{\ell + 1}) }}{\prod_{\ell=0}^{s-1} (n_{\ell+1} - n_{\ell})!} H_{(n_{1}, \ldots, n_{s})}(\alpha_{1}, \ldots, \alpha_{s-1})
        \\
        & \qquad  \times \sum_{\sigma \in S_{n_{s}}} \frac{e^{2 \pi i \alpha_{1} \sum_{\ell=1}^{n_{1}} \lambda_{s, \sigma(\ell)}} \ldots e^{2 \pi i \alpha_{s-1} \sum_{\ell=1}^{n_{s-1}} \lambda_{s, \sigma(\ell)}  }}{ \prod_{a<b}^{s} \prod_{i \in S_{a}} \prod_{j \in S_{b}} \sh (\lambda_{s, \sigma(i)} - \lambda_{s, \sigma(j)})}  \\
        & = \frac{i^{ - \sum_{\ell=0}^{s-2} (n_{\ell+1} - n_{\ell})(n_{s} - n_{\ell + 1}) }  H_{(n_{1}, \ldots, n_{s})}(\alpha_{1}, \ldots, \alpha_{s-1})}{ \prod_{\ell=0}^{s-1} (n_{\ell+1} - n_{\ell})!}  \int \left(  \prod_{\ell=1}^{n_{s}} d \lambda_{s, \ell}   \right) \\
        & \qquad \times e^{2 \pi i \alpha_{1} \sum_{\ell=1}^{n_{1}} \lambda_{s, \ell}} \ldots e^{2 \pi i \alpha_{s} \sum_{\ell=1}^{n_{s}} \lambda_{s, \ell}}  
        \prod_{i=0}^{s-1} \prod_{n_{i}<k < \ell}^{n_{i+1}} \textnormal{sh}^{2} \left( \lambda_{s, k} - \lambda_{s, \ell} \right) \\
        & \qquad \times \prod_{a<b}^{s} \prod_{i \in S_{a}} \prod_{j \in S_{b}} \textnormal{sh} \left( \lambda_{s, i} - \lambda_{s, j} \right)  \prod_{k=1}^{n_{s}} \prod_{\ell=1}^{n_{s+1}} \frac{1}{\textnormal{ch}(\lambda_{s, k} - \lambda_{s+1, \ell})} \: .
    \end{split}
\end{equation}
But the integral appearing in this expression is of the form encountered in our previous claim, and is thus given by
\begin{equation}
    \begin{split}
        \mathcal{I}_{n_{1}, \ldots, n_{s+1}}(\alpha_{1}, \ldots, \alpha_{s}) & = \frac{i^{-n_{s} (n_{s+1} - n_{s})}}{(n_{s+1}-n_{s})!} \tilde{H}^{s}_{n_{1}, \ldots, n_{s+1}}(\alpha_{1}, \ldots, \alpha_{s}) \\
        & \qquad \sum_{\sigma \in S_{n_{s+1}}} \frac{\prod_{a=1}^{s} e^{2 \pi i \alpha_{a} \sum_{\ell=1}^{n_{a}} \lambda_{s+1, \sigma(\ell)}}}{\prod_{a<b}^{s+1} \prod_{k \in S_{a}} \prod_{\ell \in S_{b}} \sh(\lambda_{s+1, \sigma(k)} - \lambda_{s+1, \sigma(\ell)})} \: ,
    \end{split}
\end{equation}
so
\begin{equation}
    \begin{split}
        \mathcal{I}_{s} 
        & = \frac{i^{ - n_{s}(n_{s+1}-n_{s}) - \sum_{\ell=0}^{s-2} (n_{\ell+1} - n_{\ell})(n_{s} - n_{\ell + 1}) }  H_{(n_{1}, \ldots, n_{s})}(\alpha_{1}, \ldots, \alpha_{s-1})}{ \prod_{\ell=0}^{s} (n_{\ell+1} - n_{\ell})!} \tilde{H}^{s}_{n_{1}, \ldots, n_{s+1}}(\alpha_{1}, \ldots, \alpha_{s})  \\
        & \qquad \sum_{\sigma \in S_{n_{s+1}}} \frac{\prod_{a=1}^{s} e^{2 \pi i \alpha_{a} \sum_{\ell=1}^{n_{a}} \lambda_{s+1, \sigma(\ell)}}}{\prod_{a<b}^{s+1} \prod_{k \in S_{a}} \prod_{\ell \in S_{b}} \sh(\lambda_{s+1, \sigma(k)} - \lambda_{s+1, \sigma(\ell)})} \: .
    \end{split}
\end{equation}
We note that
\begin{equation}
    \begin{split}
        & n_{s}(n_{s+1} - n_{s}) + \sum_{\ell=0}^{s-2}(n_{\ell+1} - n_{\ell})(n_{s} - n_{\ell+1}) \\
        & \qquad = (n_{s}-n_{s-1})(n_{s+1} - n_{s})  + \sum_{\ell=0}^{s-2}(n_{\ell+1} - n_{\ell})(n_{s+1} - n_{\ell+1}) \: ,
    \end{split}
\end{equation}
so
\begin{equation}
    \begin{split}
        \mathcal{I}_{s} 
        & = \frac{i^{ - \sum_{\ell=0}^{s-1} (n_{\ell+1} - n_{\ell})(n_{s} - n_{\ell + 1}) }  H_{(n_{1}, \ldots, n_{s+1})}(\alpha_{1}, \ldots, \alpha_{s})}{ \prod_{\ell=0}^{s} (n_{\ell+1} - n_{\ell})!}  \\
        & \qquad \sum_{\sigma \in S_{n_{s+1}}} \frac{\prod_{a=1}^{s} e^{2 \pi i \alpha_{a} \sum_{\ell=1}^{n_{a}} \lambda_{s+1, \sigma(\ell)}}}{\prod_{a<b}^{s+1} \prod_{k \in S_{a}} \prod_{\ell \in S_{b}} \sh(\lambda_{s+1, \sigma(k)} - \lambda_{s+1, \sigma(\ell)})} \: ,
    \end{split}
\end{equation}
as desired. 
\end{proof}

Using the above claim to perform all of the integrals except those with respect to the bulk zero modes, the partition function from the beginning of this subsection is
\begin{equation}
    \begin{split}
        Z[HS^{4}] & = \frac{i^{ - \sum_{\ell=0}^{N_{5}-2} (n_{\ell+1} - n_{\ell})(n_{N_{5}} - n_{\ell + 1}) }}{\prod_{\ell=0}^{N_{5}-1} (n_{\ell+1} - n_{\ell})!} \lim_{\alpha_{1}, \ldots, \alpha_{N_{5}-1} \rightarrow 0} H_{(n_{1}, \ldots, n_{N_{5}})}(\alpha_{1}, \ldots, \alpha_{N_{5}-1})
        \\
        & \qquad \int \left( \prod_{i=1}^{N} d \lambda_{i} \right) e^{- \frac{4 \pi^{2}}{g_{\textnormal{YM}}^{2}} \sum_{i=1}^{N} \lambda_{i}^{2}} \prod_{a=1}^{N_{5}-1} e^{2 \pi i \alpha_{a} \sum_{\ell=1}^{n_{a}} \lambda_{\ell}} \prod_{i < j}^{N} (\lambda_{i} - \lambda_{j})  
        \\
        & \qquad \prod_{a=1}^{N_{5}} \prod_{\substack{i, j \in S_{a} \\ i < j}} \: \textnormal{sh} (\lambda_{i} - \lambda_{j}) \: .
    \end{split}
\end{equation}
We may as well take $\alpha_{1} = \ldots = \alpha_{N_{5}-1}$ before taking the limit. We may therefore write
\begin{equation}
\begin{split}
    Z[HS^{4}] & =  \frac{i^{ - \sum_{\ell=0}^{N_{5}-2} (n_{\ell+1} - n_{\ell})(n_{N_{5}} - n_{\ell + 1}) }}{\prod_{\ell=0}^{N_{5}-1} (n_{\ell+1} - n_{\ell})!} \lim_{\alpha, a \rightarrow 0} \lim_{b \rightarrow 2 \pi} a^{-N(N-1)/2} H_{(n_{1}, \ldots, n_{N_{5}})}(\alpha) \\
    & \qquad \int \left( \prod_{i=1}^{N} d \lambda_{i} \right) e^{- \frac{1}{2s} \sum_{i=1}^{N} \lambda_{i}^{2}}  \prod_{c=1}^{N_{5}-1} e^{2 \pi i \alpha_{c} \sum_{\ell=1}^{n_{c}} \lambda_{\ell}} 
    \\
    & \qquad \prod_{i < j}^{N} \sh \left( \frac{a(\lambda_{i} - \lambda_{j})}{2 \pi} \right) \prod_{c=1}^{N_{5}} \prod_{\substack{i, j \in S_{c} \\ i < j}} \sh \left( \frac{b(\lambda_{i} - \lambda_{j})}{2 \pi} \right)  \: ,
    \end{split}
\end{equation}
where we let $s \equiv \frac{g_{\textnormal{YM}}^{2}}{8 \pi^{2}}$ and $H_{(n_{1}, \ldots, n_{N_{5}})}(\alpha) \equiv H_{(n_{1}, \ldots, n_{N_{5}})}(\alpha, \ldots, \alpha)$. 
Again using the identity (\ref{eq:sinhid}), 
we may express 
\begin{equation}
    \begin{split}
        \mathcal{I} & \equiv \int \left( \prod_{i=1}^{N} d \lambda_{i} \right) e^{- \frac{1}{2s} \sum_{i=1}^{N} \lambda_{i}^{2}}  \prod_{c=1}^{N_{5}-1} e^{2 \pi i \alpha_{c} \sum_{\ell=1}^{n_{c}} \lambda_{\ell}} 
        \\
        & \qquad \prod_{i < j}^{N} \sh \left( \frac{a(\lambda_{i} - \lambda_{j})}{2 \pi} \right) \prod_{c=1}^{N_{5}} \prod_{\substack{i, j \in S_{c} \\ i < j}} \sh \left( \frac{b(\lambda_{i} - \lambda_{j})}{2 \pi} \right) 
    \end{split}
\end{equation}
as 
\begin{equation}
    \begin{split}
        \mathcal{I} & = \int \left( \prod_{i=1}^{N} d \lambda_{i} \right) e^{- \frac{1}{2s} \sum_{i=1}^{N} \lambda_{i}^{2}}   \prod_{c=1}^{N_{5}-1} e^{2 \pi i \alpha_{c} \sum_{\ell=1}^{n_{c}} \lambda_{\ell}} 
        \\
        & \qquad \left( \sum_{\sigma \in S_{N}} (-1)^{\sigma} \prod_{j=1}^{N} e^{a \left( \frac{N+1}{2} - \sigma_{j} \right) \lambda_{j}} \right) \prod_{c=1}^{N_{5}} \left( \sum_{\sigma_{c} \in S_{n_{c} - n_{c-1}}} (-1)^{\sigma_{c}} \prod_{j \in S_{c}} e^{b \left( \frac{n_{c}-n_{c-1}+1}{2} - \sigma_{c, j-n_{c-1}} \right) \lambda_{j}} \right)  \: .
    \end{split}
\end{equation}
Performing the Gaussian integrals, one finds
\begin{equation}
    \begin{split}
        \mathcal{I} & = (2 \pi s)^{N/2} \sum_{\sigma \in S_{N}} (-1)^{\sigma} \sum_{\sigma_{1}, \ldots, \sigma_{m+1}} (-1)^{\sigma_{1} + \ldots + \sigma_{m}} \\
        & \qquad \prod_{c=1}^{N_{5}} \left( \prod_{j \in S_{c}} e^{\frac{s}{2} \left[ a \left( \frac{N+1}{2} - \sigma_{j} \right) + b \left( \frac{n_{c} - n_{c-1} + 1}{2} - \sigma_{c, j-n_{c-1}} \right) + 2 \pi i (N_{5}-c) \alpha \right]^{2}} \right)  \: ,
    \end{split}
\end{equation}
and thus, defining $\ell_{c} \equiv n_{c}-n_{c-1}$ (the linking numbers in the case with only NS5-branes),
\begin{equation}
    \begin{split}
        \mathcal{I} 
        & = (2 \pi s)^{N/2} e^{\frac{s a^{2} N (N-1) (N+1)}{24} + \frac{s b^{2} }{24} \sum_{c=1}^{N_{5}} \ell_{c} (\ell_{c}-1)(\ell_{c}+1) -2 \pi^{2} s \alpha^{2} \sum_{c=1}^{N_{5}} (N_{5}-c)^{2} \ell_{c}} \\
        & \qquad e^{\pi i (N+1) s \alpha a \sum_{c=1}^{N_{5}} (N_{5}-c) \ell_{c}} \sum_{\sigma_{1}, \ldots, \sigma_{N_{5}}} (-1)^{\sigma_{1} + \ldots + \sigma_{N_{5}}} \sum_{\sigma \in S_{N}} (-1)^{\sigma}   \prod_{j=1}^{N} e^{- \mu_{j} \sigma_{j}} \: ,
    \end{split}
\end{equation}
where
\begin{equation}
    \begin{split}
        \mu_{j} & \equiv s a b \left( \frac{\ell_{c} + 1}{2} - \sigma_{c, j - n_{c-1}} \right) + 2 \pi i s a \alpha (N_{5}-c) \qquad \textnormal{for} \qquad j \in S_{c} \: .
    \end{split}
\end{equation}
That is, using the identity (\ref{eq:sinhid}) above, the Gaussian integral gives
\begin{equation}
    \begin{split}
         \mathcal{I} 
        & = (2 \pi s)^{N/2} e^{\frac{s a^{2} N (N-1) (N+1)}{24} + \frac{s b^{2} }{24} \sum_{c=1}^{N_{5}} \ell_{c} (\ell_{c}-1)(\ell_{c}+1) -2 \pi^{2} s \alpha^{2} \sum_{c=1}^{N_{5}} (N_{5}-c)^{2} \ell_{c}} \\
        & \qquad \sum_{\sigma_{1}, \ldots, \sigma_{N_{5}}} (-1)^{\sigma_{1} + \ldots + \sigma_{N_{5}}} \prod_{i < j}^{N} \sh\left( \frac{\mu_{i} - \mu_{j}}{2 \pi} \right) \: .
    \end{split}
\end{equation}
We therefore have
\begin{equation}
\begin{split}
    Z[HS^{4}] & =  \frac{i^{ - \sum_{\ell=0}^{N_{5}-2} (n_{\ell+1} - n_{\ell})(n_{N_{5}} - n_{\ell + 1}) }}{\prod_{\ell=0}^{N_{5}-1} (n_{\ell+1} - n_{\ell})!} \lim_{\alpha, a \rightarrow 0} \lim_{b \rightarrow 2 \pi} a^{-N(N-1)/2} H_{(n_{1}, \ldots, n_{N_{5}})}(\alpha) \\
    & \qquad (2 \pi s)^{N/2} e^{\frac{s a^{2} N (N-1) (N+1)}{24} + \frac{s b^{2} }{24} \sum_{c=1}^{N_{5}} \ell_{c} (\ell_{c}-1)(\ell_{c}+1) -2 \pi^{2} s \alpha^{2} \sum_{c=1}^{N_{5}} (N_{5}-c)^{2} \ell_{c}} \\
    & \qquad \sum_{\sigma_{1}, \ldots, \sigma_{N_{5}}} (-1)^{\sigma_{1} + \ldots + \sigma_{N_{5}}} \left( \prod_{c=1}^{N_{5}} \prod_{\substack{i, j \in S_{c} \\ i < j}} \sh \left( \frac{s a b ( \sigma_{c, j - n_{c-1}} - \sigma_{c, i - n_{c-1}})}{2 \pi} \right) \right) \\
    & \qquad \left( \prod_{c<d}^{N_{5}} \prod_{i \in S_{c}} \prod_{j \in S_{d}} \sh \left( \frac{s a b (\sigma_{d, j - n_{d-1}} - \sigma_{c, i - n_{c-1}})}{2 \pi} + \frac{sab(\ell_{c} - \ell_{d})}{4 \pi} +  i s a \alpha (d-c) \right) \right) \: .
    \end{split}
\end{equation}
Taking the $a \rightarrow 0$ and $b \rightarrow 2 \pi$ limits gives
\begin{equation}
\begin{split}
    Z[HS^{4}] & =  i^{ - \sum_{\ell=0}^{N_{5}-2} (n_{\ell+1} - n_{\ell})(n_{N_{5}} - n_{\ell + 1}) }  \left( \frac{\gym^{2}}{4 \pi} \right)^{ \frac{N^{2}}{2}} e^{ \frac{\gym^{2} }{48} \sum_{c=1}^{N_{5}} \ell_{c} (\ell_{c}-1)(\ell_{c}+1)} \\
    & \qquad \times \prod_{c=1}^{N_{5}} G_{2}(\ell_{c}+1) \lim_{\alpha \rightarrow 0} H_{(n_{1}, \ldots, n_{N_{5}})}(\alpha) e^{ -2 \pi^{2} s \alpha^{2} \sum_{c=1}^{N_{5}} (N_{5}-c)^{2} \ell_{c}} \\
    & \qquad \times \left( \prod_{c < d}^{N_{5}} \prod_{i =1}^{\ell_{c}} \prod_{j = 1}^{\ell_{d}} \left( (j - i) + \frac{(\ell_{c} - \ell_{d})}{2} + i \alpha (d-c) \right) \right) \: .
    \end{split}
\end{equation}

Now, we claim that
\begin{equation}
\begin{split}
    & \lim_{\alpha \rightarrow 0} \tilde{H}^{s}_{n_{1}, \ldots, n_{s+1}}(\alpha) \prod_{d=1}^{s} \prod_{i=1}^{(n_{c} - n_{c-1})} \prod_{j=1}^{(n_{s+1} - n_{s})} \left( (j-i) - \frac{(n_{s+1} - n_{s}) - (n_{c} - n_{c-1})}{2} + i \alpha (s + 1 - c) \right) \\
    & \qquad = 2^{- n_{s}} \prod_{c=1}^{s} \left[ \left( (\ell_{s + 1} - \ell_{c})!! \right)^{\ell_{c}} \prod_{k=1}^{\ell_{c} - 1} \left( \frac{\ell_{s+1} - \ell_{c}}{2} + k \right)^{\ell_{c} - k} \right]^{2}  \\
    & \qquad \qquad \times \prod_{\substack{ c \in \{1 , \ldots, s\} \\ (\ell_{c} - \ell_{s+1}) \equiv 0 \: (\textnormal{mod 2}) }} \left( \frac{i}{\pi} \right)^{\ell_{c}} (-1)^{\frac{\ell_{c}}{2}(\ell_{s+1}-1)} 2^{-(\ell_{s+1}-\ell_{c}) \ell_{c}} \\
    & \qquad \qquad \times \prod_{\substack{ c \in \{1 , \ldots, s\}  \\ (\ell_{c} - \ell_{s+1}) \equiv 1 \: (\textnormal{mod 2})} } (-1)^{\frac{\ell_{c} \ell_{s+1}}{2}} 2^{-(\ell_{s+1} - \ell_{c}+1) \ell_{c}} \: .
\end{split}
\end{equation}

\begin{proof}
We may verify this by induction. The base case $s=1$ is straightforward to verify individually for the cases $\ell_{1} - \ell_{2}$ even and odd.

Now suppose that the claim holds for some $s = p-1$; then by induction hypothesis, we have
\begin{equation}
\begin{split}
    & \lim_{\alpha \rightarrow 0} \tilde{H}^{p}_{n_{1}, \ldots, n_{p+1}}(\alpha) \prod_{c=1}^{p} \prod_{i=1}^{(n_{c} - n_{c-1})} \prod_{j=1}^{(n_{p+1} - n_{p})} \left( (j-i) + \frac{(n_{c} - n_{c-1})- (n_{p+1} - n_{p})}{2} + i \alpha (p + 1 - c) \right) \\
    & \qquad = \lim_{\alpha \rightarrow 0} h_{n_{1} + n_{p} + n_{p+1}}(p \alpha)^{n_{1}} \prod_{i=1}^{\ell_{1}} \prod_{j=1}^{\ell_{p+1}} \left( (j-i) + \frac{\ell_{1} - \ell_{p+1}}{2} + i \alpha p \right) \\
    & \qquad \qquad \times 2^{- (n_{p} - n_{1})} \prod_{c=2}^{p} \left[ \left( (\ell_{p+1} - \ell_{c})!! \right)^{\ell_{c}} \prod_{k=1}^{\ell_{c} - 1} \left( \frac{\ell_{p+1} - \ell_{c}}{2} + k \right)^{\ell_{c} - k} \right]^{2}  \\
    & \qquad \qquad \times \prod_{\substack{ c \in \{2 , \ldots, p\} \\ (\ell_{c} - \ell_{p+1}) \equiv 0 \: (\textnormal{mod 2}) }} \left( \frac{i}{\pi} \right)^{\ell_{c}} (-1)^{\frac{\ell_{c}}{2}(\ell_{p+1}-1)} 2^{-(\ell_{p+1}-\ell_{c}) \ell_{c}} \\
    & \qquad \qquad \times \prod_{\substack{ c \in \{2 , \ldots, p\}  \\ (\ell_{c} - \ell_{p+1}) \equiv 1 \: (\textnormal{mod 2})} } (-1)^{\frac{\ell_{c} \ell_{p+1}}{2}} 2^{-(\ell_{p+1} - \ell_{c}+1) \ell_{c}} \: .
\end{split}
\end{equation}
If $n_{1} + n_{p} + n_{p+1}$ is odd, then $\ell_{1} - \ell_{p+1} = n_{1} + n_{p} - n_{p+1}$ is odd, so
\begin{equation}
    \begin{split}
        & \lim_{\alpha \rightarrow 0} h_{n_{1} + n_{p} + n_{p+1}}(p \alpha)^{n_{1}} \prod_{i=1}^{\ell_{1}} \prod_{j=1}^{\ell_{p+1}} \left( (j-i) + \frac{\ell_{1} - \ell_{p+1}}{2} + i \alpha p \right) \\
        & \qquad = 2^{-n_{1}} (-1)^{\frac{\ell_{1}}{2}(\ell_{p+1}-\ell_{1} + 1)} \left( \left( \frac{1}{2} \right) \times \left( \frac{3}{2} \right) \times \ldots \times \left( \frac{\ell_{p+1} - \ell_{1}}{2} - 1 \right) \times \left( \frac{\ell_{p+1} - \ell_{1}}{2} \right) \right)^{2 \ell_{1}} \\
        & \qquad \qquad \times (-1)^{\frac{\ell_{1}}{2} (\ell_{1} - 1)} \prod_{k=1}^{\ell_{1}-1} \left( \frac{\ell_{p+1} - \ell_{1}}{2} + k \right)^{2 (\ell_{1} - k)} \: ,
    \end{split}
\end{equation}
and thus
\begin{equation}
\begin{split}
    & \lim_{\alpha \rightarrow 0} \tilde{H}^{p}_{n_{1}, \ldots, n_{p+1}}(\alpha) \prod_{c=1}^{p} \prod_{i=1}^{(n_{c} - n_{c-1})} \prod_{j=1}^{(n_{p+1} - n_{p})} \left( (j-i) + \frac{(n_{c} - n_{c-1})- (n_{p+1} - n_{p})}{2} + i \alpha (p + 1 - c) \right) \\
    & \qquad =  2^{- n_{p} } \prod_{c=1}^{p} \left[ \left( (\ell_{p+1} - \ell_{c})!! \right)^{\ell_{c}} \prod_{k=1}^{\ell_{c} - 1} \left( \frac{\ell_{p+1} - \ell_{c}}{2} + k \right)^{\ell_{c} - k} \right]^{2}  \\
    & \qquad \qquad \times \prod_{\substack{ c \in \{1 , \ldots, p\} \\ (\ell_{c} - \ell_{p+1}) \equiv 0 \: (\textnormal{mod 2}) }} \left( \frac{i}{\pi} \right)^{\ell_{c}} (-1)^{\frac{\ell_{c}}{2}(\ell_{p+1}-1)} 2^{-(\ell_{p+1}-\ell_{c}) \ell_{c}} \\
    & \qquad \qquad \times \prod_{\substack{ c \in \{1 , \ldots, p\}  \\ (\ell_{c} - \ell_{p+1}) \equiv 1 \: (\textnormal{mod 2})} } (-1)^{\frac{\ell_{c} \ell_{p+1}}{2}} 2^{-(\ell_{p+1} - \ell_{c}+1) \ell_{c}} \: ,
\end{split}
\end{equation}
which is of the desired form. On the other hand, if $n_{1} + n_{p} + n_{p+1}$ is even, then
\begin{equation}
    \begin{split}
        & \lim_{\alpha \rightarrow 0} h_{n_{1} + n_{p} + n_{p+1}}(p \alpha)^{n_{1}} \prod_{i=1}^{\ell_{1}} \prod_{j=1}^{\ell_{p+1}} \left( (j-i) + \frac{\ell_{1} - \ell_{p+1}}{2} + i \alpha p \right) \\
        & \qquad = \left( \frac{i}{2 \pi} \right)^{-n_{1}} (-1)^{\frac{\ell_{1}}{2}(\ell_{p+1}-\ell_{1})} \left( 1 \times 2 \times \ldots \times \left( \frac{\ell_{p+1} - \ell_{1}}{2} - 1 \right) \times \left( \frac{\ell_{p+1} - \ell_{1}}{2} \right) \right)^{2 \ell_{1}} \\
        & \qquad \qquad \times (-1)^{\frac{\ell_{1}}{2} (\ell_{1} - 1)} \prod_{k=1}^{\ell_{1}-1} \left( \frac{\ell_{p+1} - \ell_{1}}{2} + k \right)^{2 (\ell_{1} - k)} \: ,
    \end{split}
\end{equation}
and thus
\begin{equation}
\begin{split}
    & \lim_{\alpha \rightarrow 0} \tilde{H}^{p}_{n_{1}, \ldots, n_{p+1}}(\alpha) \prod_{c=1}^{p} \prod_{i=1}^{(n_{c} - n_{c-1})} \prod_{j=1}^{(n_{p+1} - n_{p})} \left( (j-i) + \frac{(n_{c} - n_{c-1})- (n_{p+1} - n_{p})}{2} + i \alpha (p + 1 - c) \right) \\
    & \qquad =  2^{- n_{p} } \prod_{c=1}^{p} \left[ \left( (\ell_{p+1} - \ell_{c})!! \right)^{\ell_{c}} \prod_{k=1}^{\ell_{c} - 1} \left( \frac{\ell_{p+1} - \ell_{c}}{2} + k \right)^{\ell_{c} - k} \right]^{2}  \\
    & \qquad \qquad \times \prod_{\substack{ c \in \{1 , \ldots, p\} \\ (\ell_{c} - \ell_{p+1}) \equiv 0 \: (\textnormal{mod 2}) }} \left( \frac{i}{\pi} \right)^{\ell_{c}} (-1)^{\frac{\ell_{c}}{2}(\ell_{p+1}-1)} 2^{-(\ell_{p+1}-\ell_{c}) \ell_{c}} \\
    & \qquad \qquad \times \prod_{\substack{ c \in \{1 , \ldots, p\}  \\ (\ell_{c} - \ell_{p+1}) \equiv 1 \: (\textnormal{mod 2})} } (-1)^{\frac{\ell_{c} \ell_{p+1}}{2}} 2^{-(\ell_{p+1} - \ell_{c}+1) \ell_{c}} \: ,
\end{split}
\end{equation}
again of the desired form. This establishes the claim. 
\end{proof}

We can use the above claim in an inductive argument to establish
\begin{equation}
    \begin{split}
        & \lim_{\alpha \rightarrow 0} H_{(n_{1}, \ldots, n_{N_{5}})}(\alpha)  \left( \prod_{c < d}^{N_{5}} \prod_{i =1}^{\ell_{c}} \prod_{j = 1}^{\ell_{d}} \left( (j - i) + \frac{(\ell_{c} - \ell_{d})}{2} + i \alpha (d-c) \right) \right) \\
        & \qquad = 2^{-\sum_{i=1}^{N_{5}-1} n_{i}} \prod_{c<d} \left[ \left( (\ell_{d} - \ell_{c})!! \right)^{\ell_{c}} \prod_{k=1}^{\ell_{c}-1} \left( \frac{\ell_{d} - \ell_{c}}{2} + k \right)^{\ell_{c} - k} \right]^{2}
        \\
        & \qquad \qquad
        \times \prod_{\{c<d :\ell_{cd} \equiv 0 \: (\textnormal{mod 2}) \}} \left( \frac{i}{\pi} \right)^{\ell_{c}} (-1)^{\frac{\ell_{c}}{2}(\ell_{d}-1)} 2^{-(\ell_{d}-\ell_{c}) \ell_{c}} \\
        & \qquad \qquad \times \prod_{\{ c< d : \ell_{cd} \equiv 1 \: (\textnormal{mod 2}) \}}  (-1)^{\frac{\ell_{c} \ell_{d}}{2}} 2^{-(\ell_{d} - \ell_{c}+1) \ell_{c}} 
    \end{split}
\end{equation}
Indeed, the base case $N_{5} = 2$ coincides with the base case of the previous claim. 
Now, suppose that the claim holds for some $N_{5}$. Then we have by induction hypothesis 
\begin{equation}
    \begin{split}
        & \lim_{\alpha \rightarrow 0} H_{(n_{1}, \ldots, n_{N_{5} + 1})}(\alpha)  \left( \prod_{c < d}^{N_{5}+1} \prod_{i =1}^{\ell_{c}} \prod_{j = 1}^{\ell_{d}} \left( (j - i) + \frac{(\ell_{c} - \ell_{d})}{2} + i \alpha (d-c) \right) \right) \\
        & \qquad = 
        2^{-\sum_{i=1}^{N_{5}-1} n_{i}} \prod_{c<d} \left[ \left( (\ell_{d} - \ell_{c})!! \right)^{\ell_{c}} \prod_{k=1}^{\ell_{c}-1} \left( \frac{\ell_{d} - \ell_{c}}{2} + k \right)^{\ell_{c} - k} \right]^{2}
        \\
        & \qquad \qquad 
        \times \prod_{\{c<d :\ell_{cd} \equiv 0 \: (\textnormal{mod 2}) \}} \left( \frac{i}{\pi} \right)^{\ell_{c}} (-1)^{\frac{\ell_{c}}{2}(\ell_{d}-1)} 2^{-(\ell_{d}-\ell_{c}) \ell_{c}} \\
        & \qquad \qquad \times \prod_{\{ c< d : \ell_{cd} \equiv 1 \: (\textnormal{mod 2}) \}}  (-1)^{\frac{\ell_{c} \ell_{d}}{2}} 2^{-(\ell_{d} - \ell_{c}+1) \ell_{c}} \\
        & \qquad \qquad \times \lim_{\alpha \rightarrow 0} \tilde{H}^{N_{5}}_{n_{1}, \ldots, n_{N_{5}+1}}(\alpha) \prod_{c=1}^{N_{5}} \prod_{i=1}^{\ell_{c}} \prod_{j=1}^{\ell_{N_{5}+1}} \left( (j-i) + \frac{\ell_{d} - \ell_{c}}{2} + i \alpha (N_{5} + 1 - c) \right) \: ,
    \end{split}
\end{equation}
so the previous claim provides the desired result. 
We may therefore deduce
\begin{equation}
\begin{split}
    Z[HS^{4}] & =  i^{ - \sum_{\ell=0}^{N_{5}-2} (n_{\ell+1} - n_{\ell})(n_{N_{5}} - n_{\ell + 1}) }  \left( \frac{\gym^{2}}{4 \pi} \right)^{ \frac{N^{2}}{2}} e^{ \frac{\gym^{2} }{48} \sum_{c=1}^{N_{5}} \ell_{c} (\ell_{c}-1)(\ell_{c}+1)} \\
    & \qquad \times \left( \prod_{c=1}^{N_{5}} G_{2}(\ell_{c}+1) \right) 2^{-\sum_{i=1}^{N_{5}-1} n_{i}} \prod_{c<d}^{N_{5}} \left[ \left( (\ell_{d} - \ell_{c})!! \right)^{\ell_{c}} \prod_{k=1}^{\ell_{c}-1} \left( \frac{\ell_{d} - \ell_{c}}{2} + k \right)^{\ell_{c} - k} \right]^{2}
    \\
    & \qquad 
    \times \prod_{\{c<d :\ell_{cd} \equiv 0 \: (\textnormal{mod 2}) \}} \left( \frac{i}{\pi} \right)^{\ell_{c}} (-1)^{\frac{\ell_{c}}{2}(\ell_{d}-1)} 2^{-(\ell_{d}-\ell_{c}) \ell_{c}} \\
    & \qquad \times \prod_{\{ c< d : \ell_{cd} \equiv 1 \: (\textnormal{mod 2}) \}}  (-1)^{\frac{\ell_{c} \ell_{d}}{2}} 2^{-(\ell_{d} - \ell_{c}+1) \ell_{c}}  \: .
    \end{split}
\end{equation}
If we denote
\begin{equation}
    \epsilon_{cd} \equiv \begin{cases}
        0 & \ell_{cd} \equiv 0 \: \textnormal{(mod 2)} \\
        1 & \ell_{cd} \equiv 1 \: \textnormal{(mod 2)}
    \end{cases} \: ,
\end{equation}
then we can write
\begin{equation}
\begin{split}
    Z[HS^{4}] & =  i^{ - \sum_{\ell=0}^{N_{5}-2} (n_{\ell+1} - n_{\ell})(n_{N_{5}} - n_{\ell + 1}) }  \left( \frac{\gym^{2}}{4 \pi} \right)^{ \frac{N^{2}}{2}} e^{ \frac{\gym^{2} }{48} \sum_{c=1}^{N_{5}} \ell_{c} (\ell_{c}-1)(\ell_{c}+1)} \\
    & \qquad \times \left( \prod_{c=1}^{N_{5}} G_{2}(\ell_{c}+1) \right) 2^{-\sum_{i=1}^{N_{5}-1} n_{i}} \prod_{c<d}^{N_{5}} \Big[ \left( \left( (\ell_{d} - \ell_{c})!! \right)^{\ell_{c}} \prod_{k=1}^{\ell_{c}-1} \left( \frac{\ell_{d} - \ell_{c}}{2} + k \right)^{\ell_{c} - k} \right)^{2}
    \\
    & \qquad \times (-1)^{\frac{\ell_{c} \ell_{d}}{2}} \pi^{- (1 - \epsilon_{cd}) \ell_{c}} 2^{-(\ell_{d}-\ell_{c} + \epsilon_{cd}) \ell_{c}} \Big] \: ,
    \end{split}
\end{equation}
that is,
\begin{equation}
\begin{split}
    Z[HS^{4}] & =  (2 \pi)^{-\sum_{i=1}^{N_{5}-1} n_{i}} \left( \frac{\gym^{2}}{4 \pi} \right)^{ \frac{N^{2}}{2}} e^{ \frac{\gym^{2} }{48} \sum_{c=1}^{N_{5}} \ell_{c} (\ell_{c}-1)(\ell_{c}+1)}  \left( \prod_{c=1}^{N_{5}} G_{2}(\ell_{c}+1) \right) \\
    & \qquad \times  \prod_{c<d}^{N_{5}} \left[ 2^{-(\ell_{d} - \ell_{c}) \ell_{c}} \left( \frac{\pi}{2} \right)^{\epsilon_{cd} \ell_{c}} \left( \left( (\ell_{d} - \ell_{c})!! \right)^{\ell_{c}} \prod_{k=1}^{\ell_{c}-1} \left( \frac{\ell_{d} - \ell_{c}}{2} + k \right)^{\ell_{c} - k} \right)^{2}  \right] \: .
    \end{split}
\end{equation}

\section{Statistics of Boundary \texorpdfstring{$F$}{}: Details}
\label{sec:StatDetails}


To understand the behaviour of $F_{\partial}^{\textnormal{SUGRA}}$, which is easier to analyze analytically than $F_{\partial}$ and provides a good approximation for large $N$ and suitable linking numbers, we will momentarily consider the contribution to the $\lambda$-independent term in $F_{\partial}^{\textnormal{SUGRA}}$, proportional to
\begin{equation}
    F_{0}(p_{A}) \equiv \sum_{A, B} \Big[ (p_{A} + p_{B})^{2} \ln \big( (p_{A} + p_{B})^{2} \big) - (p_{A} - p_{B})^{2} \ln \big( (p_{A} - p_{B})^{2} \big) \Big] \: ,
\end{equation}
where $p_{A} = L_{A} / N$ for D5-branes or $p_{A} = K_{A} / N$ for NS5-branes. 
Using concavity of the logarithm, we find inequality
\begin{equation}
    \begin{split}
        F_{0}(p_{A}) & \geq \sum_{A, B} \left[ (p_{A} + p_{B})^{2} \ln \big( (p_{A} + p_{B})^{2} \big) - (p_{A} - p_{B})^{2} \ln \big( (p_{A} + p_{B})^{2} \big) \right] \\
        & = 8 \sum_{A, B} p_{A} p_{B} \ln  (p_{A} + p_{B}) \geq 8 \sum_{A, B} p_{A} p_{B} \left( \ln 2 + \frac{1}{2} \ln p_{A} + \frac{1}{2} \ln p_{B} \right) \\
        & = 8 \ln 2 + 8 \sum_{A} p_{A} \ln p_{A} = 8 \ln 2 - 8 S (p_{A}) \: ,
    \end{split}
\end{equation}
where $S(p_{A})$ is the classical entropy of the probability distribution. The smallest possible value for the right hand side of our inequality is $8 \ln (2 / N)$, realized on the maximum entropy distribution
\begin{equation}
    p_{1} = \ldots = p_{N} = \frac{1}{N} \: .
\end{equation}
And in fact, for this particular distribution, the inequality is saturated and one finds
\begin{equation}
    F_{0}\left( p_{1} = \ldots = p_{N} = \frac{1}{N} \right) = 8 \ln (2 / N) \: .
\end{equation}
We may therefore deduce that $F_{0}(p_{A})$ is minimized for the maximum entropy probability distribution. 

On the other hand, we note that if $p_{1}, p_{2} \leq \frac{1}{2}$, then
\begin{equation}
    \begin{split}
        0 \geq \Big[ (p_{1} + p_{2})^{2} \ln \big( (p_{1} + p_{2})^{2} \big) - (p_{1} - p_{2})^{2} \ln \big( (p_{1} - p_{2})^{2} \big) \Big] \: ,
    \end{split}
\end{equation}
while if $\frac{1}{2} \leq p_{1} \leq 1$ and $0 < p_{2} \leq 1 - p_{1}$ then
\begin{equation}
    \begin{split}
        8 p_{1} p_{2} \ln (4 p_{1}^{2}) \geq 2 \Big[ (p_{1} + p_{2})^{2} \ln \big( (p_{1} + p_{2})^{2} \big) - (p_{1} - p_{2})^{2} \ln \big( (p_{1} - p_{2})^{2} \big) \Big] + 4 p_{2}^{2} \ln (4 p_{2}^{2}) \: .
    \end{split}
\end{equation}
Consequently, one finds that if the distribution $\{p_{A}\}$ has $p_{1}, \ldots, p_{N} \leq \frac{1}{2}$, then
\begin{equation}
    F_{0}(p_{A}) \leq 0 \: ,
\end{equation}
whereas if $p_{1} \geq \frac{1}{2}$ and $p_{2}, \ldots, p_{N} \leq \frac{1}{2}$, then
\begin{equation}
     \begin{split}
         F_{0}(p_{A}) & \leq 4 p_{1}^{2} \ln (4 p_{1}^{2}) + 2 \sum_{A > 1} \Big[ (p_{1} + p_{A})^{2} \ln \big( (p_{1} + p_{A})^{2} \big) - (p_{1} - p_{A})^{2} \ln \big( (p_{1} - p_{A})^{2} \big) \Big] \\
         & \qquad + 4 \sum_{A > 1} p_{A}^{2} \ln (4 p_{A}^{2}) + 2 \sum_{B > A > 1} \Big[ (p_{A} + p_{B})^{2} \ln \big( (p_{A} + p_{B})^{2} \big) - (p_{A} - p_{B})^{2} \ln \big( (p_{A} - p_{B})^{2} \big) \Big] \\
         & \leq 4 p_{1}^{2} \ln (4 p_{1}^{2}) + 4 \sum_{A > 1} p_{A}^{2} \ln (4 p_{A}^{2})  \\
         & \qquad + 2 \sum_{A > 1} \Big[ (p_{1} + p_{A})^{2} \ln \big( (p_{1} + p_{A})^{2} \big) - (p_{1} - p_{A})^{2} \ln \big( (p_{1} - p_{A})^{2} \big) \Big] \\
         & \leq 4 p_{1}^{2} \ln (4 p_{1}^{2}) + 8 p_{1} (1 - p_{1}) \ln (4 p_{1}^{2}) = 4 p_{1}(2-p_{1}) \ln (4 p_{1}^{2}) \: .
     \end{split}
\end{equation}
The right hand side of this inequality is a monotonically increasing function, so it is maximized at $p_{1} = 1$, where it is equal to $4 \ln 4$. In fact, the minimum entropy distribution
\begin{equation}
    p_{1} = 1 \: , \qquad p_{2} = \ldots = p_{N} = 0
\end{equation}
saturates this inequality, and one can see that
\begin{equation}
    F_{0}\left( p_{1} = 1, p_{2} = \ldots = p_{N} = 0 \right) = 4 \ln 4 \: .
\end{equation}
Thus, $F_{0}(p_{A})$ is maximized for the minimum entropy probability distribution. 

We can apply these considerations to determine for which boundary conditions consisting of D5-branes only or NS5-branes only $F_{\partial}^{\textnormal{SUGRA}}$ will be maximized or minimized. For D5-brane boundary conditions, we found
\begin{equation}
    F_{\partial}^{\textnormal{SUGRA}} = \frac{N^{2}}{4} \left( \frac{3}{2} + \ln \left( \frac{\lambda}{4 \pi^{2} N^{2}} \right) \right) - \frac{\pi^{2} N^{4}}{3 \lambda} \sum_{A} p_{A}^{3} - \frac{N^{2}}{16} F_{0}(p_{A}) \: .
\end{equation}
The term in parentheses is independent of the choice of boundary condition, while the remaining terms are both minimized (maximized) on the minimum (maximum) entropy probability distributions. Thus, we can conclude that $F_{\partial}^{\textnormal{SUGRA}}$ is minimized (maximized) on the minimum (maximum) entropy probability distributions. 
Similarly, for NS5-brane boundary conditions, we found
\begin{equation}
    F_{\partial}^{\textnormal{SUGRA}} = \frac{N^{2}}{4} \left( \frac{3}{2} + \ln \left( \frac{4}{\lambda} \right) \right) - \frac{\lambda N^{2}}{48} \sum_{A} p_{A}^{3} - \frac{N^{2}}{16} F_{0}(p_{A}) \: ,
\end{equation}
so $F_{\partial}^{\textnormal{SUGRA}}$ is again minimized (maximized) on the minimum (maximum) entropy probability distributions.

\section{Calculation of Boundary \texorpdfstring{$F$}{} in a Bottom-Up Model}
\label{BottomUp}

In this appendix, we will compute the boundary $F$ in a bottom-up holographic model of a BCFT where the boundary in the CFT gives rise to an end-of-the-world (ETW) brane with tension $T$. Here, the vacuum solution may be described as a portion of pure AdS spacetime described by $x/z < \frac{T}{\sqrt{1 - T^{2}}}$ in Fefferman-Graham coordinates, with an ETW brane at $x/z = \frac{T}{\sqrt{1 - T^{2}}}$ \cite{Takayanagi:2011zk}. Defining $z = w \cos(\theta)$ and $x = w \sin(\theta)$, we can write the metric as
\be
ds^2 = {L^{2} d \theta^2 \over \cos^2(\theta)} + {L^2 \over w^2 \cos^2 \theta}(dw^2 - dt^2 + dx_\perp^2) \; ,
\ee
and the ETW now appears at $\theta = \arcsin(T)$. 
The extremal surface corresponds to the hemisphere $w = R$; using the result (\ref{AdS4area}), we have that the regulated area of the extremal surface is
\be
L^3 \int_{-\pi/2 + \arcsin(\epsilon/R)}^{\arcsin(T)} {d \theta \over \cos^3 \theta} 2 \pi \left( {R \cos(\theta) \over \epsilon} - 1 \right)
\ee
From this, we need to subtract off half the regulated area of the hemispherical surface in pure AdS corresponding to a boundary ball of radius $R$. This area is
\bea
\label{subtract}
{\rm Area_{AdS}} &=& L^3 \int_0^{\cos^{-1}{\epsilon \over R}} d \theta {4 \pi \sin^2 \theta \over \cos^3 \theta} \cr
&=& L^3 \left( {2 \pi R^2 \over \epsilon^2} - 2 \pi \ln {2 R \over \epsilon} - \pi + {\cal O}(\epsilon^2)\right) \; .
\eea
Using these results and applying the definitions (\ref{defF1},\ref{defF2},\ref{defF3}), we find that
\be
F_{\partial} = {L^3 \pi \over 4G}\left( {T \over 1 - T^2} + {1 \over 2} \ln{1 + T \over 1 - T}\right) \; .
\ee
This gives a monotonic relation between boundary $F$ and the tension parameter $T$, where $F_{\partial}$ is an odd function of $T$ and where $F_{\partial} \to \pm \infty$ for $T \to \pm 1$.

\bibliographystyle{JHEP}
\bibliography{BCFT}

\providecommand{\href}[2]{#2}\begingroup\raggedright\begin{thebibliography}{10}

\bibitem{Zamolodchikov:1986gt}
A.~Zamolodchikov, \emph{{Irreversibility of the Flux of the Renormalization
  Group in a 2D Field Theory}}, {\emph{JETP Lett.} {\bf 43} (1986) 730--732}.

\bibitem{Jafferis:2011zi}
D.~L. Jafferis, I.~R. Klebanov, S.~S. Pufu and B.~R. Safdi, \emph{{Towards the
  F-Theorem: N=2 Field Theories on the Three-Sphere}},
  \href{http://dx.doi.org/10.1007/JHEP06(2011)102}{\emph{JHEP} {\bf 06} (2011)
  102}, [\href{https://arxiv.org/abs/1103.1181}{{\tt 1103.1181}}].

\bibitem{Casini:2012ei}
H.~Casini and M.~Huerta, \emph{{On the RG running of the entanglement entropy
  of a circle}},
  \href{http://dx.doi.org/10.1103/PhysRevD.85.125016}{\emph{Phys. Rev. D} {\bf
  85} (2012) 125016}, [\href{https://arxiv.org/abs/1202.5650}{{\tt
  1202.5650}}].

\bibitem{Cardy:1988cwa}
J.~L. Cardy, \emph{{Is There a c Theorem in Four-Dimensions?}},
  \href{http://dx.doi.org/10.1016/0370-2693(88)90054-8}{\emph{Phys. Lett. B}
  {\bf 215} (1988) 749--752}.

\bibitem{Komargodski:2011vj}
Z.~Komargodski and A.~Schwimmer, \emph{{On Renormalization Group Flows in Four
  Dimensions}}, \href{http://dx.doi.org/10.1007/JHEP12(2011)099}{\emph{JHEP}
  {\bf 12} (2011) 099}, [\href{https://arxiv.org/abs/1107.3987}{{\tt
  1107.3987}}].

\bibitem{Giombi:2014xxa}
S.~Giombi and I.~R. Klebanov, \emph{{Interpolating between $a$ and $F$}},
  \href{http://dx.doi.org/10.1007/JHEP03(2015)117}{\emph{JHEP} {\bf 03} (2015)
  117}, [\href{https://arxiv.org/abs/1409.1937}{{\tt 1409.1937}}].

\bibitem{Klebanov:2011gs}
I.~R. Klebanov, S.~S. Pufu and B.~R. Safdi, \emph{{F-Theorem without
  Supersymmetry}}, \href{http://dx.doi.org/10.1007/JHEP10(2011)038}{\emph{JHEP}
  {\bf 10} (2011) 038}, [\href{https://arxiv.org/abs/1105.4598}{{\tt
  1105.4598}}].

\bibitem{Affleck:1991tk}
I.~Affleck and A.~W.~W. Ludwig, \emph{{Universal noninteger 'ground state
  degeneracy' in critical quantum systems}},
  \href{http://dx.doi.org/10.1103/PhysRevLett.67.161}{\emph{Phys. Rev. Lett.}
  {\bf 67} (1991) 161--164}.

\bibitem{Friedan:2003yc}
D.~Friedan and A.~Konechny, \emph{{On the boundary entropy of one-dimensional
  quantum systems at low temperature}},
  \href{http://dx.doi.org/10.1103/PhysRevLett.93.030402}{\emph{Phys. Rev.
  Lett.} {\bf 93} (2004) 030402},
  [\href{https://arxiv.org/abs/hep-th/0312197}{{\tt hep-th/0312197}}].

\bibitem{Jensen:2013lxa}
K.~Jensen and A.~O'Bannon, \emph{{Holography, Entanglement Entropy, and
  Conformal Field Theories with Boundaries or Defects}},
  \href{http://dx.doi.org/10.1103/PhysRevD.88.106006}{\emph{Phys. Rev. D} {\bf
  88} (2013) 106006}, [\href{https://arxiv.org/abs/1309.4523}{{\tt
  1309.4523}}].

\bibitem{Estes:2014hka}
J.~Estes, K.~Jensen, A.~O'Bannon, E.~Tsatis and T.~Wrase, \emph{{On Holographic
  Defect Entropy}},
  \href{http://dx.doi.org/10.1007/JHEP05(2014)084}{\emph{JHEP} {\bf 05} (2014)
  084}, [\href{https://arxiv.org/abs/1403.6475}{{\tt 1403.6475}}].

\bibitem{Cardy:1984bb}
J.~L. Cardy, \emph{{Conformal Invariance and Surface Critical Behavior}},
  \href{http://dx.doi.org/10.1016/0550-3213(84)90241-4}{\emph{Nucl. Phys. B}
  {\bf 240} (1984) 514--532}.

\bibitem{McAvity:1995zd}
D.~McAvity and H.~Osborn, \emph{{Conformal field theories near a boundary in
  general dimensions}},
  \href{http://dx.doi.org/10.1016/0550-3213(95)00476-9}{\emph{Nucl. Phys. B}
  {\bf 455} (1995) 522--576},
  [\href{https://arxiv.org/abs/cond-mat/9505127}{{\tt cond-mat/9505127}}].

\bibitem{Liendo:2012hy}
P.~Liendo, L.~Rastelli and B.~C. van Rees, \emph{{The Bootstrap Program for
  Boundary CFT\_d}},
  \href{http://dx.doi.org/10.1007/JHEP07(2013)113}{\emph{JHEP} {\bf 07} (2013)
  113}, [\href{https://arxiv.org/abs/1210.4258}{{\tt 1210.4258}}].

\bibitem{Mazac:2018biw}
D.~Maz\'a\v{c}, L.~Rastelli and X.~Zhou, \emph{{An analytic approach to
  BCFT$_{d}$}}, \href{http://dx.doi.org/10.1007/JHEP12(2019)004}{\emph{JHEP}
  {\bf 12} (2019) 004}, [\href{https://arxiv.org/abs/1812.09314}{{\tt
  1812.09314}}].

\bibitem{Nozaki:2012qd}
M.~Nozaki, T.~Takayanagi and T.~Ugajin, \emph{{Central Charges for BCFTs and
  Holography}}, \href{http://dx.doi.org/10.1007/JHEP06(2012)066}{\emph{JHEP}
  {\bf 06} (2012) 066}, [\href{https://arxiv.org/abs/1205.1573}{{\tt
  1205.1573}}].

\bibitem{Gaiotto:2014gha}
D.~Gaiotto, \emph{{Boundary F-maximization}},
  \href{https://arxiv.org/abs/1403.8052}{{\tt 1403.8052}}.

\bibitem{Kobayashi:2018lil}
N.~Kobayashi, T.~Nishioka, Y.~Sato and K.~Watanabe, \emph{{Towards a
  $C$-theorem in defect CFT}},
  \href{http://dx.doi.org/10.1007/JHEP01(2019)039}{\emph{JHEP} {\bf 01} (2019)
  039}, [\href{https://arxiv.org/abs/1810.06995}{{\tt 1810.06995}}].

\bibitem{Giombi:2020rmc}
S.~Giombi and H.~Khanchandani, \emph{{CFT in AdS and boundary RG flows}},
  \href{https://arxiv.org/abs/2007.04955}{{\tt 2007.04955}}.

\bibitem{Casini:2016fgb}
H.~Casini, I.~Salazar~Landea and G.~Torroba, \emph{{The g-theorem and quantum
  information theory}},
  \href{http://dx.doi.org/10.1007/JHEP10(2016)140}{\emph{JHEP} {\bf 10} (2016)
  140}, [\href{https://arxiv.org/abs/1607.00390}{{\tt 1607.00390}}].

\bibitem{Jensen:2015swa}
K.~Jensen and A.~O'Bannon, \emph{{Constraint on Defect and Boundary
  Renormalization Group Flows}},
  \href{http://dx.doi.org/10.1103/PhysRevLett.116.091601}{\emph{Phys. Rev.
  Lett.} {\bf 116} (2016) 091601},
  [\href{https://arxiv.org/abs/1509.02160}{{\tt 1509.02160}}].

\bibitem{Casini:2018nym}
H.~Casini, I.~Salazar~Landea and G.~Torroba, \emph{{Irreversibility in quantum
  field theories with boundaries}},
  \href{http://dx.doi.org/10.1007/JHEP04(2019)166}{\emph{JHEP} {\bf 04} (2019)
  166}, [\href{https://arxiv.org/abs/1812.08183}{{\tt 1812.08183}}].

\bibitem{Gaiotto:2008sa}
D.~Gaiotto and E.~Witten, \emph{{Supersymmetric Boundary Conditions in N=4
  Super Yang-Mills Theory}},
  \href{http://dx.doi.org/10.1007/s10955-009-9687-3}{\emph{J. Statist. Phys.}
  {\bf 135} (2009) 789--855}, [\href{https://arxiv.org/abs/0804.2902}{{\tt
  0804.2902}}].

\bibitem{Gaiotto:2008ak}
D.~Gaiotto and E.~Witten, \emph{{S-Duality of Boundary Conditions In N=4 Super
  Yang-Mills Theory}},
  \href{http://dx.doi.org/10.4310/ATMP.2009.v13.n3.a5}{\emph{Adv. Theor. Math.
  Phys.} {\bf 13} (2009) 721--896},
  [\href{https://arxiv.org/abs/0807.3720}{{\tt 0807.3720}}].

\bibitem{DHoker:2007zhm}
E.~D'Hoker, J.~Estes and M.~Gutperle, \emph{{Exact half-BPS Type IIB interface
  solutions. I. Local solution and supersymmetric Janus}},
  \href{http://dx.doi.org/10.1088/1126-6708/2007/06/021}{\emph{JHEP} {\bf 06}
  (2007) 021}, [\href{https://arxiv.org/abs/0705.0022}{{\tt 0705.0022}}].

\bibitem{DHoker:2007hhe}
E.~D'Hoker, J.~Estes and M.~Gutperle, \emph{{Exact half-BPS Type IIB interface
  solutions. II. Flux solutions and multi-Janus}},
  \href{http://dx.doi.org/10.1088/1126-6708/2007/06/022}{\emph{JHEP} {\bf 06}
  (2007) 022}, [\href{https://arxiv.org/abs/0705.0024}{{\tt 0705.0024}}].

\bibitem{Aharony:2011yc}
O.~Aharony, L.~Berdichevsky, M.~Berkooz and I.~Shamir, \emph{{Near-horizon
  solutions for D3-branes ending on 5-branes}},
  \href{http://dx.doi.org/10.1103/PhysRevD.84.126003}{\emph{Phys. Rev.} {\bf
  D84} (2011) 126003}, [\href{https://arxiv.org/abs/1106.1870}{{\tt
  1106.1870}}].

\bibitem{Assel:2011xz}
B.~Assel, C.~Bachas, J.~Estes and J.~Gomis, \emph{{Holographic Duals of D=3 N=4
  Superconformal Field Theories}},
  \href{http://dx.doi.org/10.1007/JHEP08(2011)087}{\emph{JHEP} {\bf 08} (2011)
  087}, [\href{https://arxiv.org/abs/1106.4253}{{\tt 1106.4253}}].

\bibitem{Assel:2012cp}
B.~Assel, J.~Estes and M.~Yamazaki, \emph{{Large N Free Energy of 3d N=4 SCFTs
  and $AdS_4/CFT_3$}},
  \href{http://dx.doi.org/10.1007/JHEP09(2012)074}{\emph{JHEP} {\bf 09} (2012)
  074}, [\href{https://arxiv.org/abs/1206.2920}{{\tt 1206.2920}}].

\bibitem{Estes:2018tnu}
J.~Estes, D.~Krym, A.~O'Bannon, B.~Robinson and R.~Rodgers, \emph{{Wilson
  Surface Central Charge from Holographic Entanglement Entropy}},
  \href{http://dx.doi.org/10.1007/JHEP05(2019)032}{\emph{JHEP} {\bf 05} (2019)
  032}, [\href{https://arxiv.org/abs/1812.00923}{{\tt 1812.00923}}].

\bibitem{Goto:2020per}
K.~Goto, L.~Nagano, T.~Nishioka and T.~Okuda, \emph{{Janus interface entropy
  and Calabi's diastasis in four-dimensional $\mathcal{N}=2$ superconformal
  field theories}},  \href{https://arxiv.org/abs/2005.10833}{{\tt 2005.10833}}.

\bibitem{Casini:2011kv}
H.~Casini, M.~Huerta and R.~C. Myers, \emph{{Towards a derivation of
  holographic entanglement entropy}},
  \href{http://dx.doi.org/10.1007/JHEP05(2011)036}{\emph{JHEP} {\bf 05} (2011)
  036}, [\href{https://arxiv.org/abs/1102.0440}{{\tt 1102.0440}}].

\bibitem{Myers:2010xs}
R.~C. Myers and A.~Sinha, \emph{{Seeing a c-theorem with holography}},
  \href{http://dx.doi.org/10.1103/PhysRevD.82.046006}{\emph{Phys. Rev. D} {\bf
  82} (2010) 046006}, [\href{https://arxiv.org/abs/1006.1263}{{\tt
  1006.1263}}].

\bibitem{Myers:2010tj}
R.~C. Myers and A.~Sinha, \emph{{Holographic c-theorems in arbitrary
  dimensions}}, \href{http://dx.doi.org/10.1007/JHEP01(2011)125}{\emph{JHEP}
  {\bf 01} (2011) 125}, [\href{https://arxiv.org/abs/1011.5819}{{\tt
  1011.5819}}].

\bibitem{Kawano:2014moa}
T.~Kawano, Y.~Nakaguchi and T.~Nishioka, \emph{{Holographic Interpolation
  between $a$ and $F$}},
  \href{http://dx.doi.org/10.1007/JHEP12(2014)161}{\emph{JHEP} {\bf 12} (2014)
  161}, [\href{https://arxiv.org/abs/1410.5973}{{\tt 1410.5973}}].

\bibitem{Calabrese:2009qy}
P.~Calabrese and J.~Cardy, \emph{{Entanglement entropy and conformal field
  theory}}, \href{http://dx.doi.org/10.1088/1751-8113/42/50/504005}{\emph{J.
  Phys. A} {\bf 42} (2009) 504005},
  [\href{https://arxiv.org/abs/0905.4013}{{\tt 0905.4013}}].

\bibitem{Fursaev:2016inw}
D.~V. Fursaev and S.~N. Solodukhin, \emph{{Anomalies, entropy and boundaries}},
  \href{http://dx.doi.org/10.1103/PhysRevD.93.084021}{\emph{Phys. Rev. D} {\bf
  93} (2016) 084021}, [\href{https://arxiv.org/abs/1601.06418}{{\tt
  1601.06418}}].

\bibitem{Herzog:2015ioa}
C.~P. Herzog, K.-W. Huang and K.~Jensen, \emph{{Universal Entanglement and
  Boundary Geometry in Conformal Field Theory}},
  \href{http://dx.doi.org/10.1007/JHEP01(2016)162}{\emph{JHEP} {\bf 01} (2016)
  162}, [\href{https://arxiv.org/abs/1510.00021}{{\tt 1510.00021}}].

\bibitem{Yamaguchi:2002pa}
S.~Yamaguchi, \emph{{Holographic RG flow on the defect and g theorem}},
  \href{http://dx.doi.org/10.1088/1126-6708/2002/10/002}{\emph{JHEP} {\bf 10}
  (2002) 002}, [\href{https://arxiv.org/abs/hep-th/0207171}{{\tt
  hep-th/0207171}}].

\bibitem{Takayanagi:2011zk}
T.~Takayanagi, \emph{{Holographic Dual of BCFT}},
  \href{http://dx.doi.org/10.1103/PhysRevLett.107.101602}{\emph{Phys. Rev.
  Lett.} {\bf 107} (2011) 101602}, [\href{https://arxiv.org/abs/1105.5165}{{\tt
  1105.5165}}].

\bibitem{Fujita:2011fp}
M.~Fujita, T.~Takayanagi and E.~Tonni, \emph{{Aspects of AdS/BCFT}},
  \href{http://dx.doi.org/10.1007/JHEP11(2011)043}{\emph{JHEP} {\bf 11} (2011)
  043}, [\href{https://arxiv.org/abs/1108.5152}{{\tt 1108.5152}}].

\bibitem{Wang:2020seq}
Y.~Wang, \emph{{Taming Defects in $\mathcal{N}=4$ Super-Yang-Mills}},
  \href{https://arxiv.org/abs/2003.11016}{{\tt 2003.11016}}.

\bibitem{Constable:1999ac}
N.~R. Constable, R.~C. Myers and O.~Tafjord, \emph{{The Noncommutative bion
  core}}, \href{http://dx.doi.org/10.1103/PhysRevD.61.106009}{\emph{Phys. Rev.
  D} {\bf 61} (2000) 106009}, [\href{https://arxiv.org/abs/hep-th/9911136}{{\tt
  hep-th/9911136}}].

\bibitem{Hanany:1996ie}
A.~Hanany and E.~Witten, \emph{{Type IIB superstrings, BPS monopoles, and
  three-dimensional gauge dynamics}},
  \href{http://dx.doi.org/10.1016/S0550-3213(97)00157-0}{\emph{Nucl. Phys. B}
  {\bf 492} (1997) 152--190}, [\href{https://arxiv.org/abs/hep-th/9611230}{{\tt
  hep-th/9611230}}].

\bibitem{Ryu:2006bv}
S.~Ryu and T.~Takayanagi, \emph{{Holographic derivation of entanglement entropy
  from AdS/CFT}},
  \href{http://dx.doi.org/10.1103/PhysRevLett.96.181602}{\emph{Phys. Rev.
  Lett.} {\bf 96} (2006) 181602},
  [\href{https://arxiv.org/abs/hep-th/0603001}{{\tt hep-th/0603001}}].

\bibitem{Hubeny:2007xt}
V.~E. Hubeny, M.~Rangamani and T.~Takayanagi, \emph{{A Covariant holographic
  entanglement entropy proposal}},
  \href{http://dx.doi.org/10.1088/1126-6708/2007/07/062}{\emph{JHEP} {\bf 07}
  (2007) 062}, [\href{https://arxiv.org/abs/0705.0016}{{\tt 0705.0016}}].

\bibitem{Pestun:2016zxk}
V.~Pestun et~al., \emph{{Localization techniques in quantum field theories}},
  \href{http://dx.doi.org/10.1088/1751-8121/aa63c1}{\emph{J. Phys. A} {\bf 50}
  (2017) 440301}, [\href{https://arxiv.org/abs/1608.02952}{{\tt 1608.02952}}].

\bibitem{Pestun:2007rz}
V.~Pestun, \emph{{Localization of gauge theory on a four-sphere and
  supersymmetric Wilson loops}},
  \href{http://dx.doi.org/10.1007/s00220-012-1485-0}{\emph{Commun. Math. Phys.}
  {\bf 313} (2012) 71--129}, [\href{https://arxiv.org/abs/0712.2824}{{\tt
  0712.2824}}].

\bibitem{Gomis:2011pf}
J.~Gomis, T.~Okuda and V.~Pestun, \emph{{Exact Results for 't Hooft Loops in
  Gauge Theories on S$^4$}},
  \href{http://dx.doi.org/10.1007/JHEP05(2012)141}{\emph{JHEP} {\bf 05} (2012)
  141}, [\href{https://arxiv.org/abs/1105.2568}{{\tt 1105.2568}}].

\bibitem{Pestun:2009nn}
V.~Pestun, \emph{{Localization of the four-dimensional N=4 SYM to a two-sphere
  and 1/8 BPS Wilson loops}},
  \href{http://dx.doi.org/10.1007/JHEP12(2012)067}{\emph{JHEP} {\bf 12} (2012)
  067}, [\href{https://arxiv.org/abs/0906.0638}{{\tt 0906.0638}}].

\bibitem{Hama:2012bg}
N.~Hama and K.~Hosomichi, \emph{{Seiberg-Witten Theories on Ellipsoids}},
  \href{http://dx.doi.org/10.1007/JHEP09(2012)033}{\emph{JHEP} {\bf 09} (2012)
  033}, [\href{https://arxiv.org/abs/1206.6359}{{\tt 1206.6359}}].

\bibitem{Hosomichi:2016flq}
K.~Hosomichi, \emph{{${{{\mathcal N}}=2}$ SUSY gauge theories on S$^4$}},
  \href{http://dx.doi.org/10.1088/1751-8121/aa7775}{\emph{J. Phys. A} {\bf 50}
  (2017) 443010}, [\href{https://arxiv.org/abs/1608.02962}{{\tt 1608.02962}}].

\bibitem{Kapustin:2009kz}
A.~Kapustin, B.~Willett and I.~Yaakov, \emph{{Exact Results for Wilson Loops in
  Superconformal Chern-Simons Theories with Matter}},
  \href{http://dx.doi.org/10.1007/JHEP03(2010)089}{\emph{JHEP} {\bf 03} (2010)
  089}, [\href{https://arxiv.org/abs/0909.4559}{{\tt 0909.4559}}].

\bibitem{Kapustin:2010xq}
A.~Kapustin, B.~Willett and I.~Yaakov, \emph{{Nonperturbative Tests of
  Three-Dimensional Dualities}},
  \href{http://dx.doi.org/10.1007/JHEP10(2010)013}{\emph{JHEP} {\bf 10} (2010)
  013}, [\href{https://arxiv.org/abs/1003.5694}{{\tt 1003.5694}}].

\bibitem{Benvenuti:2011ga}
S.~Benvenuti and S.~Pasquetti, \emph{{3D-partition functions on the sphere:
  exact evaluation and mirror symmetry}},
  \href{http://dx.doi.org/10.1007/JHEP05(2012)099}{\emph{JHEP} {\bf 05} (2012)
  099}, [\href{https://arxiv.org/abs/1105.2551}{{\tt 1105.2551}}].

\bibitem{Nishioka:2011dq}
T.~Nishioka, Y.~Tachikawa and M.~Yamazaki, \emph{{3d Partition Function as
  Overlap of Wavefunctions}},
  \href{http://dx.doi.org/10.1007/JHEP08(2011)003}{\emph{JHEP} {\bf 08} (2011)
  003}, [\href{https://arxiv.org/abs/1105.4390}{{\tt 1105.4390}}].

\bibitem{Sugishita:2013jca}
S.~Sugishita and S.~Terashima, \emph{{Exact Results in Supersymmetric Field
  Theories on Manifolds with Boundaries}},
  \href{http://dx.doi.org/10.1007/JHEP11(2013)021}{\emph{JHEP} {\bf 11} (2013)
  021}, [\href{https://arxiv.org/abs/1308.1973}{{\tt 1308.1973}}].

\bibitem{Gava:2016oep}
E.~Gava, K.~Narain, M.~Muteeb and V.~Giraldo-Rivera, \emph{{$N = 2$ gauge
  theories on the hemisphere $HS^4$}},
  \href{http://dx.doi.org/10.1016/j.nuclphysb.2017.04.007}{\emph{Nucl. Phys. B}
  {\bf 920} (2017) 256--297}, [\href{https://arxiv.org/abs/1611.04804}{{\tt
  1611.04804}}].

\bibitem{Gupta:2019qlg}
R.~Kumar~Gupta, C.~P. Herzog and I.~Jeon, \emph{{Duality and Transport for
  Supersymmetric Graphene from the Hemisphere Partition Function}},
  \href{http://dx.doi.org/10.1007/JHEP05(2020)023}{\emph{JHEP} {\bf 05} (2020)
  023}, [\href{https://arxiv.org/abs/1912.09225}{{\tt 1912.09225}}].

\bibitem{Drukker:2010jp}
N.~Drukker, D.~Gaiotto and J.~Gomis, \emph{{The Virtue of Defects in 4D Gauge
  Theories and 2D CFTs}},
  \href{http://dx.doi.org/10.1007/JHEP06(2011)025}{\emph{JHEP} {\bf 06} (2011)
  025}, [\href{https://arxiv.org/abs/1003.1112}{{\tt 1003.1112}}].

\bibitem{Dedushenko:2018tgx}
M.~Dedushenko, \emph{{Gluing II: Boundary Localization and Gluing Formulas}},
  \href{https://arxiv.org/abs/1807.04278}{{\tt 1807.04278}}.

\bibitem{Komatsu:2020sup}
S.~Komatsu and Y.~Wang, \emph{{Non-perturbative Defect One-Point Functions in
  Planar $\mathcal{N}=4$ Super-Yang-Mills}},
  \href{https://arxiv.org/abs/2004.09514}{{\tt 2004.09514}}.

\bibitem{Dedushenko:2020vgd}
M.~Dedushenko and D.~Gaiotto, \emph{{Algebras, traces, and boundary correlators
  in $\mathcal{N}=4$ SYM}},  \href{https://arxiv.org/abs/2009.11197}{{\tt
  2009.11197}}.

\bibitem{Karch:2000ct}
A.~Karch and L.~Randall, \emph{{Locally localized gravity}},
  \href{http://dx.doi.org/10.1088/1126-6708/2001/05/008}{\emph{JHEP} {\bf 05}
  (2001) 008}, [\href{https://arxiv.org/abs/hep-th/0011156}{{\tt
  hep-th/0011156}}].

\bibitem{Marolf:2000cb}
D.~Marolf, \emph{{Chern-Simons terms and the three notions of charge}},  in
  \emph{{International Conference on Quantization, Gauge Theory, and Strings:
  Conference Dedicated to the Memory of Professor Efim Fradkin}}, pp.~312--320,
  6, 2000.
\newblock \href{https://arxiv.org/abs/hep-th/0006117}{{\tt hep-th/0006117}}.

\end{thebibliography}\endgroup

\end{document}